\documentclass{SCIS2024}
\usepackage{multirow}
\usepackage{makecell}
\usepackage{longtable}

\begin{document}
\bstctlcite{BSTcontrol}
\ArticleType{REVIEW}
\Year{2025}
\Month{}
\Vol{}
\No{}
\DOI{}
\ArtNo{}
\ReceiveDate{}
\ReviseDate{}
\AcceptDate{}
\OnlineDate{}

\title{Wireless large AI model: shaping the AI-empowered future of 6G and beyond}{WLAM}

\author[1\textdagger]{Fenghao ZHU}{}
\author[1\textdagger]{Xinquan WANG}{}
\author[3]{Siming JIANG}{}
\author[1]{Xinyi LI}{}
\author[1]{Maojun ZHANG}{}
\author[1]{\\Yixuan CHEN}{}
\author[1,2*]{Chongwen HUANG}{{chongwenhuang@zju.edu.cn}} 
\author[1,2]{Zhaohui YANG}{}
\author[1,2]{Xiaoming CHEN}{}
\author[1,2]{\\Zhaoyang ZHANG}{}
\author[1,2]{Richeng JIN}{}
\author[4]{Yongming HUANG}{}
\author[5]{Wei FENG}{} 
\author[7]{Tingting YANG}{}
\author[8]{\\Baoming BAI}{}
\author[6]{Feifei GAO}{}
\author[9,10]{Kun YANG}{}
\author[11]{Yuanwei LIU}{}
\author[12]{Sami MUHAIDAT}{}
\author[13]{\\Chau YUEN}{}
\author[11]{Kaibin HUANG}{}
\author[14,15]{Kai-Kit WONG}{}
\author[16]{Dusit NIYATO}{}
\author[17]{\\Ying-Chang LIANG}{}
\author[18]{Mérouane DEBBAH}{}

\AuthorMark{F. ZHU, X. WANG}

\AuthorCitation{F. ZHU, X. WANG} 

\contributions{Fenghao ZHU and Xinquan WANG contributed equally to this work.} 

\address[1]{College of Information Science and Electronic Engineering, Zhejiang University, Hangzhou {\rm 310027}, China}
\address[2]{Zhejiang Provincial Key Laboratory of Multi-Modal Communication Networks and Intelligent Information Processing, \\ and National Key Laboratory of Millimeter-Wave and Terahertz Remote Sensing, Hangzhou {\rm 310027}, China}
\address[3]{Guangdong Tobacco Maoming Co., Ltd., \rm{525000}, Maoming, China}
\address[4]{Information and Communications Engineering, Southeast University, Nanjing \rm{210096}, China}
\address[5]{Department of Electronic Engineering, Tsinghua University, Beijing \rm{100084}, China}
\address[6]{Department of Automation, Tsinghua University, Beijing \rm{100084}, China}
\address[7]{Pengcheng Laboratory, Shenzhen \rm{518066}, China}
\address[8]{State Key Laboratory of Integrated Service Networks, Xidian University, Xi'an \rm{710071}, China}
\address[9]{State Key Laboratory of Novel Software Technology, Nanjing University, Nanjing \rm{210008}, China}
\address[10]{ School of Intelligent Software and Engineering, Nanjing University (Suzhou Campus), Suzhou, \rm{215163}, China.}
\address[11]{ Department of Electrical and Computer Engineering, the University of Hong Kong, Hong Kong, China}
\address[12]{ KU 6G Research Center, Computer and Communication Engineering, Khalifa University, Abu Dhabi \rm{127788}, UAE}
\address[13]{ School of Electrical and Electronic Engineering, Nanyang Technological University, Singapore, Singapore}
\address[14]{ Department of Electronic and Electrical Engineering, University College London, Torrington Place, \rm{WC1E 7JE}, UK}
\address[15]{ Department of Electronic Engineering, Kyung Hee University, Yongin-si, Gyeonggi-do \rm{17104}, Korea}
\address[16]{ College of Computing and Data Science, Nanyang Technological University, Singapore \rm{639798}, Singapore}
\address[17]{ Institute for Fundamental and Frontier Sciences, University of Electronic Science and Technology of China, Chengdu, China}
\address[18]{ Research Institute for Digital Future, Khalifa University, \rm{127788} Abu Dhabi, UAE} 

\abstract{The emergence of sixth-generation and beyond communication systems is expected to fundamentally transform digital experiences through introducing unparalleled levels of intelligence, efficiency, and connectivity. A promising technology poised to enable this revolutionary vision is a wireless large AI model (WLAM), characterized by its exceptional capabilities in data processing, inference, and decision-making. In light of these remarkable capabilities, this paper provides a comprehensive survey of WLAM, explaining its fundamental principles, diverse applications, critical challenges, and future research opportunities. We begin by introducing the background of WLAM and analyzing the key synergies with wireless networks, emphasizing the mutual benefits. Subsequently, we explore the foundational characteristics of WLAM, delving into their unique relevance in wireless environments. Then, the role of WLAM in optimizing wireless communication systems across various use cases and the reciprocal benefits are systematically investigated. Furthermore, we discuss the integration of WLAM with emerging technologies, highlighting their potential to enable transformative capabilities and breakthroughs in wireless communication. Finally, we thoroughly examine the high-level challenges and discuss pivotal future research directions.}

\keywords{Large AI model, 6G communications, beyond 6G, edge intelligence, intelligent wireless communications}

\maketitle

\section{Introduction}
The advent of sixth-generation (6G) and beyond communication systems indicates a paradigm shift in wireless communications, envisioning a future characterized by unprecedented levels of intelligence, efficiency, and seamless connectivity \cite{dang2020should, 10054381, you2021towards}. To realize this ambitious vision and to navigate the escalating complexity of future wireless networks, novel technological paradigms are urgently needed. Among these, a wireless large AI model (WLAM) emerges as a pivotal technology, holding the potential to fundamentally reshape wireless communications \cite{shahid2025largescaleaitelecomcharting}. Distinguished by its sophisticated architectures and parameters at massive scales, WLAM offers unparalleled capabilities in data processing, inference, and decision-making, specifically tailored for the unique challenges and opportunities of wireless environments. By leveraging its inherent adaptability and generalization capabilities, WLAM can move beyond current narrow applications, offering the promise of establishing truly AI-empowered wireless networks capable of handling multifaceted tasks and evolving demands \cite{chen2024big}.
\par
To elaborate on this vision, AI-empowered 6G and beyond refers to future communication systems where AI is not merely an add-on feature but is fundamentally integrated into the core design, operation, and optimization of the network fabric. Unlike previous generations where AI might address specific tasks or optimize isolated functions, AI-empowered networks intrinsically leverage AI across all scenarios and designs. As envisioned in the comprehensive overview of 6G AI \cite{cui2025overview}, this deep integration aims to enhance wireless performance through AI applications such as optimization and management in physical, network and semantic layers, while utilizing the wireless infrastructure to efficiently support AI operations by enabling edge intelligence and supporting AI security. This holistic and AI-empowered approach targets unprecedented levels of network automation, self-optimization, resilience, dynamic resource allocation, and the delivery of highly personalized and context-aware services, defining the capabilities expected for the 6G and beyond.
\par
This survey provides a comprehensive exploration into how WLAM can shape the AI-empowered future of communications, and how it can be effectively developed and deployed to realize this potential. The analysis delves into its fundamental principles, diverse applications across wireless communication, existing challenges, and promising future research directions. We aim to provide a holistic understanding of this burgeoning field and its profound implications for the future of wireless communications.

\begin{figure*}[t]
	\begin{center}
		\centerline{\includegraphics[width=0.8\linewidth]{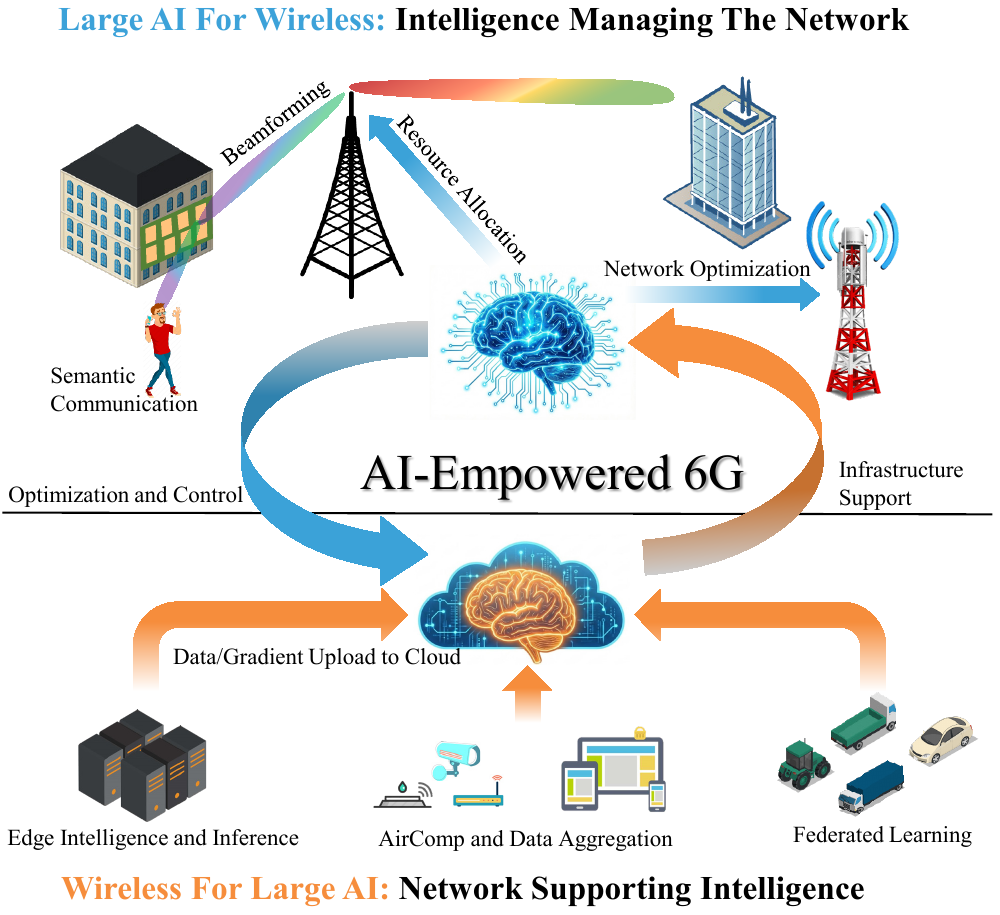}}
		\captionsetup{font=footnotesize, name={Fig.}, labelsep=period} 
		\caption{\, The synergistic relationship between large AI for wireless and wireless for large AI. The top half illustrates the top-down optimization where AI acts as the network brain, while the bottom half depicts the bottom-up infrastructure support provided by the wireless network for AI training and inference.}
		\label{synergy}
		\vspace{-8mm}
	\end{center}
\end{figure*}

\subsection{Background}
Wireless communication has become an indispensable part of modern society, supporting a vast array of applications ranging from personal devices to critical infrastructure. Moreover, large-scale AI is rapidly transforming industries by providing advanced solutions to complex problems. The convergence of these two powerful domains presents a transformative vision for the future, where wireless networks become not only more interconnected but also significantly more intelligent, efficient, and adaptive \cite{chen2026overview, 10750803}.

\begin{figure}[t]
\begin{center} 
\centerline{\includegraphics[width=0.65\linewidth]{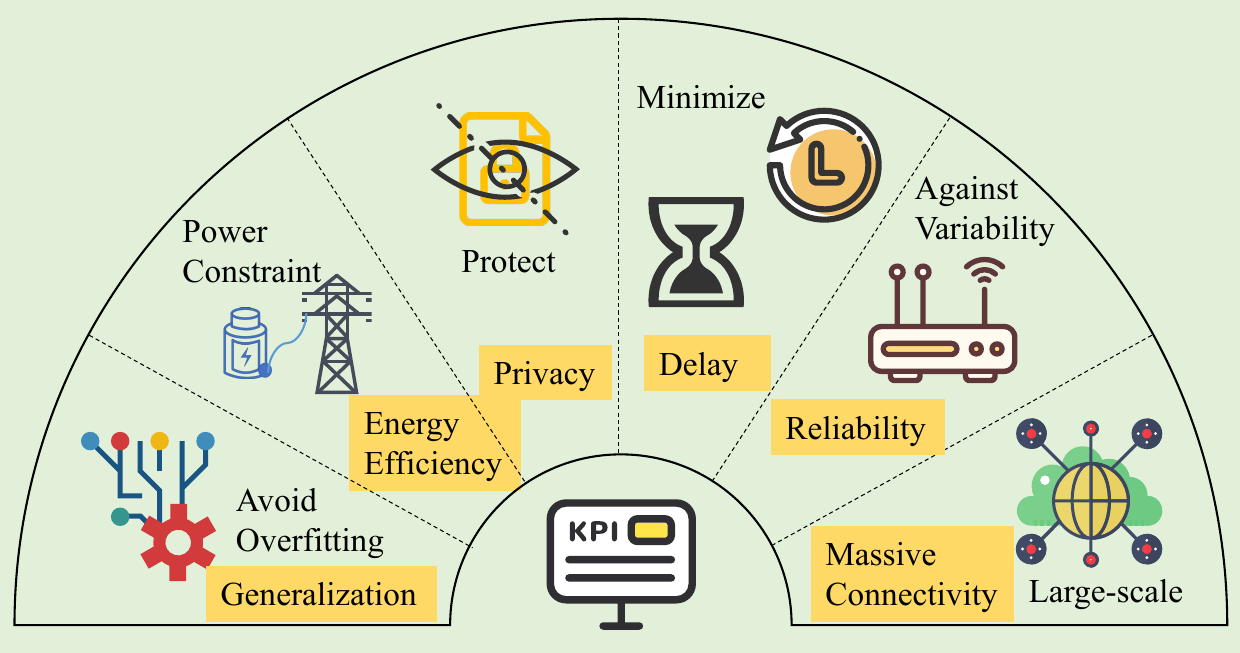}} 
\captionsetup{font=footnotesize, name={Fig.}, labelsep=period} 
\caption{The KPIs for WLAM. WLAM evaluation requires a holistic balance, which involves navigating the trade-offs among computation-communication delay, energy efficiency on edge devices, reliability under noise, robust generalization capabilities, strict privacy protection, and the capacity for massive connectivity in decentralized networks.} 
\label{kpi}
\vspace{-8mm} 
\end{center} 
\end{figure}

Conventionally, AI models in wireless networks are designed for specific tasks and scenarios, making them highly dependent on extensive data collection and customized training for each application. While effective in isolated cases, these conventional approaches face fundamental limitations that hinder their scalability. First, the task-specific paradigm creates isolated intelligence silos, where a model trained for channel estimation cannot generalize to beamforming, leading to excessive management overhead. Second, these models suffer from poor generalization in dynamic wireless environments, requiring frequent and costly retraining when deployment scenarios change. Third, conventional AI lacks the capability to process multi-modal data or interpret high-level human intents, restricting the network from realizing true intent-based autonomy. To address these systemic bottlenecks, WLAMs have emerged as a groundbreaking solution. A WLAM refers to large-scale AI models that seamlessly integrate with wireless communication systems, enhancing network performance through advanced intelligence levels. Unlike conventional models, a WLAM serves as a versatile foundation model that can be repurposed for various wireless applications with minimal adaptation \cite{chen2024big}. By leveraging techniques such as prompt engineering and fine-tuning, WLAMs can adapt to evolving user requirements and emerging wireless technologies without undergoing exhaustive retraining cycles \cite{bariah2024large}. This shift paves the way for next-generation intelligent wireless services, offering enhanced efficiency, robustness, and scalability across a wide range of applications.
\par
Large AI models and wireless communications are forming a deeply synergistic ecosystem, propelling technological advancement through a bidirectional interplay. As illustrated in Fig. \ref{synergy}, this relationship can be categorized into two distinct yet interconnected paradigms: ``Large AI for Wireless'' and ``Wireless for Large AI''. In the top-down paradigm of Large AI for Wireless, the AI model functions as the intelligent ``brain'' of the network. By offering adaptable, generalizable frameworks that outperform traditional task-specific models, WLAMs optimize critical functions ranging from physical layer signal processing to network management \cite{du2024distributed}. For instance, they enable predictive resource allocation and, in the context of semantic communication (SemCom), boost efficiency by transmitting only essential information to reduce bandwidth consumption \cite{jiang2024large}. Conversely, the bottom-up paradigm of Wireless for Large AI depicts the wireless network as the essential infrastructure supporting intelligence. Advanced wireless technologies address the immense computational and data demands of large models. Edge intelligence moves computation closer to data sources to cut latency \cite{zeng2024implementation}, federated learning supports decentralized training while preserving privacy \cite{chen2024role}, and Over-the-air computation (AirComp) minimizes data aggregation overhead by merging communication and computation \cite{wang2024over}. This closed-loop interaction where the network supports AI operations and AI in turn optimizes the network is key to unlocking the potential of AI-empowered 6G applications.

\subsection{Key performance indicators}
As large AI models integrate with wireless communication technologies, evaluating system performance and efficiency becomes crucial. This combination presents unique challenges in optimization, energy consumption, and communication efficiency. In WLAM, key performance indicators (KPIs) assess the effectiveness of system, stability, and training progress, providing insights into real-world application performance. In this subsection, we expand upon the definitions of the six core KPIs shown in Fig. \ref{kpi}. We explicitly highlights their distinctions from traditional algorithms, the novel aspects introduced by WLAM, the inherent trade-offs, and critical areas for future optimization.

\begin{table*}[t] 
    \centering 
    \caption{Comparison of our work with existing related surveys.}
    \label{tab:surveys}

   \resizebox{\linewidth}{!}{
        \setlength{\tabcolsep}{4pt}
        \renewcommand{\theadfont}{\normalsize} 
        \setcellgapes{4pt} 
        \makegapedcells   
        \begin{tabular}{|c|c|c|c|c|c|c|c|c|c|c|c|c|c|c|c|c|c|c|c|c|}
        \hline
        \multicolumn{1}{|c|}{} 
        & \multicolumn{6}{c|}{Fundamentals for WLAM} 
        & \multicolumn{5}{c|}{Large AI Model for Wireless} 
        & \multicolumn{5}{c|}{Wireless for Large AI Model} 
        & \multicolumn{4}{c|}{Emerging Technology for WLAM} \\
        \hline
        Ref. &
        \makecell{Traditional\\Architecture} &
        \makecell{Diffusion\\Models} &
        MoE &
        Dataset &
        \makecell{Super-\\vised\\Learning} &
        RL &
        \makecell{PE\\FT} &
        \makecell{Physical\\Layer} &
        \makecell{Network\\Layer} &
        \makecell{Sem\\Com} &
        \makecell{Wireless\\Agents} &
        \makecell{Edge\\Intelli-\\gence} &
        \makecell{DC\\ML} &
        \makecell{Air\\Comp} &
        PLS &
        HDC &
        \makecell{Quantum\\Computing} &
        PINN &
        HNs &
        \makecell{NextGen\\Sequence\\Modeling} \\
        \hline
        \cite{shen2024large} & & & & \checkmark & \checkmark & & & \checkmark & & \checkmark & & \checkmark & & & & & & & &
        \\ \hline
        
        \cite{lin2023pushing} & & & & & \checkmark & & \checkmark & \checkmark & & & & \checkmark & \checkmark & & & & & & &
        \\ \hline
        
        \cite{xu2024when} & & & & \checkmark & & \checkmark & & \checkmark & \checkmark & \checkmark & \checkmark & \checkmark & & & & & & & &
        \\ \hline
        
        \cite{tarkoma2023ai} & & & & \checkmark & \checkmark & & & \checkmark & \checkmark & \checkmark & \checkmark & & \checkmark & & \checkmark & & & & &
        \\ \hline
        
        \cite{chen2024big} & & & & \checkmark & \checkmark & & & \checkmark & & \checkmark & & \checkmark & \checkmark & \checkmark & & & & & &                   
        \\ \hline
        
        \cite{bariah2024large} & & & & & & \checkmark & & \checkmark & & \checkmark & \checkmark & \checkmark & & & & & & & &                   
        \\ \hline
        
        \cite{xu2024large} & & & & & \checkmark & & & \checkmark & \checkmark & \checkmark & \checkmark & & & & & & & & &                   \\ \hline
        
        \cite{zhou2024large} & \checkmark & & & \checkmark & \checkmark & \checkmark & \checkmark & \checkmark & & & \checkmark & \checkmark & & & & & & & & 
        \\ \hline
        
        \makecell{This\\Work} & \checkmark & \checkmark & \checkmark & \checkmark & \checkmark & \checkmark & \checkmark & \checkmark & \checkmark & \checkmark & \checkmark & \checkmark & \checkmark & \checkmark & \checkmark & \checkmark & \checkmark & \checkmark & \checkmark & \checkmark        
        \\ \hline
        \end{tabular}
    } 
\end{table*}

\begin{figure*}[t]
	\begin{center}
		\centerline{\includegraphics[width=0.92\linewidth]{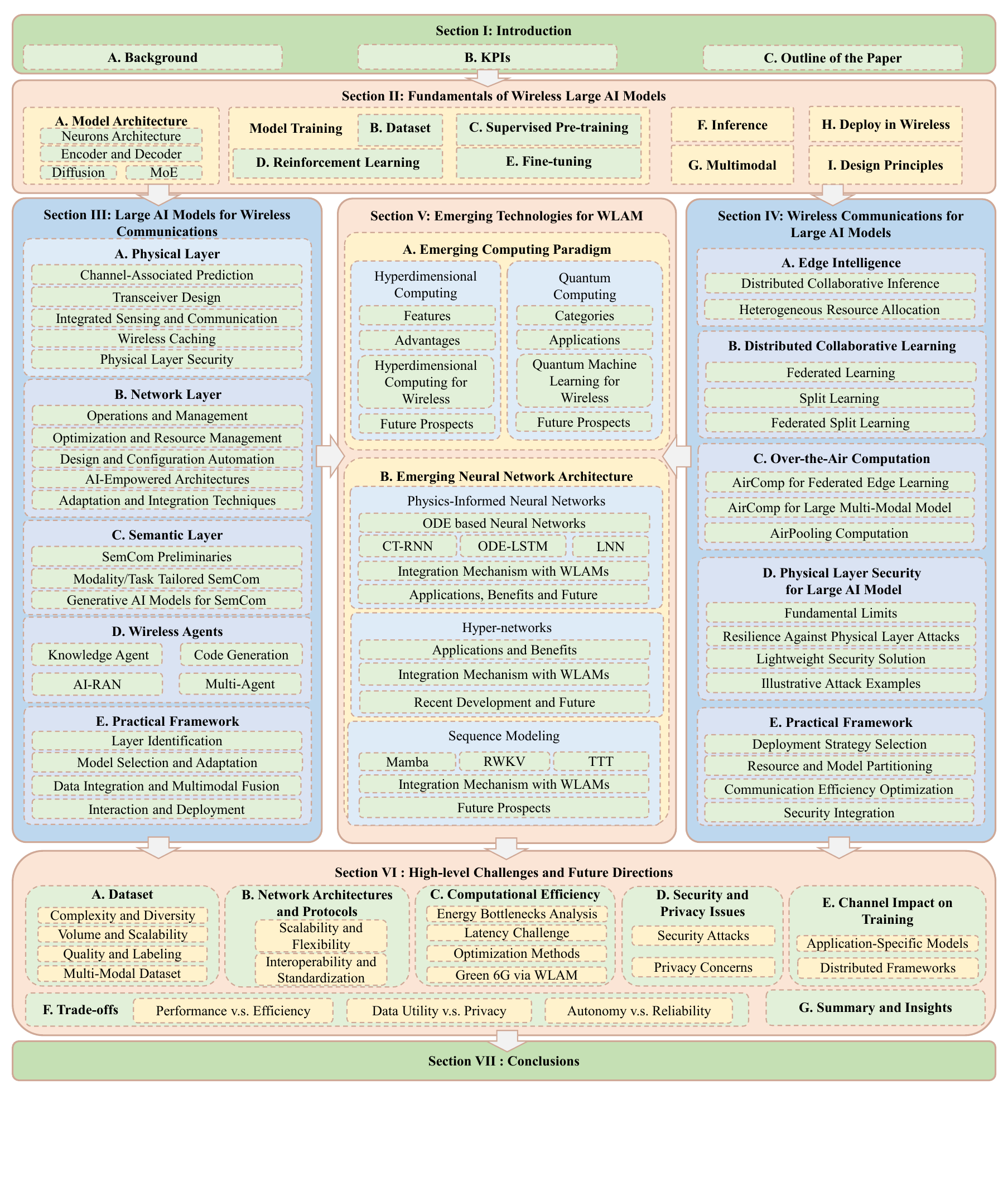}}
		\captionsetup{font=footnotesize, name={Fig.}, labelsep=period} 
		\caption{\, The outline of this survey.}
		\label{outline}
		\vspace{-8mm}
	\end{center}
\end{figure*}

\subsubsection{\textbf{Delay}}
Delay refers to the total time required from the start of local training on the client to the final acquisition of the updated global model. Unlike traditional algorithms where delay is dominated by communication, WLAM includes both the local computation time and communication delays \cite{yang2020delay}. The immense processing time for such massive models is a potential drawback. Using delay as a metric essential for WLAMs. Therefore, evaluating delay as a key metric is essential for WLAMs. A careful trade-off between computation and communication is necessary to optimize system responsiveness and performance, requiring future optimizations including hardware-aware scheduling and efficient parameter compression to mitigate these novel latency bottlenecks \cite{yang2020delay}.

\subsubsection{\textbf{Energy efficiency}}
Energy efficiency is crucial in WLAM, as energy consumption can impact the sustainability and scalability of the system \cite{you2025ai}. This includes both the energy consumed for local computation during model training and for data transmission between clients and the central server. The computational complexity and memory requirements of WLAMs may impose an impractical energy burden on edge devices, which remains a potential drawback. It is vital to design systems that optimize energy usage while balancing between computational load and communication needs. This necessitates future research into green-quantized frameworks that utilize numeric metrics and measurment methods for both local training and wireless transmission to optimally balance energy efficiency, convergence rate, and learning accuracy \cite{kim2023green, zhao2023energy}.

\subsubsection{\textbf{Reliability}}
Reliability evaluates how consistently the system performs under variable wireless conditions, encompassing both transmission reliability and the robustness of the WLAM outputs. Given the unpredictability of wireless communication environment, ensuring that model updates are robust against the noise is vital for the overall performance of WLAM. The inherently stochastic nature of generative models may lead to unpredictable, collapsed, or non-compliant network configurations, which remains a potential drawback. The system should be designed to handle communication errors, device failures, adversarial attacks and other reliability issues while maintaining stable model training and updates. This necessitates future research to guarantee deterministic, standard-compliant, and safe actions from WLAM agents.

\subsubsection{\textbf{Massive connectivity}}
Wireless systems must manage the communication needs of a large number of distributed devices, extending beyond traditional sensor data transmission to support decentralized multi-agent systems synchronizing high-dimensional model parameters. The system should be designed to handle massive connectivity, ensuring that communication between devices remains efficient. Addressing connectivity issues without introducing significant delays is crucial for maintaining system efficiency as the number of devices grows, necessitating future optimizations like over-the-air computation and semantic communication architectures to aggregate massive updates efficiently \cite{yang2020federated, shi2020joint}.

\renewcommand{\arraystretch}{1.2}
\begin{table}[t]
\centering
\caption{Key abbreviations.} 
\label{tab:abbreviations_swapped} 
\footnotesize
\begin{tabular}{|l|l|l|l|}
\hline
\textbf{Abbreviation} & \textbf{Full term} & \textbf{Abbreviation} & \textbf{Full term} \\
\hline
AI & Artificial intelligence & LoRA & Low-rank adaptation \\
\hline
CoT & Chain of thought & MARL & Multi-agent RL \\
\hline
CSI & Channel state information & MoE & Mixture of experts \\
\hline
CT & Continuous-time & ODE & Ordinary differential equation \\
\hline
DiT & Diffusion transformer & PEFT & Parameter-efficient fine-tuning \\
\hline
DPO & Direct preference optimization & PINN & Physics-inspired neural network \\
\hline
DCML & Distributed collaborative machine learning & PLS & Physical layer security \\
\hline
DDIM & Denoising diffusion tmplicit model & PPO & Proximal policy optimization \\
\hline
DDPM & Denoising diffusion probabilistic model & QML & Quantum machine learning \\
\hline
FL & Federated learning & RAN & Radio access network \\
\hline
FSL & Federated split learning & RAG & Retrieval augmented generation \\
\hline
GAN & Generative adversarial network & RL & Reinforcement learning \\
\hline
GPT & Generative pre-trained transformer & RLHF & RL from human feedback \\
\hline
GRPO & Group relative policy optimization & RF & Radio frequency \\
\hline
HDC & Hyperdimensional computing & RWKV & Receptance weighted key value \\
\hline
HN & Hyper-networks & SemCom & Semantic communication \\
\hline
ISAC & Integrated sensing and communications & 6G & Sixth-generation \\
\hline
IoT & Internet of things & SL & Split learning \\
\hline
KPI & Key performance indicator & TTT & Test-time training \\
\hline
LLM & Large language model & URLLC & Ultra-reliable low-latency communication \\
\hline
LNN & Liquid neural network & WLAM & Wireless large AI model \\
\hline
\end{tabular}
\end{table}

\subsubsection{\textbf{Privacy}}
Privacy protection is fundamental in WLAM, evolving from traditional cryptographic bitstream encryption to preventing massive model weights from memorizing and reconstructing sensitive local user features. The system should be designed with privacy-preserving techniques to ensure that sensitive data is not exposed during model training and updates. While decentralized frameworks naturally preserve data locality, their high vulnerability to novel AI-specific threats, such as model inversion and gradient leakage attacks, remains a critical drawback. Ensuring privacy while maintaining performance is an ongoing challenge in WLAM systems \cite{dwork2014algorithmic, kalapaaking2023blockchain}.

\subsubsection{\textbf{Generalization}}
Generalization refers to the ability of a model to perform well on new, unseen data \cite{wei2022emergent}. It is crucial for large AI models to be designed with good generalization ability, ensuring that the model does not overfit to the training data but can adapt to new environments and conditions. Moreover, because edge devices typically possess limited local data and constrained computational capacities, they cannot afford extensive fine-tuning for every new scenario. Therefore, WLAM architectures must inherently possess exceptional generalization to handle dynamic network conditions directly. Effective designs must focus on maintaining high performance across contexts necessitates future research into continuous pre-training on heterogeneous telecom datasets and rigorous instruction tuning \cite{soldati2024design, al2020generalizing, zhurobustnetwork}.

\subsection{Motivation, contributions, and outline}
The recent proliferation of large AI models has ignited extensive research and discussion in wireless communications. While numerous overviews have explored the capabilities and applications of these models, a notable gap exists in the literature concerning their synergistic integration with wireless communication systems. As evidenced by the comparative analysis in Table \ref{tab:surveys}, dedicated reviews on this emerging relation remain scarce. To bridge this gap and illuminate the revolutionary potential of WLAM, this survey systematically delineates their definition, fundamental technologies, key application scenarios, and prospective future directions. Key abbreviations used throughout this paper are defined in Table \ref{tab:abbreviations_swapped} for reader convenience. The main contributions of this survey are summarized as follows:
\begin{itemize}
\item Foundational Paradigm and Rigorous Taxonomy: We establish a comprehensive foundation for WLAMs. Moving beyond the conventional parameter-scale definition, we uniquely characterize WLAMs through their structural pillars, such as RF-specific attention mechanisms, signal waveform tokenization, and physics-informed biases. Furthermore, we systematically review the entire WLAM lifecycle, from data collection to advanced inference strategies.
\item Cross-Layer Applications and Practical Implementation Framework: We provide an in-depth analysis of how WLAMs revolutionize wireless systems from the physical layer to multimodal semantic communications. Notably, we highlight the paradigm shift toward Agentic AI-RAN. To bridge theory and practice, we synthesize these applications into a step-by-step practical framework, guiding practitioners through problem formulation, model adaptation, and multimodal data fusion.
\item Infrastructure Support and Systematic Deployment Framework: We systematically investigate the bottom-up network infrastructure required to empower massive models, focusing on edge intelligence, distributed learning, and Over-the-Air computation. To facilitate real-world operation, we propose a comprehensive deployment framework addressing resource allocation, communication efficiency, and security integration. Crucially, we analyze extreme engineering bottlenecks, proposing dual-loop architectures to reconcile WLAM inference delays with strict latency requirements.
\item Emerging Technologies and Strategic Future Roadmaps: We pioneer the discussion on integrating WLAMs with emerging paradigms. Finally, we outline a strategic roadmap to overcome fundamental trade-offs. This includes trading computational energy for RF radiated energy to align with the ``Green 6G" vision, and proposing multi-layered formal verification frameworks to ensure the reliability of WLAMs in critical infrastructure.
\end{itemize}
\par
The structure of this survey, visually depicted in Fig. \ref{outline}, is organized as follows: Section II lays the foundational groundwork by elucidating the core principles of WLAM. Sections III and IV then delve into the dual facets of WLAM integration, respectively exploring the application of large AI models to enhance wireless communications and the role of wireless communications in enabling large AI models, encompassing architectural considerations and illustrative application scenarios. Section V broadens the scope to investigate the convergence of WLAM with other emerging technologies. Section VI critically analyzes the overarching challenges and outlines promising future research trajectories for WLAM. Finally, Section VII concludes this survey.

\section{Fundamentals of large AI models}\label{section2}
Large AI models, characterized fundamentally by their massive scale in parameters, extensive training datasets, and sophisticated architectures, represent a significant leap in general AI capabilities. However, to establish a rigorous taxonomy in the context of this survey, we emphasize that a Wireless Large AI Model (WLAM) must be defined beyond mere parameter scale. What distinguishes a WLAM from a general purpose LAM is its inherently architectural adaptations. Unlike generic models that process discrete semantic text tokens, a WLAM fundamentally bridges the physical and digital worlds through three distinct structural pillars. The first pillar is signal waveform tokenization. WLAMs incorporate specialized tokenizer layers capable of transforming continuous complex valued RF signals into high dimensional discrete latent representations. For instance, recent pioneering architectures like RF-GPT \cite{zou2026rfgptteachingaiwireless} leverage short time Fourier transforms to convert raw in phase and quadrature samples into spectrograms, which are subsequently patchified and linearly projected into dedicated RF tokens that language models can process. The second pillar involves RF specific attention mechanisms. Instead of standard sequence level self attention, WLAM architectures utilize multidimensional spatial temporal frequency attention tailored to physical antenna array topologies and fading characteristics, allowing the model to capture non stationary wireless channel dynamics. The final pillar is the physics informed structural bias. To ensure outputs adhere to electromagnetic propagation laws and strict telecommunication protocol constraints, WLAMs embed physics guided operations and telecom specific output heads mapping latent states directly to executable continuous variables like beamforming matrices.
\par
This architectural taxonomy ensures that WLAMs are not just generic foundation models applied to wireless datasets, but a distinct class of models engineered for the telecommunications domain. Empowered by these foundational structural adaptations alongside their massive scale, WLAMs can inherently encompass multi-modality by fusing heterogeneous data types including RF signals, text, and images. This allows them to achieve cross task learning across diverse wireless tasks, such as channel estimation and network optimization, without task specific retraining \cite{wei2022emergent}. Understanding these core principles is essential for leveraging their power across the increasingly complex landscape of wireless communications.
\par
\begin{figure}[t]
	\begin{center}
		\centerline{\includegraphics[width=0.65\linewidth]{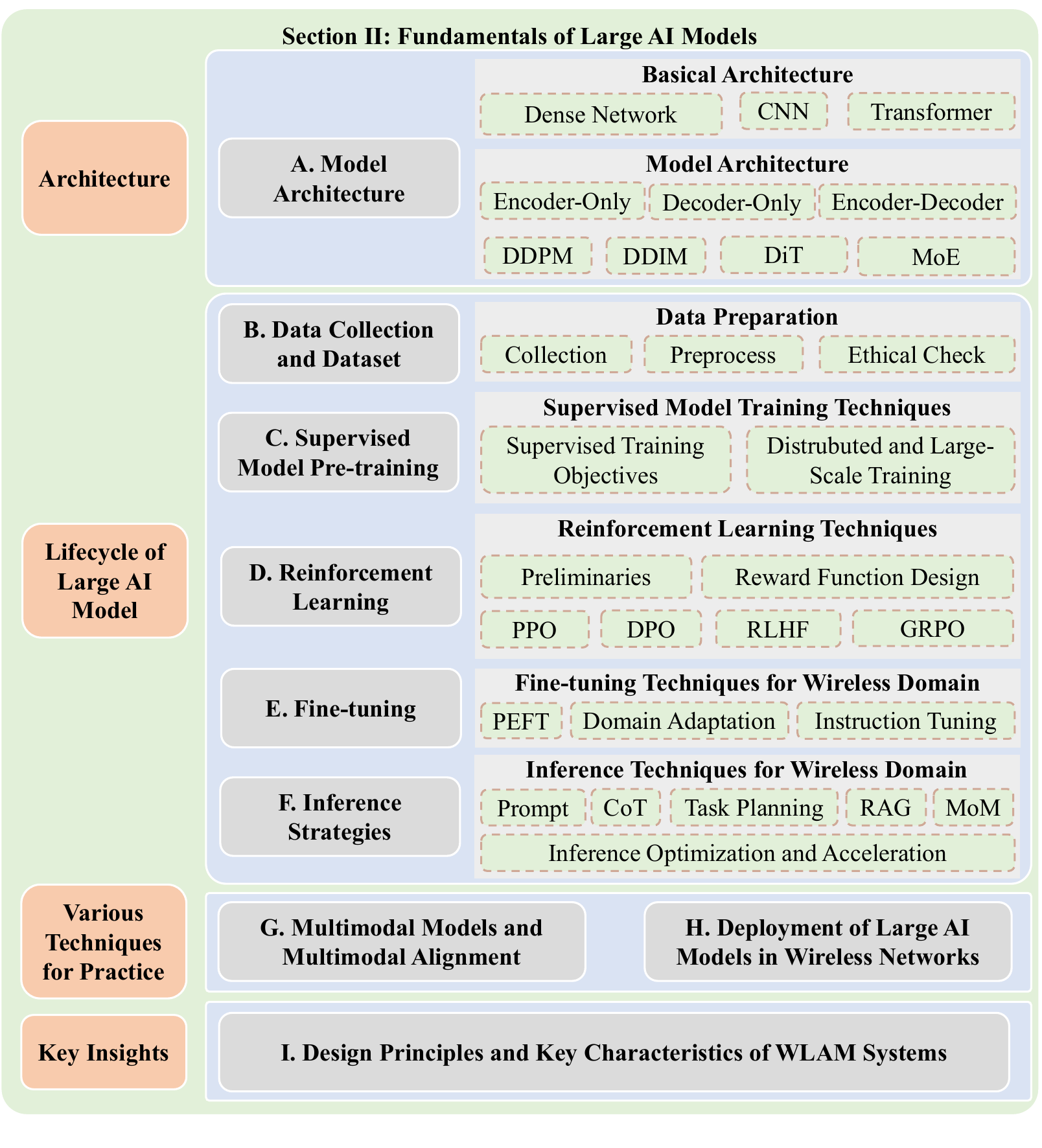}}
		\captionsetup{font=footnotesize, name={Fig.}, labelsep=period} 
		\caption{\, The outline of Section 2.}
		\label{outline:sec2}
		\vspace{-8mm}
	\end{center}
\end{figure}

This section provides a comprehensive overview of large AI model fundamentals, progressing logically from concept to practice, as summarized in Fig. \ref{outline:sec2}. The analysis begins with the foundational model architectures before proceeding through the essential lifecycle of a large AI model, which covers data collection, supervised pre-training, refinement via reinforcement learning and fine-tuning, and finally, inference strategies. Building upon this lifecycle, the discussion explores crucial practical techniques, including multimodal alignment and deployment. The section culminates by synthesizing these diverse technical elements into a cohesive set of key design principles for WLAM systems.

\subsection{Model architecture}
The choice of model architecture is crucial in determining how well a system can process, interpret, and generate data. Different architectures are designed to handle specific types of data, such as spatial, temporal, or sequential information, and to optimize performance for tasks like signal processing, resource allocation, and network optimization. In this subsection, we explore various model architectures, ranging from dense networks to specialized approaches like transformers, convolutional neural networks (CNNs), and diffusion models, all of which offer unique advantages in wireless communication systems.

\subsubsection{\textbf{Dense network}}
Dense (fully-connected) neural network is the most basic architecture where each neuron in one layer connects to all neurons in the next layer. Though simple, dense network provides the foundation for learning arbitrary mappings. However, it is generally less efficient for spatial or sequential data common in wireless signals. Moreover, scaling parameters in dense network can lead to increased computational complexity and longer training times, especially with large datasets common in wireless applications.

\subsubsection{\textbf{Convolutional neural network}}
CNN is a class of deep learning models that have proven highly effective for tasks involving spatial data, such as image and signal processing \cite{alexnet}. CNNs use convolutional layers that apply a series of filters to the input data, capturing local patterns and hierarchies of features. This structure makes CNNs suited for processing and analyzing complex patterns in wireless signals, such as those found in channel estimation and signal classification. By leveraging the spatial locality and translational invariance properties of CNNs, wireless communication systems can significantly enhance their performance in tasks like interference mitigation, beamforming, and modulation recognition. Increasing the number of parameters in CNNs, such as depth or number of filters, can improve their ability to learn complex features but may also slow down training convergence and increase the risk of overfitting, especially with limited or biased data from dynamic wireless environments. The computational load during inference also increases with larger CNNs, potentially leading to higher latency under dynamic wireless conditions.

\subsubsection{\textbf{Transformer}}
Transformer \cite{vaswani2017attention}, initially developed for natural language processing, has revolutionized the field of AI with its ability to handle sequential data through self-attention mechanisms. Unlike CNN, transformer does not rely on convolutions or recurrence, allowing them to capture long-range dependencies and relationships in the data. This characteristic is particularly beneficial for wireless communication \cite{wang2022transformer}, where understanding the temporal dynamics and dependencies between transmitted and received signals is crucial. Transformers can be employed for tasks such as sequence prediction, channel state information (CSI) feedback, and end-to-end communication system design \cite{wu2023vision,wu2024transformer}. While transformers may exhibit poor performance on a small scale due to high computational and memory requirements, they demonstrate outstanding scalability. As the scale of the system increases, their performance improves significantly, making them highly effective for large-scale wireless communication networks.

\subsubsection{\textbf{Encoder and decoder related architectures}}
The architectural paradigm of a large AI model significantly influences its ability to process, interpret, and generate data. Among the most widely adopted structures are encoder-only, decoder-only, and encoder-decoder models. 
\paragraph{Encoder-only}
Encoder-only architectures are designed to encode a complete input sequence into a rich, contextualized representation, leveraging attention mechanisms to incorporate bidirectional context. An example is bidirectional encoder representations from transformers (BERT) \cite{devlin2018bert}, which masks portions of the input and learns to reconstruct them using the surrounding context. This strategy enables the model to develop deep semantic understanding of the input sequence, making it exceptionally effective for discriminative tasks. In wireless systems, encoder-only models are naturally suited to interpret structured data or signals where global context matters. For instance, such models can be employed to analyze CSI across multiple antennas or subcarriers, identifying spatial and frequency-domain patterns to detect anomalies, perform user classification, or conduct package loss detection \cite{bertwireless}. 
\paragraph{Decoder-only}
Decoder-only models (e.g., generative pre-trained transformer (GPT) series \cite{radford2019language, brown2020language, achiam2023gpt} and LLaMA series \cite{touvron2023llama, touvron2023llama2, dubey2024llama}) perform autoregressive generation and excel at sequence prediction. They are well-suited for forecasting future wireless traffic patterns or beamforming sequences in time-varying environments. Their in-context learning capabilities also enable real-time adaptation without retraining.
The basic principle underlying GPT models is to compress the world knowledge into the decoder-only transformer model by language modeling, such that it can recover (or memorize) the semantics of world knowledge and serve as a general-purpose task solver \cite{zhao2023survey}. Compared to GPT-3, LLaMA incorporates several specific enhancements to maintain similar performance while significantly reducing the number of parameters \cite{touvron2023llama}. For example, in order to enhance training stability, LLaMA normalizes the input of each sub-layer instead of normalizing the output. However, LLaMA cannot generate responsive text \cite{touvron2023llama}, and extra fine-tuning is still required. 
\paragraph{Encoder-decoder}
Encoder-decoder models combine the strengths of both encoders and decoders, offering a two-stage pipeline. The encoder first processes the input into a compressed latent representation, and the decoder subsequently generates the target output conditioned on this representation. Models like text-to-text transfer transformer (T5) \cite{2020t5}, bidirectional and auto-regressive transformers (BART) \cite{bart}, and transformer-based sequence-to-sequence architectures exemplify this class. These models are effective for tasks involving input-output transformations, such as translation, summarization, and question-answering. Therefore, this structure aligns with tasks like end-to-end communication system modeling or semantic data compression, where input-output transformations are needed. However, they face challenges in scaling and efficiency, especially with longer inputs.

\subsubsection{\textbf{Diffusion related architectures}}
Diffusion-based architectures have become a powerful class of generative models, known for their ability to produce high-quality data through a controlled process of iterative refinement \cite{10812969}. In this subsection, we discuss these diffusion-based models and their application in wireless systems.

\paragraph{Basic principle of diffusion models}
Diffusion models have emerged as a leading paradigm in generative modeling, distinguished by their capacity for high-fidelity sample generation and stable training dynamics. The core principle is inspired by non-equilibrium thermodynamics, which describes how systems evolve from order to disorder. This concept is reflected in two opposing processes: a fixed forward diffusion process and a learned reverse denoising process. 
In the forward process, real data $\mathbf{x}_0$ is gradually perturbed by Gaussian noise through $T$ discrete steps:
\begin{equation}
    q(\mathbf{x}_t \mid \mathbf{x}_{t-1}) = \mathcal{N}(\sqrt{\alpha_t}\mathbf{x}_{t-1}, (1-\alpha_t)\mathbf{I}),
\end{equation}
where $\alpha_t \in (0,1)$ controls the noise level at step $t$, and $\bar{\alpha}_t = \prod_{s=1}^t \alpha_s$ denotes the cumulative product.

\par
By iterating this process, we obtain $\mathbf{x}_T \approx \mathcal{N}(\mathbf{0},\mathbf{I})$, which contains no information from $\mathbf{x}_0$. 
The reverse process aims to recover $\mathbf{x}_0$ from $\mathbf{x}_T$ by learning a parameterized model $p_\theta$:
\begin{equation}
    p_\theta(\mathbf{x}_{t-1} \mid \mathbf{x}_t) = \mathcal{N}\big(\boldsymbol{\mu}_\theta(\mathbf{x}_t, t), \sigma_t^2 \mathbf{I}\big),
\end{equation}
where $\boldsymbol{\mu}_\theta$ is predicted from the noisy sample $\mathbf{x}_t$ and the step index $t$. Training minimizes the discrepancy between the true noise and the predicted noise.
This framework serves as the foundation for various diffusion-based architectures.

\paragraph{Denoising diffusion probabilistic models}
Denoising diffusion probabilistic models (DDPMs) are a class of generative models that create data samples through an iterative denoising process \cite{ddpm}. They work by gradually corrupting real data with noise during training, and then learning to reverse this process. The model is typically trained to predict either the original data or the noise added at various levels of degradation. The loss function is denoted as 
\begin{equation}
    \mathcal{L}_{\text{DDPM}} = \mathbb{E}_{t, \mathbf{x}_0, \boldsymbol{\epsilon}} \left[ || \boldsymbol{\epsilon} - \boldsymbol{\epsilon}_\theta(\sqrt{\bar{\alpha}_t}\mathbf{x}_0 + \sqrt{1-\bar{\alpha}_t}\boldsymbol{\epsilon}, t) ||^2 \right],
\end{equation}
where $\boldsymbol{\epsilon}_\theta$ is trained to predict the noise $\boldsymbol{\epsilon}$ that was added to the original data $\mathbf{x}_0$ to create the noisy version $\mathbf{x}_t = \sqrt{\bar{\alpha}_t}\mathbf{x}_0 + \sqrt{1-\bar{\alpha}_t}\boldsymbol{\epsilon}$ at timestep $t$. $\bar{\alpha}_t$ is a factor derived from the noise schedule. Minimizing this objective teaches the model the reverse denoising process.
This step-by-step refinement leads to high-quality generation and stable training, often outperforming adversarial models in fidelity and diversity. DDPMs are suited for tasks where control in generation are crucial. Scaling DDPMs by increasing denoising steps or network size can improve the quality of outputs but significantly increases inference latency, which can be problematic for real-time applications in dynamic wireless environments.

\paragraph{Denoising diffusion implicit models}
Denoising diffusion implicit models (DDIMs) are a variant of DDPMs that focus on improving generation efficiency \cite{ddim}. While DDPMs rely on stochastic sampling, DDIMs use a deterministic reverse process to generate data, denoted as
\begin{equation}
        \mathbf{x}_{t-1} = \sqrt{\bar{\alpha}_{t-1}} \left( \frac{\mathbf{x}_t - \sqrt{1-\bar{\alpha}_t}\boldsymbol{\epsilon}_\theta(\mathbf{x}_t, t)}{\sqrt{\bar{\alpha}_t}} \right)
        + \sqrt{1-\bar{\alpha}_{t-1}} \boldsymbol{\epsilon}_\theta(\mathbf{x}_t, t), 
\end{equation}
which shows the deterministic update step ($\sigma=0$ case) for generating $\mathbf{x}_{t-1}$ from $\mathbf{x}_t$. It uses the predicted noise $\boldsymbol{\epsilon}_\theta(\mathbf{x}_t, t)$ to first estimate the original data $\hat{\mathbf{x}}_0$ (the term in parenthesis) and then deterministically computes the previous state $\mathbf{x}_{t-1}$ using the noise schedule constants $\bar{\alpha}_t$ and $\bar{\alpha}_{t-1}$. This allows faster sampling compared to DDPM. This allows for fewer steps during inference, significantly speeding up sample generation without retraining the model. DDIMs are compatible with models trained using DDPM techniques and maintain comparable quality. The deterministic property also enables applications like latent-space interpolation and semantic editing. In wireless contexts, DDIMs can be used to rapidly synthesize signal waveforms or channel responses, making them attractive for low-latency or real-time systems.

\paragraph{Diffusion transformers}
Diffusion transformers (DiTs) are used by transforming input images into a sequence of patches, which are then processed by transformer blocks \cite{peebles2023scalable}. To use DiTs, the images need to undergo a patchifying step, which breaks them into smaller tokens. These tokens are then passed through multiple DiT blocks, which use self-attention mechanisms and optional conditioning techniques, such as adaptive layer normalization (adaLN) to handle various input factors like noise timestamp $t$ or class labels. The core computation within a DiT block can be represented as:
\begin{equation}
    \mathbf{z}' = \mathbf{z} + \text{SelfAttention}(\text{adaLN}(\mathbf{z}, t, c)),
\end{equation}
\begin{equation}
    \mathbf{z}_{\text{out}} = \mathbf{z}' + \text{MLP}(\text{adaLN}(\mathbf{z}', t, c)),
\end{equation}
where $\mathbf{z}$ is the input sequence of tokens, and adaLN modulates the network based on the timestep $t$ and an optional class label $c$. This allows the model to effectively learn the denoising function across different noise levels.
DiTs can be trained with latent diffusion models to reduce computational cost, and they scale efficiently with increased model size, improving image quality as the model size or computational resources grow. DiT models like Hunyuan-DiT \cite{li2024hunyuandit} have gained attention for their outstanding capability in processing pictures.

\subsubsection{\textbf{Mixture of experts}}
Mixture of experts (MoE) is a scalable model architecture that dynamically routes inputs to different specialized subnetworks, known as experts \cite{chen2022towards}. A gating mechanism selects a subset of these experts based on the input, allowing only the chosen experts to process the data. This conditional computation enables models with massive parameter counts to operate efficiently, as only a small portion of the network is active at a time. MoE architectures have demonstrated impressive scalability, particularly in large language models (LLMs) like DeepSeek-V3 \cite{liu2024deepseek}, where they improve performance while maintaining inference cost. However, training such models with numerous experts can be challenging, particularly in ensuring the gating mechanism effectively learns to route inputs. In dynamic wireless environments, these challenges manifest as channel conditions and data distributions fluctuate. Recent advancements like DeepEP \cite{deepep2025}, DeepGEMM \cite{deepgemm2025} addressed these issue by optimizing MoE training with efficient graphic processing unit (GPU) communication and accelerated matrix operations. In wireless applications, MoEs could enable adaptive processing by activating experts tailored to specific environments, such as urban or rural channel conditions, thereby enhancing model specialization and generalization \cite{wang2025dynamicalmultimodalfusionmixtureofexperts}.

\subsection{Data collection and dataset}
High-quality, diverse, and large-scale data is the foundation of any successful large AI model. The effectiveness of a large model is largely a reflection of the diversity and coverage of its training data.
Pre-training large AI models on wireless data presents unique challenges and opportunities. In the context of wireless communication, building effective datasets involves unique challenges, such as dynamic environments, device heterogeneity, and privacy constraints. This subsection outlines key considerations in dataset construction: sourcing diverse data, preprocessing and cleaning, and navigating ethical and regulatory concerns.

\subsubsection{\textbf{Data sources and diversity}}
Pretraining large models requires diverse datasets. In wireless, this includes channel measurements, CSI matrices, beam logs, and traffic traces collected from real-world networks or high-fidelity simulators. Ensuring coverage across different frequency bands, mobility profiles, and deployment scenarios improves generalization.
Unlike conventional pre-training that relies heavily on existing static datasets sourced from the internet, wireless pre-training leverages real-time, multi-modal data collected from a diverse array of internet of things (IoT) devices, including smartphones, sensors, and autonomous as well. This dynamic and heterogeneous data reflects a more accurate and timely representation of the physical world, enhancing the ability of models to understand and predict complex patterns and interactions in various environments. Synthetic data which is generated using ray-tracing, digital twins, or data augmentation strategies to simulate rare or extreme events have also gain consideration \cite{deepmimo}. In semantic communication systems, data must also cover various modalities (text, image, video, audio) and tasks (translation, reconstruction, classification) to ensure the model can learn semantic compression and transmission strategies that generalize across use cases.

\subsubsection{\textbf{Data cleaning and preprocessing}}
Large raw datasets are often noisy, redundant, or inconsistent. Without systematic cleaning and preprocessing, the model may learn incorrect correlations or suffer from slow convergence and degraded performance. This process may include duplication removement, outlier filtering, scaling or normalization, tokenization, formatting and noise modeling.
Data quality directly affects downstream performance. High signal-to-noise ratio (SNR) training data may not generalize to low-quality operational settings. Conversely, overfitting to synthetic noise distributions may degrade performance on real-world interference. A balanced and well-curated dataset is critical.

\subsubsection{\textbf{Ethical considerations}}
Ethical challenges in constructing large-scale wireless datasets primarily concern two areas: mitigating data bias to ensure fairness, and protecting user privacy from the inherent sensitivity of communication data. Training datasets often reflect existing social, geographic, or behavioral biases; for example, a WLAM trained predominantly on data from urban deployments may underperform in rural conditions, leading to uneven service quality and digital exclusion. Mitigating this risk requires ensuring balanced representation across diverse demographics, devices, and operating environments. In addition to fairness, data privacy is a paramount concern. Technical data such as packet logs, and location traces can inadvertently contain personal identifiers or reveal behavioral patterns, and the broadcast nature of radio signals makes this information particularly vulnerable to misuse. Therefore, all data collection practices must strictly align with regulatory standards like the general data protection regulation \cite{gdpr2016} and the California consumer privacy act \cite{ccpa2018}.

\subsection{Supervised model pre-training}
Supervised pre-training serves as the foundational stage in building large AI models, allowing them to acquire generalized representations from vast datasets. Given this foundational nature and the extensive pre-training involved, such large AI models are often referred to using related terms like foundation models or pretrained foundation models. In this phase, models are exposed to structured learning objectives and are optimized over large-scale corpora using powerful training infrastructures. Though traditionally centered in natural language and vision domains, supervised pre-training is now becoming increasingly relevant to wireless communication systems, where large-scale signal data, mobility logs, and protocol traces provide a rich training ground. 
Supervised model pre-training empowers large AI models with generalized knowledge before task-specific fine-tuning. In wireless domains, this involves adapting classical language pretraining paradigms to communication signals, enabling new opportunities in system modeling, semantic understanding, and AI-empowered transceiver design.

\subsubsection{\textbf{Training objectives}}

The choice of training objective plays a critical role in shaping what a model learns during pre-training, with different objectives aligning with specific model architectures and downstream use cases. The next-token prediction objective, commonly used in GPT-style decoder-only models, trains the model to predict the next element in a sequence given the preceding tokens \cite{zhao2023survey}. This approach excels in generative tasks and is well-suited to sequence modeling problems in wireless systems, such as predicting channel evolution or traffic patterns. Masked modeling, used in encoder-only architectures like BERT, involves masking random portions of the input and training the model to reconstruct them using bidirectional context. In wireless applications, this method can be employed for tasks such as recovering missing signal samples or corrupted subcarriers. Sequence-to-sequence generation objectives, typical of encoder-decoder models like T5 or BART, transform an input sequence into a target output, which is particularly useful for end-to-end communication systems, semantic compression, or signal-to-action mappings in adaptive wireless control.

\subsubsection{\textbf{Distributed and large-scale training}}
Large model training involves billions of parameters and requires substantial computational resources. Supervised pre-training is typically carried out over distributed systems that involve multiple GPUs connected through high-speed interconnects. To efficiently scale across devices, training workflows employ parallelization strategies such as data parallelism, where each device trains on a separate batch of data with synchronized gradients, or model parallelism, where the model itself is partitioned across devices to handle larger parameter sizes than a single GPU can accommodate.
\par
In pipeline parallelism, the layers of the model are distributed across devices and trained in a staged fashion, enabling higher throughput. WLAM, particularly those involving long input sequences or multimodal inputs, benefit from such techniques. Mixed-precision training reduces memory usage and increases compute efficiency by using lower-precision arithmetic without sacrificing accuracy, and sharded state parallel \cite{fsdp} reduces the overhead of large optimizer memory footprints. Specially designed methods like Dualpipe \cite{dualpipe} also provide an innovative bidirectional pipeline parallelism algorithm and implementation.

\subsection{Reinforcement learning}
Reinforcement learning (RL) provides a powerful framework for WLAM to learn effective policies for decision-making in dynamic wireless environments. The introduction of this framework begins with its fundamental preliminaries, upon which the critical design of reward functions for wireless systems is discussed. The analysis culminates with several advanced algorithms specifically tailored for large AI model training and alignment.

\subsubsection{\textbf{Reinforcement learning preliminaries}}
RL is a computational approach where an agent, such as a network controller or autonomous base station, learns to make optimal decisions by interacting with a dynamic wireless environment. This interaction is formally modeled as a markov decision process, defined by a set of states ($\mathcal{S}$ representing the network status (e.g., channel conditions, traffic load), a set of actions ($\mathcal{A}$) available to the agent (e.g., resource allocation, power control), a reward function ($R$) tied to key performance indicators (e.g., throughput, latency), and state transition dynamics
$P(s'|s,a)$ that capture the environmental evolution. The decision-making logic of an agent is governed by a policy, $\pi(a|s)$, which specifies the probability of taking action $a$ in state $s$. The primary objective is to learn an optimal policy, $\pi ^*$, that maximizes the long-term cumulative discounted reward, effectively optimizing network performance over time. This goal is formally expressed as:
\begin{equation}\label{eq:rlobj}
\pi^* = \arg\max_{\pi} \mathbb{E}{\pi} \left[ \sum{k=0}^{\infty} \gamma^k R_{t+k+1} \mid S_t=s \right],
\end{equation}
where $\gamma\in [0,1]$ is the discount factor that balances immediate and future performance. This fundamental objective provides a common mathematical basis for diverse RL algorithms, including the policy-based, value-based, and actor-critic methods discussed later. To achieve \eqref{eq:rlobj}, RL algorithms often learn a value function to estimate the long-term utility of network states. Unlike supervised learning, which requires static labeled data, RL enables goal-directed optimization in complex, time-varying wireless systems through exploration and exploitation.

In the context of large AI models, this RL framework is particularly valuable for tasks where direct supervision is hard to define. It shares strong parallels with adaptive wireless systems, where decisions like resource scheduling must be made under uncertainty with delayed feedback. The dynamic, exploratory nature of RL makes it well-suited for non-stationary or ambiguous objectives, which are common in both advanced wireless networks and large-scale model alignment. By integrating RL with supervised pre-training, models can be fine-tuned to continuously adapt to the evolving demands of dynamic environments.
As large models scale, hybrid training strategies could further enhance efficiency. Moreover, by integrating RL with the existing supervised pre-training, models can adjust to the evolving demands of wireless environments.

\subsubsection{\textbf{Reward function design}}
Effective reward function design is paramount for applying reinforcement learning to wireless systems, as it translates high-level performance objectives into a guiding feedback signal for the agent. Adopting a problem-oriented approach, the following analysis categorizes reward design based on the fundamental wireless challenges.

\paragraph{Resource management and optimization}
The core problem in this domain is the allocation of limited network assets (e.g., bandwidth, power) among competing entities to optimize a system-level utility. The common principle for reward design is to construct a multi-objective utility function, often as a weighted sum of conflicting KPIs, to balance system efficiency with user satisfaction. For example, in dynamic resource allocation for network slicing, the goal is to allocate bandwidth efficiently across different slices. The reward function is designed as a weighted sum of system-wide spectral efficiency (SE) and quality of experience (QoE) \cite{8540003}. While SE is measured at the physical layer, it serves as a crucial system-level feedback for the network-level allocation decision. The objective is to maximize:
\begin{equation}
    \mathbb{E}\{R(w,d)\} = \mathbb{E}\{\zeta \cdot SE(w,d) + \beta \cdot QoE(w,d)\},
\end{equation}
where $w$ is the bandwidth sharing solution, $d$ represents the fluctuating demands, and the weights $\zeta$ and $\beta$ balance the importance of spectrum efficiency and user experience.

\paragraph{Mobility management and trajectory optimization}
This category focuses on decisions involving physical movement, such as unmanned aerial vehicle (UAV) trajectory planning. The core principle for reward design is to balance the communication performance gains with the physical costs of movement, such as energy consumption or time. The reward is often formulated as an efficiency metric. For instance, in a system where a UAV optimizes its three dimensional trajectory, the reward function is defined as the cumulative sum of the instantaneous energy efficiency achieved at each step of the trajectory \cite{3D-Trajectory}:
\begin{equation}
    r(s(n),a(n)) = \sum_{k=1}^{K}\sum_{n'=1}^{n+1} \frac{t_{n'}^{u} r_{k,n'}}{e_{n'}^{uav}} - p_{0},
\end{equation}
where the term inside the summations represents the instantaneous energy efficiency for user $k$ at a past time slot $n'$. Specifically, $t_{n'}^{u}$ is the duration of time slot $n'$, $r_{k,n'}$ is the data rate for ground terminal $k$, and $e_{n'}^{uav}$ is the UAV propulsion energy. The double summation aggregates this efficiency metric over all $K$ ground terminals and all time slots from the beginning to the current one. The term $p_0$ is a constant penalty applied if the data rate within a time slot falls below a required average, ensuring the overall data delivery task is met. This structure directly incentivizes the agent to learn a trajectory that maximizes energy efficiency throughout its mission.

\paragraph{Security and adversarial problems}
This category addresses scenarios involving an intelligent adversary, such as an eavesdropper or jammer. The reward design principle is often based on a differential metric that captures the zero-sum nature of the problem. The goal is not just to improve the performance of the legitimate link, but to simultaneously degrade the performance of the adversary. For instance, in physical layer security with cooperative jamming, this principle is applied by maximizing the secrecy rate \cite{he2022deep}. First, the secrecy capacity $C^s$ for a single user is explicitly defined as the differential channel capacity between the legitimate user and the eavesdropper:
\begin{equation}
    C^{s}(u_{k}, \alpha_{k}) = [C_{\alpha_{k},k} - C^{e}(\alpha_{k})]^{+},
\end{equation}
where $u_k$ is the k-th legitimate user, $\alpha_k$ is the index of the access point serving user $u_k$, $C_{\alpha_{k},k}$ is the Shannon capacity of the legitimate link, and $C^{e}(\alpha_{k})$ is the capacity of the link to the most effective eavesdropper. The operation $[x]^{+} = \max(x,0)$ ensures the secrecy capacity is non-negative. Therefore, the overall reward function is defined as the sum of these individual secrecy capacities across all $K$ users in the system:
\begin{equation}
    R(S, a) = \sum_{k=1}^{K} C^{s}(u_{k}, \alpha_{k}).
\end{equation}
The action of the agent, denoted by $a$, represents the power allocation vector for all access points. By maximizing this aggregate reward $R(S,a)$, the agent is incentivized to learn a power allocation strategy that intelligently manipulates the signal environment to increase each $C_{\alpha_{k},k}$ while simultaneously suppressing each $C^{e}(\alpha_{k})$, thereby enhancing system-wide security.

\begin{figure}[t]
	\begin{center}
		\centerline{\includegraphics[width=0.6\linewidth]{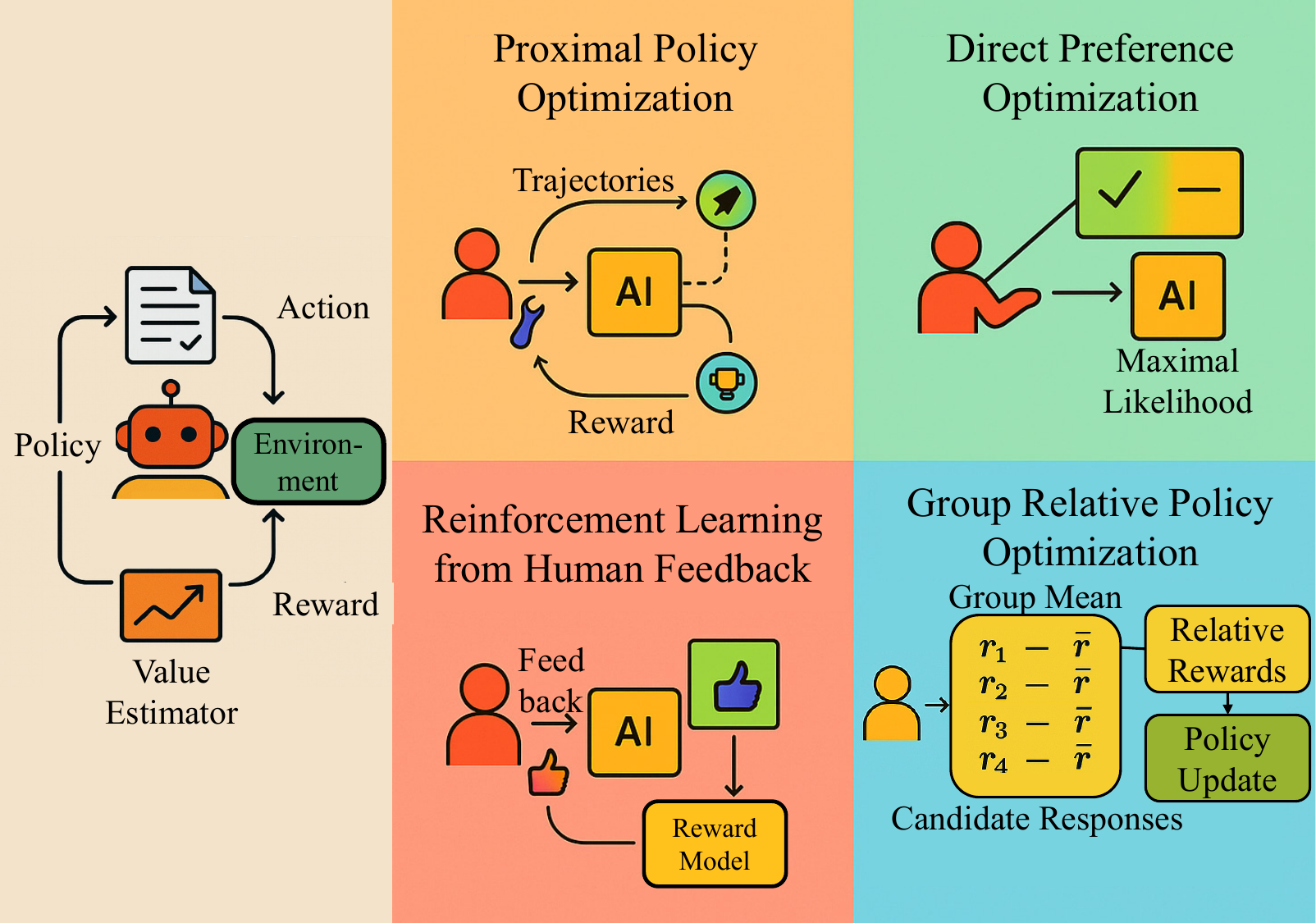}}
		\captionsetup{font=footnotesize, name={Fig.}, labelsep=period} 
		\caption{\, Advanced reinforcement learning algorithms.}
		\label{preliminary:rl}
		\vspace{-8mm}
	\end{center}
\end{figure}

\subsubsection{\textbf{Advanced RL algorithms for large model training and alignment}}
The following paragraphs present several advanced algorithms particularly suited to the scale and complexity of large models, as summarized in Fig. \ref{preliminary:rl}. The analysis begins with a foundational policy gradient method for stable training, before progressing to sophisticated paradigms designed explicitly for aligning model behavior with complex objectives.

\paragraph{Proximal policy optimization}
Proximal policy optimization (PPO) \cite{ppo} is a policy gradient RL algorithm that has gained significant popularity due to its stability and efficiency, making it a common choice for fine-tuning LLMs, especially in the context of RL from human feedback. PPO is designed to address the training instability common in earlier policy gradient methods, which are often susceptible to destructively large policy updates that can lead to a sudden collapse in performance. The core idea is to ensure that the updated policy remains proximal, or close, to the policy used for collecting data. To achieve this, PPO maximizes a surrogate objective function that incorporates a clipping mechanism. This mechanism constrains the probability ratio between the new and old policies, effectively discouraging policy updates that are too large. The clipped surrogate objective is denoted as
\begin{equation}
    L^{\text{CLIP}}(\theta) = \hat{\mathbb{E}}_t \left[ \min \left( r_t(\theta) \hat{A}_t, \text{clip}(r_t(\theta), 1-\epsilon, 1+\epsilon) \hat{A}_t \right) \right],
\end{equation}
which represents the core PPO clipped surrogate objective function $L^{\text{CLIP}}(\theta)$. The expectation $\hat{\mathbb{E}}_t$ is over a batch of transitions. $r_t(\theta) =\pi_\theta(a_t|s_t)/\pi_{\theta_{\text{old}}}(a_t|s_t)$ is the probability ratio of action $a_t$ under the current policy $\pi_\theta$ versus the policy $\pi_{\theta_{\text{old}}}$ used for data collection. $\hat{A}_t$ is the estimated advantage of taking action $a_t$ in state $s_t$. The clipping mechanism (using hyperparameter $\epsilon$) limits how much the probability ratio can change, enhancing stability. Specifically, when the advantage $\hat A_t$ is positive, this caps the objective to prevent an overly aggressive policy increase, while for a negative advantage, it sets a bound to avoid excessive penalization. By taking the minimum of the unclipped and clipped objectives, this formulation removes the overall incentive for the policy to change too drastically from its previous version, thereby ensuring more stable and reliable training updates.
PPO is widely used due to its simplicity, efficiency, and robustness when scaling to large models.

\paragraph{Reinforcement learning from human feedback}
Reinforcement learning from human feedback (RLHF) aligns LLMs with human preferences \cite{rlhf,rlhfpre}. It is especially effective in tasks where human judgment is clear but difficult to formalize. By incorporating human feedback, RLHF helps reduce biases and unsafe behavior, making it central to aligning models like InstructGPT \cite{instructgpt} and OpenAI o1 with human values. Unlike supervised learning, RLHF incorporates human judgment, refining model behavior in tasks where objectives are hard to define. The RLHF pipeline typically involves three stages: supervised fine-tuning on instruction-following examples, training a reward model with human-labeled preferences, and optimizing the model with RL, often using PPO to ensure stability and efficiency, denoted as
\begin{equation}
    \max_{\theta} \mathbb{E}_{x \sim \mathcal{D}, y \sim \pi_\theta(y|x)} \left[ R_\phi(x, y) - \beta D_{\text{KL}}(\pi_\theta(\cdot|x) || \pi_{\text{ref}}(\cdot|x)) \right],
\end{equation}
which shows the objective maximized during the RL phase of RLHF. The language model policy $\pi_\theta$ is optimized to generate responses $y$ for prompts $x$ (from distribution $\mathcal{D}$) that maximize the score from a learned reward model $R_\phi(x, y)$, which reflects human preferences. A KL-divergence penalty, weighted by $\beta$, regularizes the policy $\pi_\theta$ to stay close to a reference policy $\pi_{\text{ref}}$ (e.g., the initial supervised fine-tuned model), maintaining model capabilities and stability.
On the other hand, RLHF faces challenges, including the need for high-quality human feedback, the complexity of reward model training, and computational intensity. Additionally, there are risks of reward hacking \cite{rewardhack}, where models exploit reward signals without genuinely improving outputs. Despite these challenges, RLHF remains a powerful method for scalable LLM alignment and has potential applications in WLAM.
\par 
Such hybrid training strategies that combine supervised learning with RL can enhance model performance in dynamic environments. This hybrid approach combines data representation from supervised learning with adaptability from RL, enabling models to optimize for both structured data and uncertain environments. This results in WLAM that could be more resilient to non-stationary conditions like with unpredictable channel variations.

\paragraph{Direct preference optimization}
Direct preference optimization (DPO) \cite{dpo} is a alternative to RLHF that removes the need for an explicit reward model. DPO directly optimizes the language model using human preference data by framing alignment as a classification task. Given pairs of model outputs labeled by human preference, DPO adjusts model parameters to increase the likelihood of preferred responses, typically using a binary cross-entropy loss, as
\begin{equation}\label{dpo}
\mathcal{L}_{\text{DPO}}(\pi_\theta; \pi_{\text{ref}}) =
        - \mathbb{E}_{(x, y_w, y_l) \sim \mathcal{D}} [ \log \sigma (\beta \log \frac{\pi_\theta(y_w|x)}{\pi_{\text{ref}}(y_w|x)}
        -  \beta \log \frac{\pi_\theta(y_l|x)}{\pi_{\text{ref}}(y_l|x)}) ],
\end{equation}
which represents the DPO loss function, optimized directly on human preference data $\mathcal{D}$. Each data point consists of a prompt $x$, a preferred response $y_w$, and a dispreferred response $y_l$. The loss encourages the policy $\pi_\theta$ to assign a higher relative log-probability (compared to the reference policy $\pi_{\text{ref}}$) to the preferred response $y_w$ over the dispreferred response $y_l$. $\beta$ controls the strength of this preference margin, and $\sigma$ is the logistic sigmoid function.
This approach simplifies the pipeline, improves stability, and reduces computational overhead while achieving alignment quality comparable to or better than traditional RLHF.

\paragraph{Group relative policy optimization}
Group relative policy optimization (GRPO) is a recent RL algorithm designed for efficient fine-tuning of LLMs. Unlike PPO, which estimates the advantage using a critic to predict state values, GRPO computes relative rewards across a group of candidate responses generated for the same prompt. The reward for each response is measured relative to the group mean, and the advantage is calculated without an explicit value function, denoted as
\begin{equation}
    \hat{A}_{G}(x, y_i) = R(x, y_i) - \frac{1}{K} \sum_{j=1}^{K} R(x, y_j),
\end{equation}
which defines the group-relative advantage estimate central to GRPO. For a prompt $x$, the policy generates a group of $K$ candidate responses $\{y_j\}_{j=1}^K$. The advantage $\hat{A}_{G}$ of a specific response $y_i$ is computed as its reward $R(x, y_i)$ (obtained from a reward model or human feedback) minus the average reward across all $K$ responses in that group. This relative advantage measure replaces the value function estimate used in traditional policy gradient methods like PPO.
This approach simplifies training, reduces memory usage, and eliminates the need for large-scale critic gradient updates. Specifically, in DeepSeek-R1-Zero \cite{dsr1}, GRPO was applied without any supervised fine-tuning, yielding a model trained purely via RL. While early iterations struggled with readability, later versions like DeepSeek-R1 combined GRPO with multi-stage pretraining and alignment to achieve state-of-the-art results on reasoning tasks. The scalability of GRPO has also influenced other models such as QwQ-32B \cite{qwq32b}, which uses a GRPO-inspired two-stage RL method to achieve strong performance with just 32B parameters.
\par
GRPO is well-aligned with trends in telecom systems where low-overhead training, relative feedback mechanisms, and efficient distributed optimization are essential. Its use of group-wise advantage estimation reflects multi-agent learning setups or federated learning (FL) optimization schemes common in wireless networks. As large AI models continue to be deployed in decentralized and resource-constrained environments like satellites, GRPO offers a scalable and efficient path for adapting model behavior through lightweight on-device or near-device learning.

\subsection{Fine-tuning}
Fine-tuning pre-trained models with domain-specific data enhances performance for particular communication environments, which is essential for 6G and beyond as it eliminates the need for retraining from scratch. However, fine-tuning large AI models on edge nodes faces challenges beyond memory usage, such as high computational demands, energy efficiency, and network latency \cite{yu2024snake}. The frequent exchange of gradients or parameters can also strain distributed learning frameworks.
Additionally, techniques are necessary to adapt base models to high-level commands or specialized domains like telecommunications. 
To overcome these challenges, several promising techniques are introduced.

\subsubsection{\textbf{Parameter-efficient fine-tuning}}
Parameter-efficient fine-tuning (PEFT) is another way to reduce the cost of fine-tuning. PEFT encompasses techniques that adapt large pre-trained models with minimal parameter updates, reducing resource demands compared to full fine-tuning. In 6G networks, PEFT enables efficient deployment on edge devices by minimizing memory and computational costs, facilitating real-time applications like traffic prediction or resource allocation. Its lightweight nature ensures that advanced AI capabilities remain practical in resource-limited wireless environments.

\paragraph{Low-Rank Adaptation} Low-rank adaptation (LoRA) \cite{hu2021lora} adapts large pre-trained models by introducing low-rank updates to the weight matrices, freezing the original model parameters and training only the low-rank matrices, denoted as
\begin{equation}
    \mathbf{W} = \mathbf{W}_0 + \Delta \mathbf{W} = \mathbf{W}_0 + \mathbf{B} \mathbf{A},
\end{equation}
where $\mathbf{W}_0 \in \mathbb{R}^{d \times k}$ represents a pre-trained weight matrix which remains frozen during adaptation. The update $\Delta \mathbf{W}$ is constrained to be low-rank by decomposing it into the product of two smaller matrices, $\mathbf{B} \in \mathbb{R}^{d \times r}$ and $\mathbf{A} \in \mathbb{R}^{r \times k}$, where the rank $r \ll \min(d, k)$. Only the parameters of $\mathbf{A}$ and $\mathbf{B}$ are trained, drastically reducing the number of trainable parameters compared to updating the full $\mathbf{W}_0$ or an unconstrained $\Delta \mathbf{W}$.
This reduces the number of trainable parameters, making it ideal for resource-constrained environments. Variants like sparse LoRA \cite{zhu2023sira} enhance efficiency using sparse expert modules, while weight-decomposed LoRA \cite{liu2024dora} improves learning capacity by decomposing weights into magnitude and direction. LoRA also accelerates convergence in complex wireless datasets by simplifying the optimization landscape, preventing overfitting, and enabling faster adaptation to dynamic environments, improving model efficiency for tasks like traffic prediction and resource allocation.

\paragraph{Prefix tuning}
Prefix tuning \cite{li2021prefixtuning} is a parameter-efficient method that adds a continuous, task-specific prefix sequence to the input or hidden layers of a model. This prefix sequence is not made up of real tokens but instead consists of learnable parameters. The prefix can influence the behavior of model without modifying the entire weights. This approach allows models to retain their pre-trained parameters while optimizing only the task-specific prefix, making it memory-efficient and computationally light.

\paragraph{Prompt tuning}
Prompt tuning \cite{promptft} can be viewed as a simplified version of prefix tuning, where only a learnable prefix is added to the input text. This method has shown remarkable scaling capabilities; for sufficiently large models, prompt tuning alone can achieve performance comparable to full fine-tuning. By optimizing just the prompt parameters, prompt tuning provides an efficient way to adapt pre-trained models to specific tasks, minimizing the number of parameters that need to be updated.
In wireless networks, prompt tuning can be used to efficiently optimize models for various real-time applications, reducing memory and computational costs, and ensuring practical deployment on edge devices.

\subsubsection{\textbf{Domain adaptation}}
Domain adaptation focuses on aligning a general-purpose model with the unique data and conditions of a specific application domain. In the context of wireless systems, this includes adapting to local channel characteristics, device types, interference patterns, or mobility trends. Standard pre-trained models may perform poorly in such settings without further tuning. By continuing training on domain-specific datasets, such as network logs or user behavior traces, that models can capture key patterns relevant to the local environment. Parameter-efficient techniques like adapters or prefix tuning make domain adaptation feasible on edge devices with limited memory, enabling personalized or localized intelligence in large-scale 6G deployments.
\subsubsection{\textbf{Instruction tuning}}
Instruction tuning \cite{zhang2024instructiontuninglargelanguage} adapts a pre-trained model to better understand and follow natural language instructions across a wide range of tasks. Rather than training on task-specific data alone, the model learns from diverse instruction-response pairs, improving generalization and flexibility. In wireless networks, this allows operators to control or query the model using high-level commands such as “allocate bandwidth efficiently” or “detect congestion hotspots.” This approach reduces the need for hardcoded logic and enables more human-aligned behavior in communication environments. Instruction tuning is especially useful when tasks vary frequently or lack large labeled datasets.

\begin{figure}[t]
	\begin{center}
		\centerline{\includegraphics[width=0.6\linewidth]{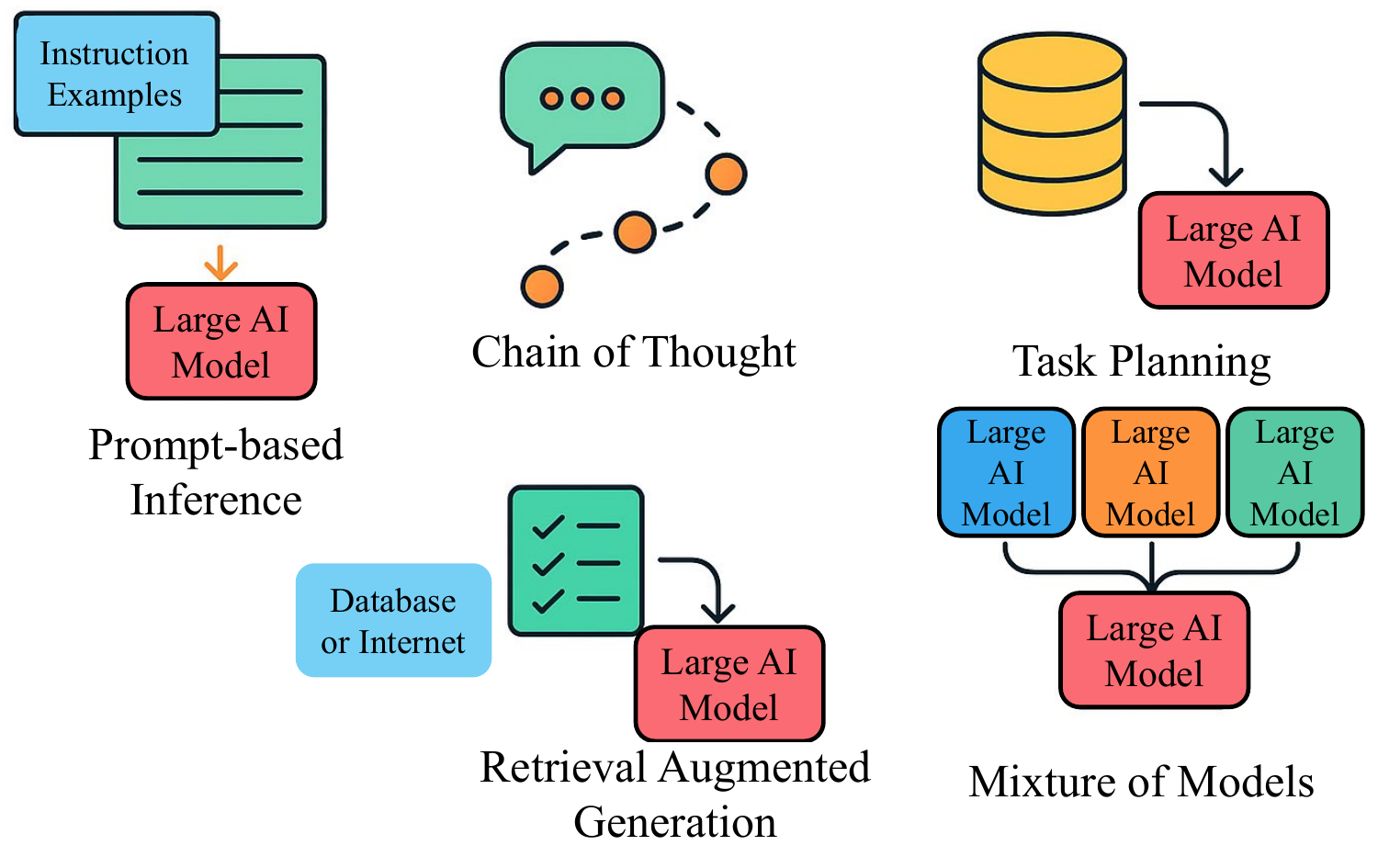}}
		\captionsetup{font=footnotesize, name={Fig.}, labelsep=period} 
		\caption{\, Inference strategies.}
		\label{preliminary:deploy}
		\vspace{-8mm}
	\end{center}
\end{figure}

\subsection{Inference strategies}
As large AI models continue to advance, their ability to perform tasks efficiently without the need for extensive retraining has opened up new opportunities for real-time, on-demand applications. In many use cases, particularly in complex fields like wireless communication, it is important to leverage the capabilities of these models to solve tasks based on available data, rather than re-training them for each new problem. This section explores various methods, such as in-context learning, prompt engineering, and other strategies that allow for efficient execution of pre-trained models in dynamic environments, as summarized in Fig. \ref{preliminary:deploy}.
\subsubsection{\textbf{Prompt-based inference}}
LLMs can solve tasks without parameter updates by using prompts with instructions and examples, as in-context learning \cite{brown2020language}. Instead of fine-tuning, the model generalizes from a few input-output demonstrations, enabling zero-shot and few-shot inference. This allows a single frozen model to adapt to various tasks by changing the prompt.
Prompt engineering guides the model by designing text prompts, which can include system instructions, user queries, or domain-specific examples. This helps the model understand wireless contexts. 
Combined with instruction tuning \cite{zhang2024instructiontuninglargelanguage}, prompt engineering enables large models to support diverse use cases such as configuration, protocol analysis, and troubleshooting.

\subsubsection{\textbf{Chain of thought}}
While basic prompts can handle simple queries, complex wireless problems benefit from multi-step reasoning. Chain of thought (CoT) prompting encourages the model to reason through intermediate steps before producing a final a answer~\cite{wei2022chain}. By appending phrases like reason step by step to the prompt, CoT enables more accurate and explainable results on tasks that require logical or numerical inference.
This technique modifies the structure of few-shot exemplars. Instead of standard 
$(input, output)$ pairs, CoT exemplars take the form of $(input, chain\_of\_thought, output)$ triples, where the $chain\_of\_thought$ is a series of sentences outlining the reasoning process.
Crucially, \cite{wei2022chain} shows that CoT is an emergent ability of model scale; it provides significant performance gains only for models with approximately 100 billion or more parameters, and can even degrade performance on smaller models. This dependency on large-scale models makes CoT a powerful but resource-intensive tool, positioning it as a strategic choice for high-value, complex problems within wireless systems.
In wireless systems, CoT is useful for anomaly detection, interference management, and resource optimization by guiding the model to iteratively explore the problem, effectively allocating more computation via token generation to tasks that require more reasoning steps.

\subsubsection{\textbf{Task planning}}
Task planning is a more advanced reasoning paradigm that builds upon, but is distinct from CoT reasoning. While CoT focuses on generating a sequence of internal reasoning steps to solve a single, self-contained query, task planning operates as a higher-level, externally-focused action framework. Its primary purpose is to create and execute a structured workflow by decomposing a high-level goal into a series of explicit, executable sub-tasks.
\par
The key distinction lies in the interaction with external tools and the environment. CoT is fundamentally an internal process, helping the model structure its thoughts to arrive at a final answer. In contrast, task planning acts as an orchestrator that not only plans the steps but also invokes external tools, application programming interfaces (APIs), or functions to perform actions at each stage. For example, to handle a high-level goal like ``optimize the network for a major public event", a task planning approach would create an executable plan, such as: first, calling a network monitoring API to get performance metrics; second, analyzing the retrieved data to confirm the bottleneck; and finally, calling a resource allocation API to reconfigure the network. Frameworks like HuggingGPT \cite{shen2023hugginggpt} exemplify this by using an LLM as a controller to manage and invoke other specialized AI models to fulfill complex user requests. This pipeline-style, tool-augmented planning is essential for creating truly autonomous wireless systems that can sense, analyze, and act upon the network environment.

\subsubsection{\textbf{Retrieval augmented generation}}
Retrieval augmented generation (RAG) is an approach that enhances inference by coupling it with an external knowledge retriever \cite{lewis2020retrieval}. In a RAG pipeline, a retriever first finds relevant documents or facts from a domain-specific datastore based on the query. 
The architecture features two main components: a neural retriever that finds relevant text passages for an input query and a generator that synthesizes an answer using the query and the retrieved content. This process treats the retrieved document $z$ as a latent variable that is marginalized to calculate the probability of generating a target sequence $y$ from an input $x$.
\par
Two primary RAG models are commonly used, differing in how they utilize retrieved documents. The RAG-Sequence model uses the same retrieved document to generate the entire output sequence. The probability of generating sequence $y$ given input $x$ is calculated by summing the likelihoods over the top-$K$ retrieved documents:
\begin{equation}
    p_{\text{RAG-Sequence}}(y|x) \approx \sum_{z \in \text{top-k}(p_{\eta}(\cdot|x))} p_{\eta}(z|x) \prod_{i=1}^{N} p_{\theta}(y_i|x,z,y_{1:i-1}). 
\end{equation}
On the other hand, the RAG-Token model can draw upon a different document for each token it generates, allowing for more dynamic synthesis of information. Its probability is calculated by marginalizing over the documents at each generation step:
\begin{equation}
    p_{\text{RAG-Token}}(y|x) = \prod_{i=1}^{N} \sum_{z \in \text{top-k}(p_{\eta}(\cdot|x))} p_{\eta}(z|x) p_{\theta}(y_i|x,z,y_{1:i-1}). 
\end{equation}
In these formulations, $x$ is the input prompt, and $y = (y_1, ..., y_N)$ is the generated output sequence of length $N$. The variable $z$ represents a retrieved document. The term $p_{\eta}(z|x)$ is the probability distribution of the retriever, which provides the likelihood of document $z$ being relevant to the input $x$. The summation $\sum_{z \in \text{top-k}(\cdot)}$ is performed over the set of top-$K$ documents returned by the retriever. Finally, $p_{\theta}(y_i|x,z,y_{1:i-1})$ is the probability of generating the next token $y_i$ given the original input $x$, the retrieved document $z$, and the previously generated tokens $y_{1:i-1}$, as modeled by the generator network with parameters $\theta$. These retrieved text snippets $z$ are provided as additional context to the generator, which integrates this evidence into its response. This mechanism effectively grounds the output in authoritative data, improving factual accuracy and reducing hallucinations \cite{lewis2020retrieval}.

\par
For example, TelecomRAG \cite{telecomrag2023} applies this paradigm to telecommunication standards by building a knowledge base from 3GPP specification documents and enabling an LLM to answer protocol questions with precise, verifiable references. In wireless communications, RAG is useful for tasks like dynamic protocol lookup, fault diagnosis, and network troubleshooting \cite{telerag}. An LLM agent can query a repository of network logs, configuration databases, or radiofrequency (RF) sensor readings and reason about network states in real-time. This yields grounded and trustworthy answers, especially important in rapidly evolving domains like wireless communications.

\subsubsection{\textbf{Mixture of models}}
Mixture of models uses multiple specialized models to tackle complex tasks by combining their expertise \cite{moa}. In practice, queries are distributed to a set of models, each offering insights based on their specialized knowledge. The outputs are then evaluated and synthesized by a coordinating model, which selects or combines the most accurate responses. This approach helps improve accuracy, mitigate biases, and ensure more reliable decision-making in domain-specific applications, such as telecommunications \cite{wang2025telemomconsensusdriventelecomintelligence}, where precise expertise is critical.

\subsubsection{\textbf{Inference optimization and acceleration}}
Practical WLAM deployment is limited by inference efficiency, as serving is memory-bound due to the large and dynamically sized key-value (KV) cache. Conventional systems store the KV cache in contiguous memory, leading to severe fragmentation that can waste over 60\% of the allocated memory, thereby limiting batch sizes and overall throughput.
To solve this, vLLM \cite{vllm} introduces PagedAttention, an algorithm that manages the KV cache in non-contiguous, fixed-size blocks, analogous to operation system paging. This design nearly eliminates memory fragmentation. The attention computation is adapted to operate on these non-contiguous blocks, where the output $o_i$ for a query $q_i$ is computed as follows:
\begin{equation}
    o_i = \sum_{j=1}^{\lceil i/B \rceil} \frac{\exp(q_i^\top K_j / \sqrt{d})}{\sum_{t=1}^{\lceil i/B \rceil} \exp(q_i^\top K_t \mathbf{1} / \sqrt{d})} V_j.
\end{equation}
Here $K_j$ and $V_j$ are blocks containing the key and value vectors, $B$ is the block size (a fixed number of tokens), and $d$ is the head dimension for scaling. 
By enabling much larger batch sizes, this method yields a 2 to 4 times throughput improvement over state-of-the-art systems.
Frameworks like SGLang \cite{sglang} further optimize complex, multi-call programs common in agentic workflows. For workloads with shared prefixes, its RadixAttention technique uses a radix tree to systematically reuse the KV cache across different requests, avoiding redundant computation. For structured generation (e.g., JSON), SGLang employs a compressed finite state machine that allows decoding of multiple deterministic tokens in a single forward pass. These co-designed runtime optimizations can increase throughput by up to 6.4 times on complex tasks.

\subsection{Multimodal models and multimodal alignment}
The increasing complexity of wireless systems with their need to handle various types of data modalities (e.g., signals, images, text, and audio) has driven the development of multimodal models. These models integrate diverse modalities into a unified system, enabling more holistic reasoning and decision-making. By processing and aligning information across multiple types of input, multimodal models unlock powerful new applications in wireless communication, such as cross-modal inference, environment-aware communication, and sensor fusion. This section explores the key architectures, strategies, and challenges for enabling seamless integration of diverse data types.

\subsubsection{\textbf{Architectures and fusion strategies}}
Multimodal models integrate information from data types like text, images, and RF signals. The fundamental approach for processing multimodal data involves a two-stage process. First, modality-specific encoders are used to independently extract high-level features from each data type; for instance, a specific Transformer might process images while another specific Transformer handles language. Second, a fusion layer or mechanism combines these disparate feature streams into a unified, cross-modal representation. Architectures such as contrastive language-image pre-training \cite{radford2021learning} and stable diffusion \cite{stadiff} use modality-specific encoders and fusion layers, while models like Hunyuan-DiT \cite{li2024hunyuandit} and Qwen-VL \cite{qwenvl} apply these concepts to generative and question-answering tasks.
\par
In wireless systems, the architectural choice reflects a trade-off between perception and efficiency. Multimodal Transformers leverage self-attention for complex perception tasks like sensing-assisted beam prediction \cite{tian2023multimodaltransformerswirelesscommunications}, while efficient models like vector-quantized variational autoencoders (VQVAE) are used for network-centric tasks such as CSI feedback optimization \cite{Bocusvqvae}.
\par
As an example, the loss function for a multimodal VQVAE (MVQVAE) with $M$ modalities is designed to train encoders and decoders using a shared codebook, which is the key to fusing information. A representative loss function $L_{\text{MVQVAE}}$, adapted from the multimodal framework in \cite{Bocusvqvae}, is expressed as:
\begin{equation}
    L_{\text{MVQVAE}} = \sum_{m=1}^{M} \|\mathbf{x}_m - \hat{\mathbf{x}}_m\|^2_2 + \sum_{m=1}^{M} \|\text{sg}(\mathbf{z}_{e,m}) - \mathbf{z}_q\|^2_2 + \beta \sum_{m=1}^{M} \|\mathbf{z}_{e,m} - \text{sg}(\mathbf{z}_q)\|^2_2. 
\end{equation}
The total loss sums three components over all $M$ modalities. The first term is the reconstruction loss between the original input $\mathbf{x}_m$ and the reconstruction $\hat{\mathbf{x}}_m$. The second term, the codebook loss, updates the shared codebook embedding $\mathbf{z}_q$ to match the encoder output $\mathbf{z}_{e,m}$. The third term, the commitment loss, is scaled by a hyperparameter $\beta$ and regularizes the encoder to produce vectors close to the codebook embeddings to minimize quantization error. The stop-gradient operator $\text{sg}(\cdot)$ isolates gradient updates to either the codebook or the encoders in the second and third terms, respectively, enabling end-to-end training.
\par
Furthermore, these architectures implement various fusion strategies based on system requirements \cite{shi2024radarcamerafusionobject}. Early fusion combines raw data at the input layer, ideal for cooperative sensing. Intermediate fusion merges features at middle layers, balancing performance and modularity. Late fusion combines final decisions from independent models, offering robustness for decentralized applications like federated learning.

\subsubsection{\textbf{Real-world challenges and alignment}}
The practical application of these methods in wireless environments faces significant hurdles from low-quality data \cite{zhang2024multimodalfusionlowqualitydata}. Wireless signals are often noisy or incomplete due to sensor errors and interference. Data can also be imbalanced, causing models to over-rely on a dominant modality, while the quality of any single sensor may vary dynamically with environmental conditions. While the architectural fusion strategies discussed previously provide a structural basis for combining modalities, simple feature concatenation often fails when inputs are severely degraded. To overcome these issues and establish robust multimodal alignment, it is imperative to implement specific technical mechanisms that project diverse wireless signals into a shared semantic space. First, cross-modal contrastive learning (architectures such as CLIP \cite{ramesh2022hierarchical}) serves as the foundational mechanism. By pulling paired heterogeneous samples (e.g., an RF CSI matrix and its corresponding environmental image or textual description) closer in a unified high-dimensional embedding space while pushing unmatched pairs apart, the model learns the intrinsic semantic correlations across domains. Second, to tackle noisy and imbalanced data, cross-attention mechanisms act as dynamic weighting filters. These allow the model to adaptively allocate attention scores based on real-time data quality, seamlessly shifting reliance to a more reliable modality (e.g., prioritizing vision when the RF signal suffers from severe fading). Finally, for handling incomplete data, masked multi-modal modeling \cite{mizrahi20234m} can be introduced to reconstruct missing sensor inputs by leveraging the mutual information embedded within the shared semantic space. In wireless contexts, actively deploying these deep alignment mechanisms is essential for applications such as vision-RF fusion and semantic signal labeling, ultimately enabling more resilient and context-aware system performance \cite{LLM4CP, alayrac2022flamingo}.

\begin{table*}[!t]
\renewcommand{\arraystretch}{1.09}
\scriptsize\centering
\caption{Progress Analysis of WLAM with Its Design Principles.}
\label{tab:wlam_principles}
\begin{tabular}{|m{0.12\textwidth}<{\centering}|m{0.22\textwidth}<{\raggedright}|m{0.29\textwidth}<{\raggedright}|m{0.3\textwidth}<{\raggedright}|}
\hline
\textbf{Design Principle} & \textbf{Current Progress \& State-of-the-Art} & \textbf{Existing Work} & \textbf{Gaps \& Future Directions} \\
\hline
\textbf{Adaptability to Dynamic Environments} &
Use of RL, adaptive architectures (MoE, LNNs), and meta-learning for real-time adaptation to varying wireless conditions. &
Discussed applications in adaptive physical layer tasks (Sec 3.1), network resource allocation (Sec 3.2), and the Agentic AI-RAN paradigm for autonomous operation (Sec 3.4). &
Handling extreme non-stationary conditions (e.g., high mobility) remains a challenge. Future work needs more efficient on-device online training and zero-shot generalization to unseen scenarios. \\
\hline
\textbf{Efficiency for Edge Deployment} &
Advances in PEFT (e.g., LoRA), model compression, and efficient architectures (e.g., Mamba) with linear complexity. &
Survey covers distributed learning to reduce client load (Sec 4.1, 4.2), AirComp for communication efficiency (Sec 4.3), and next-gen sequence models (Sec 5.2.3). &
The performance-efficiency trade-off remains a key challenge. Future directions include better hardware-software co-design for edge AI and a stronger focus on energy efficiency, not just computational cost. \\
\hline
\textbf{Multimodality and Data Diversity} &
Development of large multimodal models for aligning diverse data (signals, text, images), including vision-aided wireless and RAG for knowledge fusion. &
Applications in ISAC for sensor fusion (Sec 3.1.3) and Semantic Communications for various data types (Sec 3.3). The need for multi-modal datasets is also identified (Sec 6.1.4). &
Challenges remain in the real-time fusion of unsynchronized data and ensuring robustness to missing modalities. There is a critical need for large-scale, standard wireless multi-modal datasets to advance future research. \\
\hline
\textbf{Privacy and Security} &
Use of federated learning for privacy, enhanced by techniques like differential privacy and lightweight physical layer security. &
Survey details the dual role of physical layer security for AI security (Sec 4.4) and AI for enhancing physical layer security (Sec 3.1.5). Core privacy methods like federated learning (Sec 4.2) and key threats (Sec 6.4) are discussed. &
Federated learning vulnerabilities (e.g., poisoning) and the high overhead of cryptography persist. Future directions include better balancing the privacy-security-performance trade-off and developing methods for auditing complex WLAM systems. \\
\hline
\end{tabular}
\end{table*}

\subsection{Deployment of large AI models in wireless networks}
Deploying large AI models in wireless networks involves navigating complex trade-offs between computational efficiency, latency, privacy, and scalability. To meet the demands of real-time and resource-constrained environments, the feasibility of fine-tuning large AI models on an edge node in 6G networks primarily hinges on peak memory usage, which is dominated by model parameters, gradients, optimizer states, data batch size, intermediate activations, and fragmented memory from residual states \cite{yu2024snake}.

\subsubsection{\textbf{Architectural strategies and optimization techniques}}
Deploying large AI models in wireless communication systems requires architectural strategies that address constraints such as limited computation, stringent latency, and energy efficiency. Centralized deployments on high-performance GPU clusters that are typically hosted on public cloud platforms enable scalable inference for tasks like traffic prediction, network planning, and simulation. These platforms support serving large models via APIs to multiple users or base stations. However, latency, bandwidth cost, and privacy concerns arise when transferring telemetry to remote servers. To mitigate these limitations, hybrid solutions can cache knowledge locally or offload selective tasks closer to the data source.
\par
When deploying models closer to the network edge, compression techniques like pruning, quantization, and distillation become essential to reduce memory and compute demands \cite{han2015deep,xiao2023smoothquant}. These optimizations allow models to run on smaller-scale hardware such as edge GPUs or embedded processors. Advanced deployment schemes such as federated learning enable local model training while preserving data privacy, and split learning approaches can distribute model inference between local and remote devices \cite{kang2017neurosurgeon}. Such strategies are crucial for real-time applications like adaptive scheduling, beam tracking, and anomaly detection, which are critical for real-time applications like ultra-reliable low latency communications (URLLC). Adaptive mechanisms and continuous monitoring are also necessary to ensure robust performance under dynamic and unpredictable wireless conditions.

\subsubsection{\textbf{Hardware and specialized frameworks}}
Effective deployment also relies heavily on selecting suitable hardware and software frameworks. Inference engines must accommodate resource-constrained environments such as mobile devices, base stations, or multi-access edge computing (MEC) nodes. Specialized hardware accelerators like neural processing units enable efficient execution of AI models in these scenarios. On-device inference supports applications like personalized assistants, local RF analytics, and user-side diagnostics, offering benefits such as low-latency responses, offline functionality, and enhanced privacy \cite{deng2020edge}.
\par
To further support lightweight deployments, frameworks like LiteRT \cite{litert}, PyTorch Mobile \cite{pytorchmobile}, and ONNX Runtime \cite{onnxrun} provide infrastructure for running quantized or distilled models on constrained hardware. Efficient model architectures, such as minGPT \cite{mingpt}, complement these frameworks by reducing computational complexity without compromising significantly on performance. In hybrid deployments, devices can dynamically offload intensive tasks to more capable nodes depending on real-time conditions, maintaining a balance between autonomy and performance. Security and ethical considerations must also be integrated throughout the deployment lifecycle to safeguard sensitive data and prevent model misuse. Altogether, these strategies enable large AI models to enhance wireless networks across domains such as signal processing, resource allocation, and intelligent network management.

\begin{table*}[!t]
\renewcommand{\arraystretch}{1.4}
\scriptsize\centering
\caption{Architectural comparison: 5G + AI vs. AI-empowered 6G.}
\label{tab:ai_native_comparison}
\begin{tabular}{|m{0.12\textwidth}<{\centering}|m{0.41\textwidth}<{\raggedright}|m{0.41\textwidth}<{\raggedright}|}
\hline
\textbf{Dimension} & \textbf{5G + AI} & \textbf{AI-empowered 6G (WLAM-based)} \\
\hline
\textbf{Integration Depth} & 
AI optimizes specific tasks like channel estimation and beamforming within existing telecommunication protocols. & 
The protocol stack and air interface are designed with AI integration, such as semantic communications. \\
\hline
\textbf{Model Paradigm} & 
Task-specific small models are employed, which requires retraining for new scenarios. & 
Foundation models are trained for all tasks, enabling general knowledge transfer and few-shot learning. \\
\hline
\textbf{Data Processing} & 
Data processing is managed at its own dedicated layer, such as separate source and channel coding.& 
Multimodal data including RF, vision, and text are fused and processed in an end-to-end mode.
\\
\hline
\textbf{Deployment Strategy} & 
Models generally operate within the individual network devices while utilizing standard communication links for any necessary coordination.& 
The network architecture supports distributed deployment mechanisms such as split computing to accommodate massive models on constrained edge devices.
\\
\hline
\textbf{Operation Mode} & 
Tasks are executed by models based on strictly predefined algorithms, requiring constant human intervention and monitoring. & 
Agents autonomously perceive the environment, reason about their goals, formulate complex plans, and execute intricate workflows. \\
\hline
\end{tabular}
\end{table*}

\subsection{Design principles and key characteristics of WLAM systems}
The integration of large AI models into wireless networks offers immense potential to transform communication systems, particularly as we transition toward 6G and beyond. These models enhance flexibility, scalability, and efficiency, enabling wireless networks to meet the growing demands of increasingly complex environments and diverse applications. In this section, we explore the key characteristics and potential of WLAM, highlighting how their multi-functional, multi-resource, and scalable capabilities can address the challenges posed by modern and future wireless networks.
\par
Below, we highlight several critical aspects that must be considered to ensure the effectiveness and efficiency when designing methods that integrate large AI models with wireless communications.
First, the adaptability of the models to dynamic environments is essential. Wireless networks experience fluctuating signal strengths, mobility, and interference. Thus, AI models must adjust in real-time without extensive retraining to maintain performance in unpredictable conditions.
Given the constraints of wireless systems, such as latency, energy consumption, and computational limitations, WLAMs must be optimized for edge deployment. This involves designing models to function within the limited resources of edge devices while maintaining performance.
The integration of diverse data types is crucial. Wireless systems rely on radio-frequency signals, images, and text, so WLAM must seamlessly process and align these modalities to enable context-aware decision-making across varying network conditions.
Privacy and security concerns are critical. AI models in wireless systems must protect sensitive data, comply with regulations, and minimize biases, ensuring data security and privacy to maintain trust and meet legal requirements.
Finally, WLAM should be adaptive. The models must adapt in real-time based on network feedback, ensuring continuous performance improvement without manual intervention. This feature is vital for supporting the evolving demands of next-generation wireless systems.
\par
To further clarify the transformative role of WLAMs within these design principles, it is essential to distinguish the vision of an AI-empowered 6G system from the fifth generation of cellular network technology (5G) systems enhanced with AI capabilities. The fundamental difference lies not merely in the model size, but in the architectural philosophy and operational paradigm. In 5G, AI is typically applied as an add-on feature to optimize specific, isolated modules (e.g., AI-based CSI feedback or beam management) within a fixed protocol framework. This represents a ``one model for one task'' paradigm, where models lack cross-task generalization. In contrast, an AI-empowered 6G system implies that the air interface and network architecture are designed from scratch with WLAMs at their core \cite{fontaine2024towards}. This shift moves towards ``one foundation model for all tasks,'' enabling the network to handle unseen scenarios through the inherent generalization and reasoning capabilities of WLAMs \cite{wirelessgpt, LLM4WM}. Simultaneously, regarding data processing, the paradigm evolves from the separate source and channel coding blocks of 5G to an end-to-end approach, where WLAMs intrinsically fuse and process multimodal data including RF signals, vision, and text \cite{cheng2025large}. In terms of deployment, the system transitions from operating within the individual nodes to a collaborative edge-cloud synergy, where techniques like split computing are essential to accommodate massive model parameters. Furthermore, AI-empowered 6G introduces agentic AI, where the system possesses the autonomy to decompose high-level intents into executable network configurations \cite{jiang2025largeaimodelsagentic}. Table \ref{tab:ai_native_comparison} provides a structured comparison of these criteria.
\par
To sum up, integrating WLAM in wireless networks offers significant potential to address the growing complexity of communication systems. Specifically, the diverse enabling technologies discussed in this section map directly to critical 6G performance goals and deployment constraints: advanced neural architectures like Transformers and scaling strategies such as MoE enable the high-capacity representation learning necessary for strong generalization across diverse wireless scenarios; multimodal learning bridges heterogeneous data modalities to enhance context-awareness and sensing precision; diffusion-based models leverage iterative denoising processes to enhance robustness against non-linear physical layer distortions; reinforcement learning frameworks empower the network with dynamic adaptability and high-level autonomy; parameter-efficient fine-tuning addresses the stringent energy and storage efficiency constraints of edge deployment; and advanced inference strategies such as RAG and CoT significantly improve the reliability and factual accuracy of agentic decisions. By effectively combining multiple resources, adapting to dynamic environments, and ensuring privacy and security, WLAM can enable more efficient, scalable, and intelligent wireless systems essential for the success of 6G and beyond. We provide a summarization of the state-of-the-art technologies regarding these design principles in Table \ref{tab:wlam_principles}.

\section{Large AI models for wireless communications}\label{section3}

\begin{figure}[t]
	\begin{center}
		\centerline{\includegraphics[width=0.87\linewidth]{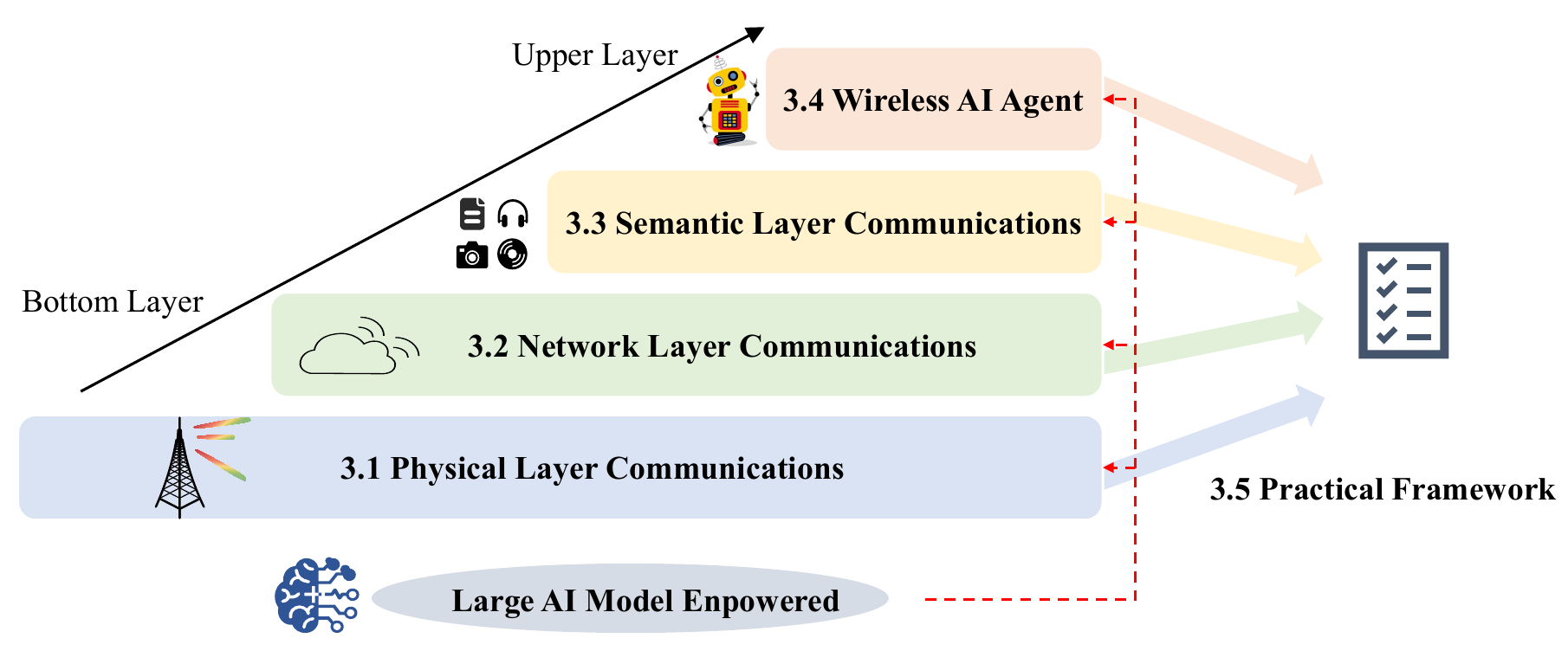}}
		\captionsetup{font=footnotesize, name={Fig.}, labelsep=period} 
		\caption{\, The outline of Section 3.}
		\label{outline:sec3}
		\vspace{-4mm}
	\end{center}
\end{figure}

The escalating complexity of modern wireless systems and the ever-increasing demand for enhanced performance present significant challenges. To overcome these, large AI models are rapidly emerging as powerful tools within the field of wireless communications. Building on the fundamentals established in Section \ref{section2}, the key to this transformative potential lies in the exceptional generalization and reasoning capabilities of WLAMs. Unlike traditional models that struggle with unseen channel distributions, the powerful generalization capability of WLAMs derived from massive-scale pre-training forms the cornerstone for adapting to the stochastic, time-varying wireless environments. With this logical foundation, this section delves into the specific applications of WLAMs by systematically progressing through the communication stack, as depicted in Fig. \ref{outline:sec3}. Our discussion begins at the physical layer, moves to the network and semantic layers, and explores the emerging concept of wireless agents. Finally, these diverse applications are synthesized into a structured framework to provide practitioners with practical implementation guidelines. An overview of the related works is provided in Table \ref{tab:AI_for_Wireless}.

\renewcommand{\arraystretch}{1.32}
\begin{table*}[!t]
\scriptsize\centering
\caption{Summary of related works on large AI models for wireless communications.}
\label{tab:AI_for_Wireless}
\begin{tabular}{|m{0.12\textwidth}<{\centering}|m{0.05\textwidth}<{\centering}|m{0.35\textwidth}<{\raggedright}|m{0.35\textwidth}<{\raggedright}|}
\hline
\textbf{Layer} &
\textbf{Ref.} &
\textbf{Scenarios} &
\textbf{Contributions} \\ \hline

\multirow{5}{*}[-4em]{\centering\parbox{0.09\textwidth}{\centering Physical Layer}} 
  & \cite{LLM4CP}
  & Adapting LLMs for high-performance wireless channel prediction in TDD and FDD systems.
  &  Propose LLM4CP, demonstrating that a fine-tuned LLM can achieve state-of-the-art performance for wireless channel prediction.
  \\ \cline{2-4}

  & \cite{LLM4WM}
  & Adapting a single pre-trained LLM to multiple, diverse wireless communication tasks.
  &  Propose LLM4WM, a general framework for efficient multi-task fine-tuning of LLMs in various wireless scenarios.
  \\ \cline{2-4} 

  & \cite{10104549}
  & End-to-end channel coding for unknown/non-differentiable channels.
  &  Leverage generative diffusion models, a key architecture in modern large AI, for end-to-end channel decoding.
  \\ \cline{2-4} 
  
  & \cite{cheng2025large} 
  & Multimodal ISAC systems, exemplified by beam prediction using multimodal sensor data (e.g., GPS, RGB images).
  & Propose an MLLM-enabled framework for multimodal ISAC to enhance communication and sensing.
  \\ \cline{2-4} 
  
  & \cite{wcmsecurity}
  & Physical layer communication security optimization problems, specifically cooperative friendly jamming scenarios.
  & Propose an MoE-enabled generative artificial intelligence framework to enhance security in cooperative jamming scenarios.
  \\ \hline

\multirow{3}{*}[-1.5em]{\centering\parbox{0.095\textwidth}{\centering Network Layer}}
  & \cite{10829820}  
  & Intelligent network operations and performance optimization in future networks.        
  &  Provide a comprehensive survey on applying LLMs to intelligent network operations and performance optimization.
  \\ \cline{2-4} 

  & \cite{NetLLM} 
  & Adapting LLMs for various networking tasks, including prediction and decision-making involving multimodal inputs.             
  &  Propose NetLLM, the first framework to efficiently adapt LLMs for networking tasks.
  \\ \cline{2-4}
  
  & \cite{chen2024netgpt}
  &  Providing personalized generative services in future AI-empowered networks via collaborative cloud-edge LLM deployment.
  & Propose NetGPT, an AI-empowered network architecture synergizing cloud and edge LLMs for network management tasks.
  \\ \hline

\multirow{3}{*}[-1.5em]{\centering\parbox{0.095\textwidth}{\centering Semantic Layer}}
  & \cite{wu2024cddm}
  & Wireless semantic communications over noisy channels, where channel noise degrades the received signal after equalization.
  & Propose CDDM as a post-equalization module to remove channel noise by learning the input signal distribution.  
  \\ \cline{2-4}
  
  & \cite{wang2024large}            
  & Semantic communication systems for text transmission over wireless channels, applying LLMs to encoding and decoding. 
  & Propose LLM-SC, the first framework using LLMs for physical layer semantic encoding and decoding. 
  \\ \cline{2-4}

  & \cite{ni2024interplay}            
  & Exploring the synergistic interplay between semantic communication and the knowledge base of large AI models.
  & Investigate the synergistic interplay where knowledge within a large model is leveraged for communication efficiency.
  \\ \hline

\multirow{5}{*}[-2.5em]{\centering\parbox{0.06\textwidth}{\centering Wireless Agents}} 
  & \cite{zou2024telecomgpt}             
  & Adapting general-purpose LLMs for telecom-specific tasks.           
  & Propose a framework and pipeline to create telecom-specific LLMs, showing strong performance on telecom tasks.
  \\ \cline{2-4}
  
  & \cite{du2024powerlargelanguagemodels}            
  & FPGA-based hardware development for advanced wireless communication signal processing algorithms.        
  & Investigate LLM assistance in FPGA-based SDR development via a case study and identified key LLM uses.
  \\ \cline{2-4}

  & \cite{xiao2025agenticainetworking6g}              
  & Agentic AI networking in 6G, supporting interaction, collaborative learning, and knowledge transfer among diverse AI agents.
  & Propose AgentNet, a novel framework for agentic AI networking, and demonstrated its potential in two application scenarios.
  \\ \cline{2-4} 

  & \cite{rezazadeh2025generative6gsimulationexperimental}
  & Automating the full lifecycle of complex 5G/6G network simulations using the ns-3 simulator. 
  & Propose an innovative multi-agent framework integrating LLMs with ns-3, and validate effectiveness through a 5G case study.
   \\ \hline
   
\end{tabular}
\end{table*}

\subsection{WLAM for physical layer communications}
\begin{figure*}[t]
	\begin{center}
		\centerline{\includegraphics[width=0.85\linewidth]{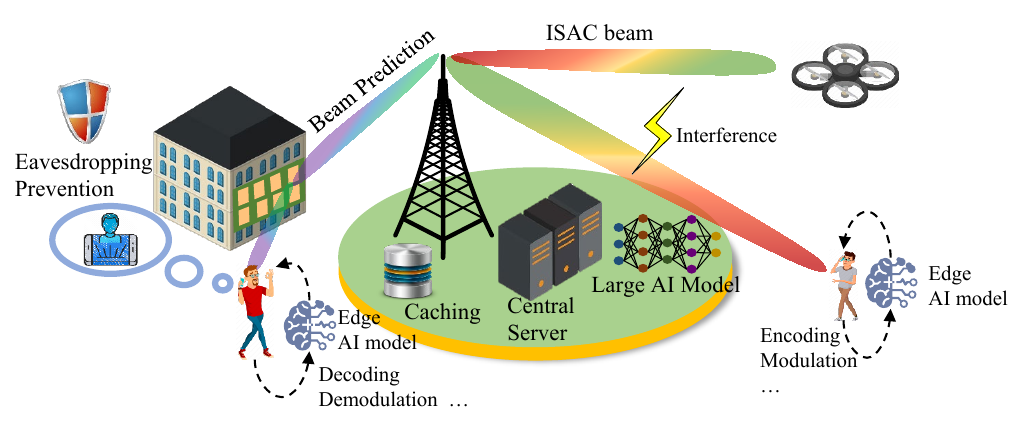}} 
		\captionsetup{font=footnotesize, name={Fig.}, labelsep=period} 
		\caption{\, Large AI models for physical layer communications. WLAMs empower the physical layer by orchestrating complex, dynamic tasks, including end-to-end transceiver design, accurate beam prediction, multimodal ISAC, predictive edge caching, and robust physical layer security against malicious interference.}
		\label{fig: Large AI models for physical layer communications}
		\vspace{-8mm}
	\end{center}
\end{figure*}

Large AI models are being explored to revolutionize physical layer communication, addressing inherent challenges and unlocking unprecedented performance levels. Traditional physical layer design often relies on signal processing techniques with limited adaptability to complex and dynamic wireless environments. Large AI models, especially deep neural networks, offer the capability to learn intricate patterns from vast datasets of wireless signals and channel characteristics. This learning ability enables the development of intelligent physical layer solutions that can dynamically optimize resource allocation, enhance signal transmission and reception, and mitigate impairments more effectively than conventional methods. The application of large AI models in the physical layer aims to achieve superior spectral efficiency, energy efficiency, and link reliability, paving the way for next generation wireless networks, as shown in Fig. \ref{fig: Large AI models for physical layer communications}.

\subsubsection{\textbf{Channel-associated prediction}}
Channel-associated prediction tasks are a significant area within physical layer communications where large AI models show remarkable potential. Accurate prediction of CSI is crucial for adaptive resource management, beamforming optimization, and interference mitigation. Traditional channel prediction methods often struggle with rapid channel variations in mobile wireless environments and complex non-linear relationships inherent in wireless propagation. Large AI models, leveraging their powerful function approximation capabilities, can learn temporal and spatial channel dynamics from historical data and environmental context. For example, in beam prediction for millimeter wave systems, LLMs are being explored to forecast optimal beam directions using past beam indices and angles of departure \cite{Sheng2025}. Vision-aided techniques are also emerging. BeamLLM from \cite{zheng2025beamllmvisionempoweredmmwavebeam} utilizes LLMs to process colored images for beam prediction, demonstrating high accuracy in vehicle-to-infrastructure scenarios. These models exhibit strong robustness and generalization capabilities, outperforming traditional learning based methods in prediction accuracy and adaptability to diverse wireless conditions.
\par
Enhancing the robustness of these predictions against the rapid channel variations and inherent uncertainties of wireless environments is important. Bayesian learning techniques, particularly Bayesian neural networks (BNNs), offer a principled framework for quantifying prediction uncertainty \cite{bayesianTCCN}. Unlike conventional models that provide point estimates, BNNs produce predictive distributions, distinguishing between aleatoric uncertainty (inherent data noise) and epistemic uncertainty (model uncertainty due to limited data). This uncertainty quantification enhances reliability, especially with limited training data or in the presence of outliers and model misspecification, common issues in wireless settings. While traditional BNN inference can be computationally intensive, approaches like real time teacher-student BNNs have been developed to enable fast uncertainty aware predictions suitable for dynamic applications like vehicle tracking \cite{bayesianJSAC}. Integrating such uncertainty awareness into WLAMs could significantly improve their robustness and decision making capabilities under fluctuating channel conditions.
\par
Furthermore, the accuracy and robustness of channel predictions can be substantially improved by fusing information from diverse data sources. Wireless systems generate heterogeneous data streams, including CSI, beam training logs, and environmental sensor data such as camera images, LiDAR point clouds, or radar signals. WLAM, often built upon transformer architectures, inherently possesses powerful mechanisms like attention for multimodal data fusion \cite{ZHOU2023126604}. 
Attention based fusion allows the model to dynamically weigh the importance of different data sources or features based on the current context, effectively integrating complementary information and enhancing prediction accuracy. For example, visual data can provide context for blockages, while CSI captures temporal dynamics. To further enhance robustness in noisy environments, models can be designed to account for data quality, or Bayesian uncertainty estimates can be used to give less weight to unreliable sources.
\par
Building upon these advancements in specific prediction tasks, general frameworks are being developed to harness large AI models for a broader range of wireless channel-associated problems. The LLM4WM framework, detailed in \cite{LLM4WM}, introduces a comprehensive approach for adapting LLMs to these challenges. This framework employs a MoE with LoRA to achieve efficient multi task fine tuning, thus enabling the transfer of pre-trained LLM knowledge to diverse wireless communication scenarios. Furthermore, the effectiveness of fine-tuned LLMs for general channel prediction is highlighted by LLM4CP, presented in \cite{LLM4CP}, which achieves state of the art performance in both time division duplex and frequency division duplex systems. By leveraging large AI models and general frameworks like LLM4WM for channel-associated prediction tasks, wireless systems can proactively adapt to channel fluctuations, optimize transmission parameters, and enhance overall network performance across multiple functionalities.

\subsubsection{\textbf{Transceiver design}}
Transceiver design represents a pivotal area in wireless communication, and large AI models are increasingly demonstrating their capability to transform traditional methodologies. Conventional transceiver designs often rely on handcrafted algorithms tailored to specific channel models. However, the adaptability and complexity of large AI models offer a paradigm shift, enabling data-driven approaches that can learn and optimize transceiver functionalities directly from communication data. This encompasses various aspects of transceiver design, from channel coding and decoding to modulation and demodulation, and even end-to-end system optimization. Recent surveys, such as the review on machine learning for channel coding in \cite{accessdecoding}, highlight the burgeoning interest in AI-driven channel coding techniques. Furthermore, research into AI-aided receivers, exemplified by \cite{TWCDecode} on adaptive orthogonal frequency division multiplexing receivers, showcases the practical application and performance gains achievable through AI in receiver design.
\par
A key paradigm enabled by WLAM is the end-to-end optimization of the entire communication system. This approach treats the transmitter and receiver as components of a single neural network, allowing for joint optimization of their parameters using gradient back-propagation. The central challenge in this framework is the physical wireless channel itself. Because the channel introduces stochastic noise and distortions in a way that is not directly differentiable with respect to the transmitter output, it prevents direct gradient flow from the receiver output back to the transmitter input. To enable end-to-end training via back-propagation, a differentiable path between transmitter and receiver is required during the training phase. WLAM offers solutions to this. For instance, a differentiable channel surrogate model can be learned explicitly or implicitly using techniques like diffusion models \cite{10104549} or generative adversarial networks (GANs). This learned, differentiable model mimics the channel behavior while allowing gradient flow, acting as a bridge during offline training. Alternatively, the system can be trained using a known, mathematically differentiable channel model (such as additive white Gaussian noise or a simplified fading model that simulates noisy conditions), similar to training an autoencoder \cite{choukroun2022error}. This end-to-end optimization allows the system to learn communication strategies potentially superior to traditional designs constrained by modular optimization.
\par
Transformers, originally developed for natural language processing, now exemplify one powerful architecture being adapted within this end to end optimization framework. A novel approach is introduced in \cite{choukroun2022error} using transformers as generative models to reduce channel error rates in end-to-end transceivers. This work demonstrates the capability of transformers to learn complex error correction codes and optimize the entire transceiver chain. Further investigation into transformer architectures for channel decoding is presented in \cite{hernandez20255g} which proposes a linear transformer architecture to efficiently decode 5G low-density parity-check (LDPC) codes, achieving competitive performance with reduced computational complexity. These studies indicate that transformers, with their powerful attention mechanisms, can capture intricate relationships within communication signals and offer a promising avenue for designing advanced AI-empowered transceivers.
\par
Diffusion models represent another class of powerful generative AI applicable to the end-to-end transceiver optimization paradigm. Authors in \cite{10104549} explore the use of diffusion models for end-to-end channel coding, demonstrating their effectiveness in learning complex channel characteristics and generating robust communication systems. Expanding on this, a joint design of diffusion models with LDPC decoding is proposed in \cite{10615729}, further enhancing error correction capabilities. Diffusion models present a distinct approach to transceiver design by learning to reverse a noise diffusion process. This capability allows them to generate intricate communication signals and optimize transceiver functionalities in a fundamentally different manner than discriminative models such as transformers. These early investigations indicate that diffusion models offer considerable potential for future transceiver designs especially in scenarios requiring robust performance in complex and changing wireless environments.

\subsubsection{\textbf{Integrated sensing and communication}}
The next generation communication system is expected to achieve both high-rate data transmission and high-accuracy sensing. Traditional methods often design communication and sensing systems separately, which is becoming inadequate for meeting these dual requirements simultaneously, especially given the increasing scarcity of spectrum resources \cite{liu2022integrated}. Integrated sensing and communication (ISAC) emerges as a crucial paradigm, aiming to reuse spectrum and hardware for both functionalities. However, conventional ISAC designs often struggle with the complexities of real-world environments and the integration of rich contextual information. Specifically, handling multimodal sensing data, which could significantly enhance both communication and sensing performance, poses a significant challenge for traditional ISAC systems that are typically tailored for unimodal data processing.
\par
Large AI models, particularly multimodal LLMs (MLLMs), offer an effective approach to address these limitations. MLLMs, trained on massive multimodal datasets, possess the ability to deeply understand and integrate semantically complex multimodal information \cite{cheng2025large}. By leveraging MLLMs, ISAC systems can move beyond unimodal operation and effectively fuse data from diverse sensors, such as RGB-D cameras, LiDAR, and radar, to create a more comprehensive and nuanced understanding of the environment. This enhanced environmental perception capability facilitates sensing-assisted communication by enabling more precise channel modeling, proactive blockage prediction, and adaptive beamforming \cite{cheng2025large, li2025large}. Conversely, communication-assisted sensing benefits from MLLMs through improved data sharing and collaborative perception among distributed agents, such as UAV swarms, leading to enhanced sensing coverage and accuracy. Furthermore, MLLMs can be employed for multi-objective optimization in ISAC systems, balancing the trade-offs between communication and sensing performance, as demonstrated in UAV networks by \cite{li2025large}. The integration of MLLMs into ISAC systems thus promises to unlock a new era of intelligent wireless systems capable of simultaneously delivering high-performance communication and sophisticated environmental perception.

\subsubsection{\textbf{Wireless caching}}
Wireless caching, a technique to store frequently accessed data closer to users, is crucial for reducing latency and improving network efficiency in wireless communications \cite{paschos2016wireless, 10292740,10606215}. Traditional wireless caching relies on static algorithms and heuristics for data placement and retrieval. These methods often struggle to adapt to the dynamic nature of wireless traffic, diverse user demands, and the sheer volume of content in modern networks. To overcome these limitations, large AI models are emerging as a promising approach to revolutionize wireless caching strategies \cite{sheraz2020artificial}.
\par
Large AI models offer advanced capabilities in understanding and predicting complex data patterns, which can be directly applied to optimize wireless caching. Unlike traditional methods, large AI models can analyze vast datasets of user access patterns, content popularity, and even contextual information about user location and network conditions. This analysis enables intelligent, proactive caching decisions that go beyond simple frequency-based heuristics. For example, large AI models could predict future data requests with higher accuracy, allowing for pre-emptive caching of content that is likely to be in demand. Furthermore, the contextual understanding of large AI models can enable personalized caching, tailoring content storage to the predicted needs of specific user groups or locations within the wireless network.
\par
The integration of large AI models into wireless caching systems holds the potential to significantly enhance wireless communication performance. By dynamically adapting caching strategies based on learned patterns and predictions, large AI models can minimize data retrieval latency, reduce backhaul traffic, and improve overall throughput. Moreover, the flexible nature of large AI models allows for optimization across various caching objectives. By adjusting the training objectives, large AI model-driven caching systems can be tailored for energy efficiency \cite{jiang2019deep}, delay reduction \cite{tsai2018mobile}, throughput maximization \cite{cheng2018localized}, or a combination of these, thereby creating highly adaptable and performant wireless caching solutions for next-generation networks.

\subsubsection{\textbf{Large AI model for physical layer security}}\label{WLAM_for_PLS}
Large AI models are rapidly transforming wireless communications, but this advancement necessitates a critical reassessment of physical layer security (PLS). While AI promises enhanced performance, the inherent complexity of these models, especially generative AI, introduces new security vulnerabilities that must be addressed for robust 6G networks \cite{tccnsecurity}. Traditional PLS techniques, while valuable, often lack the adaptability and sophistication to counter threats in AI integrated wireless systems. On the other side, Generative AI offers a powerful paradigm shift, providing tools to both analyze and enhance security at the physical layer.
\par
A comprehensive security framework for WLAM based PLS must incorporate robust threat modeling \cite{10110346}. This process involves identifying and formalizing potential attack vectors that target the AI models themselves within the physical layer context. Examples of such threats include adversarial attacks, where manipulated wireless signals deceive WLAM-based receivers, and model poisoning, where malicious data is injected during distributed training phases to corrupt the model.
\par
To counter these threats, the capabilities of large AI models themselves can be leveraged to create powerful defense mechanisms. These countermeasures include adversarial training to enhance model resilience against perturbed inputs, the use of generative models to detect anomalies deviating from normal signal behavior, and the development of robust aggregation methods in federated learning to mitigate poisoning threats \cite{tccnsecurity}. Establishing a technical framework that integrates both threat models and their corresponding countermeasures is crucial for designing and evaluating secure WLAM deployments.
\par
Besides, the dynamic nature of wireless environments and threats necessitates adaptive security mechanisms. RL presents a promising approach for enabling WLAM to dynamically adjust PLS strategies in response to real time conditions. An RL agent could monitor the communication environment, observing network state information, estimated channel quality, and outputs from threat detection modules. Based on this state, the RL agent could decide actions such as adjusting the secret key generation rate in secret key generation protocols, selecting optimal wiretap coding schemes, modifying artificial noise injection levels, or adapting parameters of authentication protocols like physical unclonable functions challenge response mechanisms \cite{mitev2023physical}. The reward function for the RL agent could balance security metrics, such as achieved secrecy rates or successful authentication counts, against communication performance indicators like throughput and latency. This allows the system to intelligently trade off security and efficiency based on the current context and detected threat level.
\par
Furthermore, the MoE architecture offers an efficient structure for implementing such adaptive security \cite{wcmsecurity}. MoE-enabled generative AI frameworks can enhance standalone models by reducing computational complexity and improving adaptability. By combining multiple specialized expert models that are each trained for a specific task (e.g., jamming resilience, eavesdropping mitigation), MoE offers flexible, context-aware security solutions. A gating mechanism, informed by an RL agent or environmental sensing, dynamically selects the most appropriate expert based on the detected scenario. This architecture facilitates targeted defenses, enhances scalability, and improves adaptability to evolving wireless security challenges, integrating well with the dynamic adjustments driven by RL. The ongoing research combining MoE, generative AI, and RL promises a new generation of intelligent and adaptive PLS for future wireless networks.

\subsection{WLAM for network layer communications}
\begin{figure*}[t]
	\begin{center}
		\centerline{\includegraphics[width=0.8\linewidth]{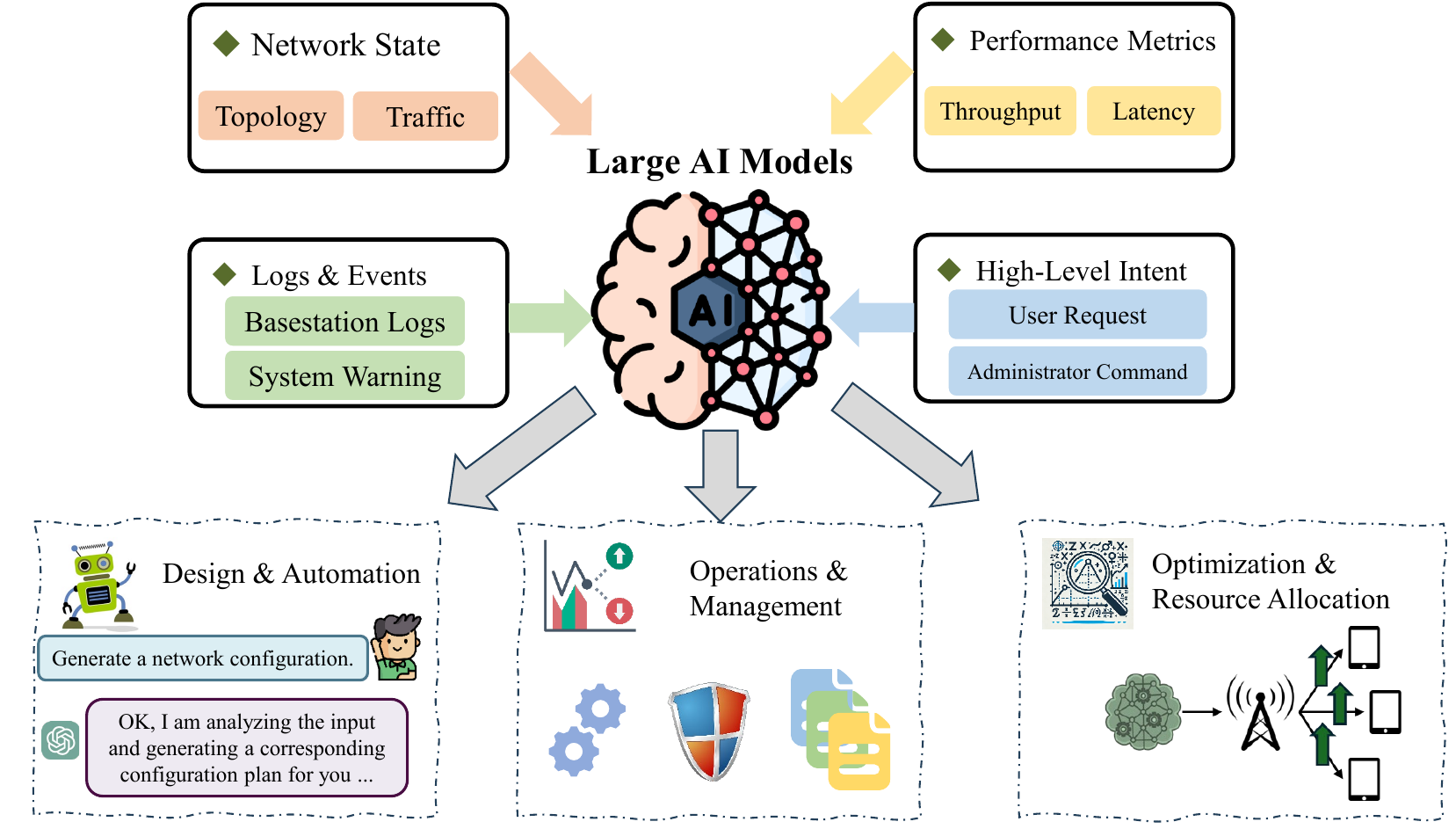}} 
		\vspace{-0mm}
		\captionsetup{font=footnotesize, name={Fig.}, labelsep=period} 
		\caption{\, Large AI models for network layer communications. WLAMs process multimodal network contexts, including topology, traffic, system logs, and human intents to intelligently drive core network layer functionalities: automated configuration, proactive fault management, and dynamic resource allocation.}
		\label{fig: Large AI models for network layer communications}
		\vspace{-8mm}
	\end{center}
\end{figure*}

The application of WLAM extends significantly into the network layer, offering potential for managing and optimizing the intricate communication pathways that underpin modern networks, particularly within the complex and dynamic environment of 6G wireless systems. WLAM can interpret high level intents, reason over complex network states, and interact with configuration and management systems, paving the way for more autonomous, efficient, and adaptable network infrastructures, aligning with visions for AI-empowered wireless networks, the roles of WLAM are summarized in Fig. \ref{fig: Large AI models for network layer communications}.

\subsubsection{\textbf{Network operations and management}}
LLMs provide powerful capabilities for enhancing wireless network operations, including real time monitoring, sophisticated fault diagnosis, and proactive security protection \cite{10829820, liuLLMNetwork}. Leveraging their advanced natural language processing abilities, LLMs can analyze diverse and voluminous data sources common in wireless environments, such as base station logs, radio access network (RAN) performance metrics, user mobility data, and security alerts. They can identify subtle anomalies indicative of wireless specific issues like interference, handover failures, or configuration drifts in RAN components. By learning patterns from historical data, LLMs can predict potential faults, enabling preemptive actions to maintain network stability and service quality \cite{10829820}. Furthermore, they can assist network administrators by diagnosing problems through interactive dialogues, interpreting complex technical information, and generating concise reports or suggesting resolution steps, thereby improving operational efficiency, particularly in complex multi vendor or Open RAN (O-RAN) settings.

\subsubsection{\textbf{Network optimization and resource allocation}}
Optimizing performance in wireless networks involves allocating scarce resources like spectrum, power, and computational capabilities across dynamic channel conditions and user mobility. LLMs offer intelligent solutions for these network layer optimization tasks \cite{10829820}. They can analyze network state information and user requirements to make informed decisions regarding radio resource allocation, dynamic spectrum allocation, network slice configuration for diverse QoS demands (e.g., URLLC, eMBB, mMTC in 6G), and traffic routing in complex topologies involving terrestrial and non terrestrial segments \cite{9537935}. Frameworks utilizing LLMs, potentially adapted using efficient techniques like data-driven low-rank adaptation, demonstrate potential in optimizing tasks such as adaptive bitrate streaming or job scheduling in mobile edge computing environments, outperforming conventional algorithms by better understanding complex system dynamics and user behavior \cite{NetLLM}. This facilitates automated, context-aware optimization that adapts to the fluid nature of wireless environments.
\par
Beyond LLMs, generative diffusion models have emerged as powerful tools for exploring complex solution spaces in non-convex optimization. Authors in \cite{11018297} and \cite{11006143} proposed DiffSG and GDSG respectively, leveraging diffusion models as generative solvers to effectively address optimization problems in network management and mobile edge computing. For multi-objective scenarios, \cite{11085045} demonstrated the potential of diffusion models in enhancing monitoring efficiency within UAV-enabled IoT systems. Furthermore, \cite{10839314} systematically explored and analyzed the feasibility of deploying diffusion models as core network optimizers.

\subsubsection{\textbf{Network configuration automation}}
LLMs can automate and enhance the design and configuration of wireless networks \cite{liuLLMNetwork}. They can assist engineers by translating high level service requirements or natural language intents into specific network configurations, such as RAN parameter settings, cell planning parameters, or network slice definitions \cite{huang2023large}. LLMs can generate configuration scripts or code for various network functions and protocols, potentially reducing manual effort and minimizing errors. Their ability to reason over complex dependencies, possibly augmented by external knowledge bases or verification tools, allows them to validate configurations, detect potential conflicts (e.g., policy violations, incorrect parameter settings), and ensure consistency across network elements, which is particularly valuable in increasingly complex and disaggregated architectures like O-RAN \cite{liuLLMNetwork, huang2023large}. LLMs can also be used to generate test scripts for validating wireless software systems, using synthetically generated test data that mirrors real world network conditions \cite{10615269}.

\subsubsection{\textbf{AI-empowered architectures for wireless networks}}
The integration of LLMs drives the evolution towards AI-empowered network architectures, where intelligence is embedded throughout the network rather than being an overlay \cite{9537935}. These architectures envision LLMs operating collaboratively across cloud, edge, and potentially device layers. Frameworks like NetGPT \cite{chen2024netgpt} exemplify this by using smaller, specialized LLMs at the wireless edge for tasks like prompt enhancement and context personalization based on local information (e.g., user location, device state), while leveraging larger cloud based LLMs for complex generative tasks, enabling efficient personalized services. Other frameworks such as NetLLM focuses on creating adaptable LLM based systems capable of handling diverse networking tasks through structured workflows involving components for analysis, planning, calculation, and interaction with network tools and environments \cite{NetLLM, huang2023large, liuLLMNetwork}. These architectures are fundamental to realizing the 6G vision of intelligent, automated, and highly flexible communication systems.

\subsubsection{\textbf{Adaptation and integration techniques for wireless LLMs}}
Bridging the gap between general purpose LLMs and the specific demands of the wireless network layer requires sophisticated adaptation and integration methods. Handling the unique data modalities in wireless communication, such as time series signal data, CSI, graph based topology information, or specific protocol formats, necessitates multimodal encoders capable of projecting this diverse information into a space understandable by the LLM \cite{NetLLM, liuLLMNetwork}. PEFT techniques, like LoRA, are critical for instilling domain specific knowledge into LLMs without the prohibitive cost of full retraining, making adaptation feasible even for resource constrained edge nodes \cite{NetLLM, chen2024netgpt}. Advanced prompt engineering, including CoT and RAG drawing upon telecom standards or operational manuals, guides LLMs towards accurate reasoning and factual responses \cite{huang2023large, liuLLMNetwork}. Crucially, seamless integration with external network tools, simulators, controllers, and verifiers is essential to enable LLMs to not just reason about the network, but also actively participate in its management and control.

\subsection{WLAM for semantic communications}

\begin{figure*}[t]
	\begin{center}
		\centerline{\includegraphics[width=1\linewidth]{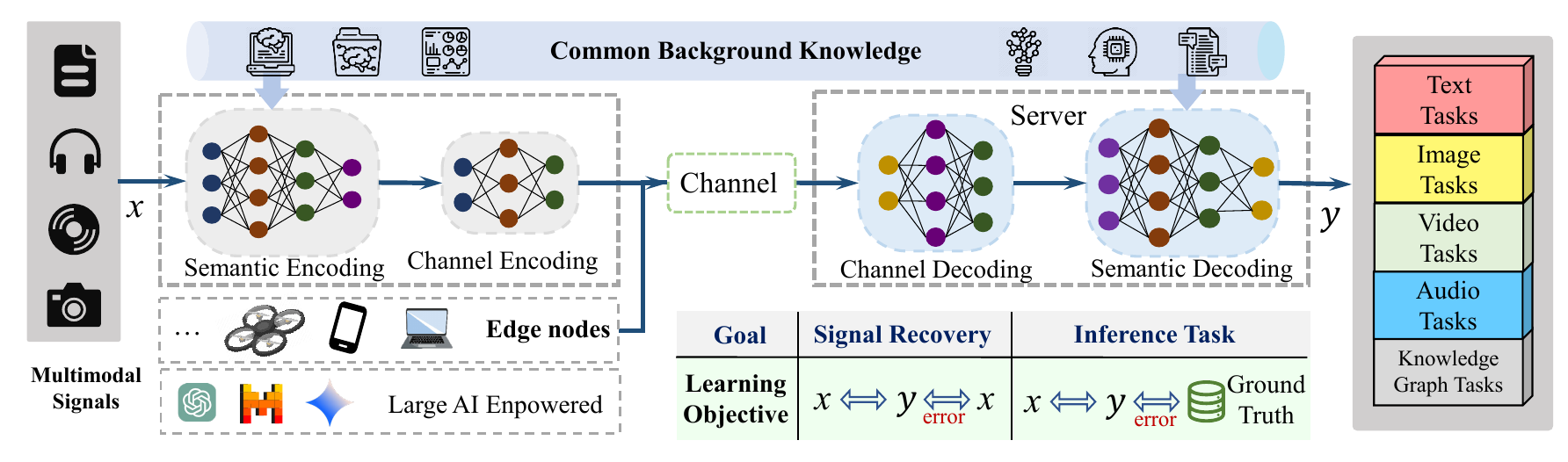}} 
		\vspace{-0mm}
		\captionsetup{font=footnotesize, name={Fig.}, labelsep=period} 
		\caption{\, Large AI models for SemCom layer communications. WLAMs drive multimodal semantic communications via joint semantic and channel coding. By leveraging shared background knowledge, they efficiently encode diverse data streams (text, image, audio, video) to achieve both high-fidelity signal recovery and accurate task-oriented inference.}
		\label{fig: Large AI models for SemCom layer communications}
		\vspace{-8mm}
	\end{center}
\end{figure*}

SemCom is a key technology for future 6G systems, promising superior communication efficiency by directly optimizing information transfer at the semantic level \cite{gunduz2022beyond,qin2021semantic,lan2021semantic, ZHAO2024107055, 10233741}. Large AI models are pivotal for realizing SemCom, providing the essential capability to extract and process semantic information from diverse data modalities like images, text, audio, video, and knowledge graphs, tailored to specific tasks such as classification and reconstruction. Key challenges for large AI models in SemCom include accurately identifying and precisely transmitting the desired meaning within communication systems, and ensuring robustness against channel variations through adaptive designs and re-transmission mechanisms. Leveraging the power of large AI models to address these challenges is anticipated to unlock substantial performance gains in semantic communication efficiency and reliability. The communication flow chart is demonstrated in Fig. \ref{fig: Large AI models for SemCom layer communications}.

\subsubsection{\textbf{Semantic communications preliminaries}} 
Understanding the theoretical underpinnings of SemCom is crucial for analyzing its performance limits and guiding system design \cite{niu2024mathematical, 9953095}. Early work reintroduced the concept of semantic entropy, positing that semantic units exhibit logical connections rather than being random \cite{gunduz2022beyond}. Building on this, the idea of utilizing shared knowledge bases between communicating parties to exchange semantic information highlighted its potential \cite{choi2022unified}. Subsequent research provided more formal definitions, such as considering multiple source units as equivalent if they share the same meaning (synonymous mapping), leading to generalized definitions of semantic entropy and insights into the fundamental limits of SemCom, particularly for many-to-one source scenarios \cite{ma2023theory}. It is recognized that this synonymous mapping, where different expressions convey identical meaning (e.g., \textit{She appeared happy} vs. \textit{She appeared joyful}), is a key factor enabling SemCom to potentially outperform traditional bit-level communication. These developments have contributed to establishing a systematic framework for semantic information theory.

\subsubsection{\textbf{Modality/Task tailored SemCom}} 
Unlike traditional bit-level communication that converts everything into bits, SemCom directly maps the data in specific modalities to channel symbols. This approach requires SemCom to be tailored for each data modality or even each task to achieve optimal performance, which are detailed as follows. 

\paragraph{Text domain}
In text domain, the authors in \cite{farsad2018deep} first proposed deep learning based joint source and channel coding (DeepJSCC) for text transmission. They developed an recurrent neural network (RNN)-based network to directly map text to channel symbols, bypassing the separate steps of data compression and interference resistance by source coding and channel coding. This method demonstrated a lower word error rate (WER) compared to conventional methods by effectively leveraging the structural feature of text. Building on the initial idea of using AI for feature extraction, a comprehensive transmission framework for text called DeepSC was proposed in \cite{xie2021deep}. DeepSC employs the advanced transformer architecture for text feature extraction and reconstruction. Training the transformer model in an end-to-end manner significantly improved performance. Additionally, they introduced a new performance metric called bilingual evaluation understudy (BLEU), which measures the similarity between two sentences in the feature space. Unlike WER, BLEU aligns more closely with human-like semantic definitions. 
To address the complexity concerns, 
a lite distributed semantic communication system called L-DeepSC was proposed in \cite{xie2020lite}, where network pruning and quantization are adopted to reduce the computation complexity of network part. Recognizing the importance of semantic-aware metric, the authors in \cite{wang2022performance} further proposed a metric of semantic similarity (MSS) that compares two sentence through their graph similarity. Based on which, they optimizes the system performance by reinforcement learning with MSS as reward. 

\paragraph{Image domain}
In image domain, substantial research efforts have been devoted to improving SemCom. Initially, the authors in \cite{bourtsoulatze2019deep} proposed the CNN models for image transmission, employing convolutional layers for encoding and transpose convolutional layers for decoding. Their experiments demonstrated the potential of DeepJSCC in image transmission by significantly outperforming the conventional schemes (i.e., JPEG+LDPC) in the low SNR regime. This approach was further refined in \cite{kurka2020deepjscc}. With the advent of powerful vision transformers, researchers began exploring transformer-based DeepJSCC networks. A vanilla transformer model for DeepJSCC was proposed in \cite{dai2022nonlinear}. This model was subsequently improved in \cite{yang2024swinjscc,wu2023vision} by replacing the original transformer block with swin transformer block. Recently, the authors in \cite{wu2024mambajscc} introduced the mamba model for DeepJSCC, achieving better performance than transformer-based JSCC.

\paragraph{Video domain}
In the video domain, the authors in \cite{tung2022deepwive} were the first to consider end-to-end DeepJSCC for video transmission. They classified video frames into key frames and non-key frames. Key frames are transmitted using the image-style JSCC model proposed in \cite{xu2021wireless}, while non-key frames are transmitted by encoding the residual information relative to the key frames. Additionally, they used reinforcement learning to achieve resource allocation at the frame level based on motion strength, demonstrating numerical results that overcome the cliff-effect present in conventional transmission schemes. This approach was further improved in \cite{zhang2023deep}.
Subsequently, the authors in \cite{wang2022wireless} proposed using the last frame as a condition for the encoding and decoding process. This paradigm allows for the full exploitation of temporal dependence and enables resource allocation for each feature embedding. Their experiments showed better performance than the conventional separate source and channel coding schemes. More recently, the authors in \cite{du2024object} considered video transmission in static scenarios, representing the movement of objects as a graph. This approach achieved better performance than conventional schemes.

\paragraph{Audio domain}
In audio domain, similarly, the structural feature of speech signal can be leveraged to reduce the transmission overhead or combat the channel noise. The authors in \cite{weng2021semantic} first proposed a deep learning based SemCom system for speech signals, called DeepSC-S. They utilised the attention mechanism to identify the important information and transmit them with larger power. This approach demonstrated their effectiveness by outperforming the conventional audio transmission scheme under the common speech signals metrics. Realizing the correlation between speech and text, DeepSC-S was further extended to speech synthesis \cite{weng2023deep}. In the speech to speech task, the authors in \cite{han2022semantic} considered two scenarios, that is, speech recognition and reconstruction. For the recognition task, the semantic-relevant information was extracted and transmitted to the receiver. For the reconstruction task, some additional information that is irrelevant to the semantic information is also transmitted for consistent speech synthesis. 
\paragraph{Knowledge graph domain}
In addition to the common modalities in media form, SemCom can also be applied to knowledge graphs, which are regarded as a standard and universal modality. Knowledge graphs can effectively express the attributes and their connections, thereby denoting semantic information \cite{zzxTCOM2025, 10550151}. The authors in \cite{zhang2023optimization} first considered adopting knowledge graphs for wireless image transmission. 
They extracted the knowledge graph from the image to be transmitted selectively, and transmitted the semantic information based on its importance and the available resource blocks. Leveraging on the feature of knowledge graph, they proposed a multimodal performance metric called image-to-graph semantic similarity (ISS) to measure the transmission accuracy of semantic information. Additionally, a multi-agent RL algorithm was proposed to minimize the sum of the average transmission accuracy while satisfying ISS requirement. Then, the transmission of knowledge graph in multi-user scenario was further considered in \cite{e26050394}. 

\subsubsection{\textbf{Generative AI models for SemCom}}
DeepJSCC has marked significant progress in SemCom, primarily by minimizing the distortion between the reconstructed data and the original source. However, just focusing on low distortion does not always guarantee optimal semantic performance or perceptual quality, reflecting the inherent rate-distortion-perception tradeoff \cite{erdemir2023generative, yilmaz2023high}. To bridge this gap, researchers are increasingly integrating generative AI models into SemCom systems. These models, capable of learning complex data distributions and generating high-fidelity samples, offer powerful tools to enhance semantic fidelity and perceptual quality beyond traditional distortion metrics. Two main categories of generative models are being explored below.

\paragraph{Traditional generative models for perceptual enhancement}
This line of research focuses on using established generative techniques like GANs and diffusion models to improve the quality of data reconstructed by DeepJSCC, particularly for perceptual tasks. One strategy involves employing generative models as post-processing modules. After DeepJSCC reconstructs the data, a generative model refines the output to look more realistic, such as using StyleGAN for face images \cite{erdemir2023generative} or diffusion models for general image restoration \cite{yilmaz2023high}. The primary challenge in wireless communications is accurately modeling the complex distortion introduced by both the DeepJSCC model and the noisy wireless channel \cite{chen2024commin}. An alternative approach leverages the inherent denoising capabilities of diffusion models more directly, using them either for preprocessing or as integrated denoisers at the receiver to combat channel impairments \cite{wu2024cddm}. Recognizing the computational and bandwidth demands of these models, techniques like integrating compression networks \cite{guo2024diffusion} and knowledge distillation \cite{pei2024latent} have been explored to improve efficiency.

\begin{figure*}[t]
	\begin{center}
		\centerline{\includegraphics[width=0.9\linewidth]{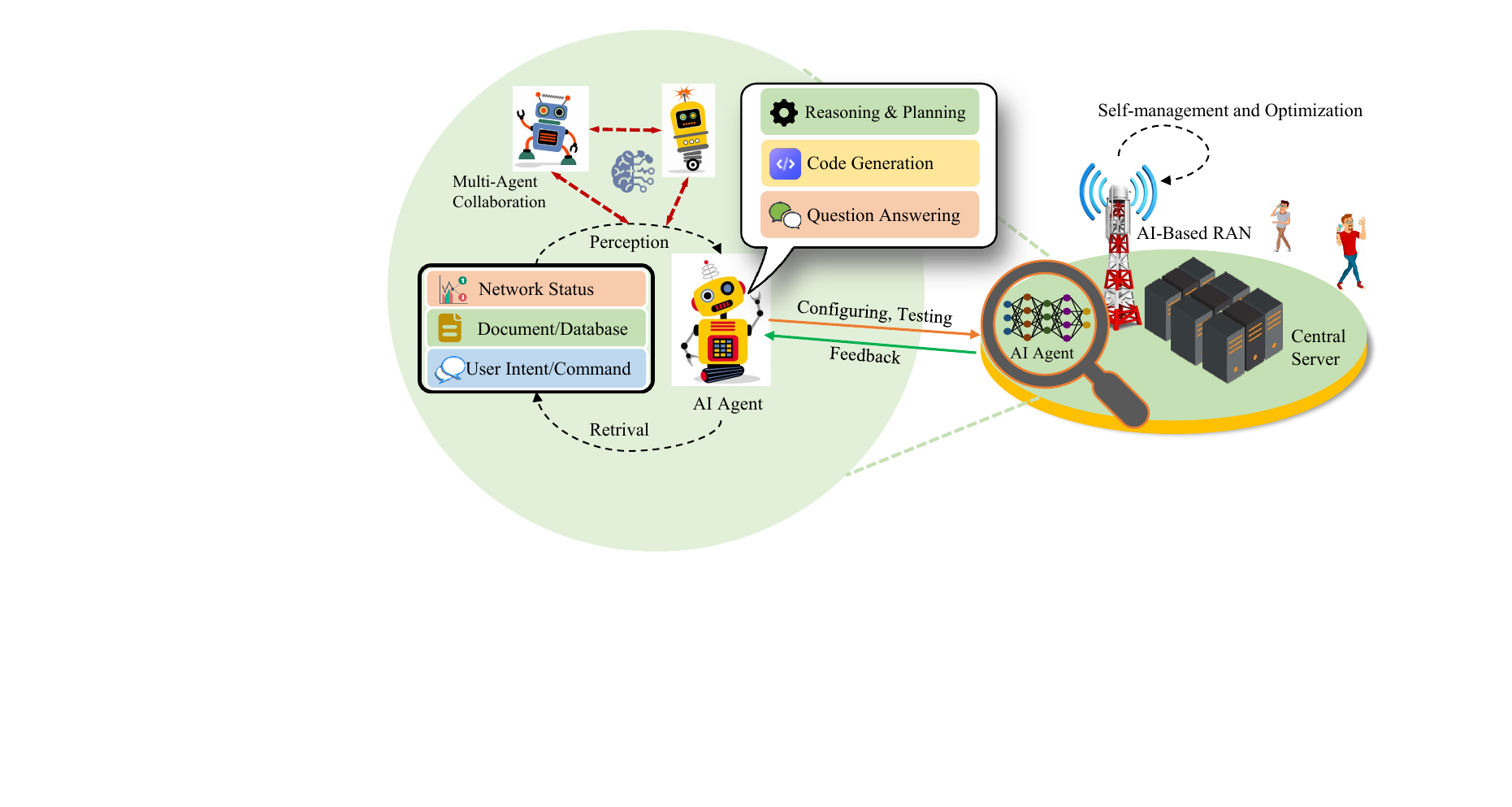}} 
		\vspace{-0mm}
		\captionsetup{font=footnotesize, name={Fig.}, labelsep=period} 
		\caption{\, Large AI models as wireless agents. WLAMs operate as autonomous entities by integrating multimodal perception, advanced reasoning, and dynamic planning. Through knowledge retrieval, code generation, and multi-agent collaboration, these agents autonomously execute complex network configuration, testing, and self-management in AI-RANs.}
		\label{fig: Large AI models as wireless agents}
		\vspace{-8mm}
	\end{center}
\end{figure*}

\paragraph{LLMs for knowledge-driven SemCom}
Leveraging the unique capabilities of LLMs opens fundamentally new avenues for SemCom, shifting focus towards deep semantic understanding and knowledge utilization. LLMs can serve as expansive knowledge bases, enabling highly efficient communication where the transmitter sends minimal, semantically crucial information, and the receiver-side LLM reconstructs the full meaning using its stored knowledge and contextual reasoning \cite{wang2024large, ni2024interplay}. Furthermore, the training objective of LLMs, related to entropy minimization \cite{deletang2023language}, makes them inherently effective source coders capable of semantic compression. The extension of LLMs to multimodal data allows for their integration into multimodal SemCom systems, enhancing information exchange across diverse formats. Finally, there is a synergistic interplay where communication can update the knowledge of LLM, and the refined knowledge, in turn, improves communication efficiency and context-awareness.

\subsection{Wireless agents}
Large AI models like LLMs are transitioning from tools for specific tasks like text generation or analysis into more autonomous entities, known as agents capable of planning reasoning and interacting with environments to achieve complex goals \cite{xiao2025agenticainetworking6g, jiang2025largeaimodelsagentic, xi2025rise, wang2025bridgingphysicaldigitalworlds}. This agentic paradigm leverages the advanced understanding reasoning and generation capabilities inherent in LLMs applying them to orchestrate complex systems like future wireless networks. In 6G communications LLMs as agents promise to manage network complexity automate operations and enable novel intelligent services by perceiving the network state making informed decisions and interacting with network functions through defined interfaces or APIs \cite{xiao2024llmagents6gorchestrator, dev2025advancedarchitecturesintegratedagentic}, as illustrated in Fig. \ref{fig: Large AI models as wireless agents}. This shift requires specific system designs model training methodologies and interaction frameworks to harness their potential effectively and reliably within the telecommunications domain. Implementing these complex agentic systems is further facilitated by the development of dedicated platforms, such as the open-source OpenManus \cite{openmanus2025} framework designed for building general AI agents.

\subsubsection{\textbf{Telecom knowledge agent}}
LLMs are increasingly being adapted to function as specialized telecom knowledge agents, capable of navigating the complexities of the telecommunications domain. TelecomGPT offers a representative framework for this adaptation, proposing a comprehensive pipeline to transform general purpose LLMs into telecom specific variants \cite{zou2024telecomgpt}. This process utilizes collected telecom data for continual pretraining, instruction dataset construction for fine tuning, and preference data for alignment tuning. The TelecomGPT framework also introduces novel benchmarks designed to evaluate critical model capabilities, including mathematical modeling, question answering, and code generation specific to the telecom context.
\par
Specialized models enhance information retrieval from technical documents, a crucial task for telecom professionals. TeleRoBERTa, for example, provides an extractive question answering model highly effective at referencing 3GPP documents \cite{10624786}. By adapting the RoBERTa base model and fine tuning it on telecom data, TeleRoBERTa delivers performance comparable to much larger foundation models on benchmarks like TeleQuAD, showcasing the efficiency gains possible through domain specific training even with fewer parameters. Complementing this, TelecomRAG leverages retrieval augmented generation to create a telecommunication standards assistant \cite{telecomrag2023}. This system focuses on providing accurate, technically detailed, and verifiable responses, directly addressing the limitations of generic models regarding precision and source traceability when dealing with standard specifications.
\par
The development of domain specific models is further supported by initiatives like the Tele-LLMs series, which provides open source LLMs adapted for telecommunications \cite{maatouk2024telellmsseriesspecializedlarge}. These models, enhanced through continual pretraining on telecom related data, demonstrate improved performance on domain relevant tasks compared to their general purpose predecessors. Together, these frameworks and specialized models highlight significant advancements in tailoring LLMs to serve as effective and reliable knowledge agents within the telecommunications field, promising greater efficiency and accessibility for professionals navigating complex technical information.

\subsubsection{\textbf{Code generation}}
LLMs demonstrate significant capability in automatic code generation, a long standing goal in computer science and software engineering. This potential extends powerfully into the wireless communication systems domain, offering methods to enhance developer productivity and automate various coding tasks. Researchers have explored several applications and evaluation dimensions concerning LLMs for creating code.
\par
General applications include translating natural language descriptions into executable code, providing context aware code completion suggestions during development, and performing automatic program repair to correct bugs in existing software \cite{10403378}. Beyond these general uses, studies have shown the applicability of these models in more specialized areas relevant to communications engineering. For instance, LLMs have been successfully employed to generate complex hardware description language code, such as Verilog, for implementing advanced wireless communication algorithms like the fast Fourier transform on field programmable gate arrays. Techniques including in context learning and chain of thought prompting were utilized to navigate the intricacies of hardware specific requirements like subtask scheduling and multi step reasoning, which are less common challenges in standard software code generation \cite{du2024powerlargelanguagemodels}. Another targeted application involves the automated generation of test scripts essential for validating telecom software systems. A proposed two stage framework first uses generative models to create synthetic yet realistic test input data based on historical network performance, and then employs an LLM to generate the actual test script code using this synthetic data combined with natural language test descriptions \cite{10615269}. This addresses the tedious nature of manual test creation and helps cover diverse scenarios, crucial for complex telecom environments like O-RAN.
\par
While the application potential is vast, the evaluation of code generated by LLMs presents ongoing challenges. Much current evaluation work focuses primarily on functional correctness, often assessed through pass rates on standardized programming benchmarks, and on identifying security vulnerabilities \cite{10403378, liu2024exploringevaluatinghallucinationsllmpowered}. However, research indicates a gap between the advancement of generation capabilities and the comprehensiveness of evaluation methodologies. There is a recognized need for evaluation metrics and processes that encompass broader software quality attributes such as code maintainability, readability, and interpretability, which receive less attention currently \cite{10403378}.
\par
A significant challenge with the reliability of LLM generated code is the occurrence of hallucinations. Similar to issues observed in natural language generation, these models can produce code that appears plausible but deviates from the user specified requirements, contradicts contextual information, contains unnecessary repetitions or non functional dead code segments, or misuses programming interfaces and identifiers based on incorrect knowledge \cite{liu2024exploringevaluatinghallucinationsllmpowered}. Studies confirm that even sophisticated models face difficulties in recognizing these hallucinations, and mitigating them effectively through prompting alone is even more challenging, highlighting a critical area for future research to ensure the trustworthiness of generated code \cite{liu2024exploringevaluatinghallucinationsllmpowered}. Furthermore, the complexity of certain domains, like hardware description language generation, introduces unique hurdles \cite{du2024powerlargelanguagemodels}. Comparing different models reveals that those specifically pretrained on large code datasets might generate better formatted or more idiomatic code compared to general purpose language models, although both can demonstrate reasoning for testing logic. Output quality also shows sensitivity to the specific prompts\cite{10615269}.
\par
In conclusion, LLMs represent a promising technology for code generation with clear benefits for wireless communication system development. However, continued research is necessary to improve evaluation techniques, address the pervasive issue of hallucinations, and refine methods for generating reliable and high quality code suitable for deployment in critical communication infrastructure.

\subsubsection{\textbf{Agentic AI-RAN}}
Agentic AI represents a significant, potentially revolutionary evolution for RAN management, shifting from traditional automation towards autonomous systems capable of pursuing complex goals with minimal human intervention \cite{jiang2025largeaimodelsagentic, xiao2025agenticainetworking6g}. Operating under a principle of bidirectional interaction where AI empowers the network and the network enhances AI, agentic AI offers a paradigm for intelligent operation, administration, and maintenance in the complex environment of 6G networks. This approach enables systems to perceive, reason, decide, and act autonomously \cite{dev2025advancedarchitecturesintegratedagentic}. Unlike conventional AI reliant on predefined rules, agentic AI RAN aims to optimize objectives like resource allocation, service assurance, and energy efficiency through dynamic adaptation to the network environment, thereby more effectively supporting emerging services such as autonomous intelligent IoT and embodied intelligence applications.
\par
The core capabilities underpinning agentic AI RAN include sophisticated multi source perception leveraging channel information, sensing data, and visual inputs to understand the network state \cite{xiao2024llmagents6gorchestrator, xiao2025agenticainetworking6g}. Based on this perception, agents employ advanced reasoning processes, perhaps using chain of thought techniques to plan sequences of actions. They then interact with the network fabric by utilizing network function APIs as tools to execute the planned configurations. Crucially, effective decision making and planning are deeply reliant on integrated retrieval mechanisms. These provide agents dynamic access to essential knowledge bases containing 3GPP standards, historical logs, network policies, and other relevant information, enabling context aware reasoning and ensuring compliance \cite{zhang2025agenticaigenerativeinformation}.
\par
Building upon these capabilities, specific frameworks and architectural concepts are proposed to realize agentic AI-RAN. For example, AgentNet \cite{xiao2025agenticainetworking6g} envisions specialized networking ecosystems supporting heterogeneous AI agents like foundation model agents and embodied agents, facilitating efficient information exchange and coordination. Generative foundation models themselves can serve as interactive knowledge resources or be specialized into agents capable of understanding tasks, orchestrating workflows, and managing network components through well defined interfaces \cite{xiao2024llmagents6gorchestrator}. This agent driven architecture offers inherent simplifications, notably through control and user plane separation, and facilitates dynamic service orchestration and intent based networking, allowing high level goals to be translated into automated network adjustments \cite{dev2025advancedarchitecturesintegratedagentic}.
\par
Similar to the hallucination issues in code generation, the stochastic nature of agentic AI introduces even more critical risks in operational environments. Specifically, agents may generate plausible but incorrect network configurations or decisions. For instance, an agent might hallucinate a resource allocation command that assigns users to a restricted frequency band (e.g., radar bands), violating spectrum regulations and causing severe interference to primary services. Besides, a temporary drop in traffic might be misinterpreted, leading to a critical macro base station being wrongly put into deep sleep. The result is an instant coverage gap and widespread call drops. To address these challenges, establishing a trustworthy agentic workflow is essential. First, leveraging RAG enables agents to cross-verify actions against up-to-date 3GPP standards and historical network logs, thereby reducing fact-conflicting hallucinations caused by outdated or missing domain knowledge \cite{liu2024hallucinationawareoptimizationlargelanguage}. Second, integrating structured causal reasoning allows agents to explicitly construct causal directed acyclic logic graphs before execution. This ensures the agent captures the correct conditional dependencies between network variables, mitigating logically inconsistent hallucinations where generated actions violate physical propagation laws \cite{li2025mitigatinghallucinationslargelanguage}. Finally, a separate AI verification layer serves as a critical safeguard. This validator model or rule-based engine audits all generated commands against physical network constraints, ensuring invalid configurations are intercepted before they can impact the live network \cite{wang2025wirelesshallucinationgenerativeaienabled}. However, for critical telecommunications infrastructure, relying solely on rule-based engines is insufficient to prevent catastrophic network failures. To address this, it is imperative to establish a rigorous, multi-layered verification framework that specifically validates every AI-generated action before execution. First, at the mathematical level, formal verification methods \cite{cohen2024llm} must be incorporated to strictly prove that the generated configurations do not violate hard safety properties (e.g., avoiding routing loops, spectrum overlapping, or antenna power overloads). Second, at the system level, integrating a digital twin-based sandbox is essential \cite{khan2022digital, bariah2023digital}. By executing the agent's decisions in a high-fidelity virtual replica of the RAN, operators can proactively observe network-wide impacts and isolate fatal hallucinations from the live infrastructure. Third, at the execution level, a human-in-the-loop authorization mechanism should be mandated for core network alterations, ensuring that autonomous agents augment rather than bypass human oversight \cite{natarajan2025human}. Together, this comprehensive verification framework strengthens the operational safety and reliability of Agentic AI in 6G networks.
\par
Beyond these reliability concerns, the performance overhead introduced by the API-based abstraction layer poses a significant challenge. Specifically, this overhead stems from the cumulative latency of model inference and the network round-trip time associated with API calling. This magnitude of delay makes direct agentic control unsuitable for ultra-low-latency data plane operations. To address this, agentic AI-RAN essentially requires a hierarchical control framework. Specifically, this hierarchical framework decouples network into a dual-loop architecture to overcome latency bottlenecks \cite{narimani2025agenticcontrol, guo2024controlagent}. The WLAM operates in the ``slow loop'' (e.g., at the timescale of seconds or minutes), taking charge of intent translation, complex reasoning, and high-level policy configuration (such as defining energy-saving strategies or network slicing priorities). Conversely, traditional lightweight algorithms and hardware accelerators constitute the ``fast loop'' (operating at sub-millisecond timescales). Once the WLAM orchestrates the overarching policy, the fast loop autonomously executes the real-time, delay-critical data plane operations, such as symbol-level beamforming or packet scheduling. This decoupling ensures that the inherent inference latency of massive AI models does not compromise the URLLC requirements of the physical layer. Furthermore, energy efficiency remains a critical constraint that warrants a profound analysis (as detailed in Section 6.3). Continued research is thus essential to develop computationally efficient AI models, secure agent interactions, and highly adaptive AI-driven processes tailored to the demanding operational realities of 6G \cite{dev2025advancedarchitecturesintegratedagentic}.

\subsubsection{\textbf{Multi-agent collaboration}}
The distributed and complex nature of 6G systems necessitates the use of multi agent systems where multiple autonomous AI agents interact and coordinate to achieve shared or individual goals \cite{zhang2025multiagentreinforcementlearningwireless, jiang2025largeaimodelsagentic}. Agentic AI networking inherently involves collaboration as agents deployed across different network locations such as user equipment edge servers or base stations, therefore, they must work together for tasks like resource allocation interference management or service orchestration \cite{xiao2025agenticainetworking6g}. Wireless distributed networks provide the necessary infrastructure platform for these multi agent interactions \cite{zhang2025multiagentreinforcementlearningwireless}.
\par
Multi-agent reinforcement learning (MARL) is a key enabling paradigm allowing agents to learn collaborative strategies through interaction. MARL frameworks address how agents learn optimal policies considering the actions and states of others moving from centralized training and execution towards more decentralized approaches like centralized training with decentralized execution or fully decentralized training and execution which better suit the distributed nature of 6G \cite{zhang2025multiagentreinforcementlearningwireless}. Effective collaboration hinges on mechanisms for communication and information sharing considering network constraints. Techniques like graph enhanced MARL information bottleneck enhanced MARL and mirror learning are being explored to improve communication efficiency robustness and scalability in multi agent settings. Generative information retrieval can also support collaboration by providing agents access to shared knowledge bases or contextual information relevant to joint tasks \cite{zhang2025agenticaigenerativeinformation}.
\par
Structured workflows and specific agent roles are key to effective collaboration in practical agentic systems. For instance, LLM agents can automate physical layer tasks through a coordinated approach, with agents dedicated to task awareness, environmental observation, system configuration, and API invocation, all responding to high-level requests and environmental perception \cite{xiao2024llmagents6gorchestrator}. Likewise, generative simulation frameworks employ agents for simulation generation, test design, execution, and result interpretation, collaborating via a central orchestrator to iteratively refine network models \cite{rezazadeh2025generative6gsimulationexperimental}. Architectures like AgentNet further enhance inter-agent collaboration, supporting learning and knowledge transfer across diverse agent types \cite{xiao2025agenticainetworking6g}. Crucially, agentic AI also facilitates human-AI collaboration, augmenting human capabilities in knowledge-intensive tasks and complex operations \cite{jiang2025largeaimodelsagentic}.
\par
While multi-agent collaboration holds immense promise, significant challenges must be addressed to fully realize its potential, particularly in areas like 6G networks. These challenges include managing communication overhead, ensuring effective coordination despite partial observability, handling non-stationarity arising from evolving agent policies, and maintaining scalability as agent numbers grow \cite{zhang2025multiagentreinforcementlearningwireless}. Overcoming these hurdles and developing robust, efficient collaboration mechanisms is paramount to achieving the vision of truly intelligent and autonomous 6G networks.

\subsection{Practical framework for applying large AI models}
To bridge the gap between the extensive capabilities of large AI models and their practical application in wireless communications, a structured implementation framework is essential. This framework elucidates a systematic approach for practitioners to harness these models effectively.

\subsubsection{\textbf{Problem formulation and layer identification}}
The process begins with problem formulation and layer identification, where the specific challenge, such as channel associated prediction or network resource management, is clearly defined and mapped to its corresponding communication layer \cite{LLM4WM, NetLLM}. This initial step includes identifying the scope and objectives of the AI integration. For instance, tasks like transceiver design fundamentally reside at the physical layer \cite{10104549}, while network optimization and management are network layer concerns \cite{10829820}. Semantic communications introduces its own layer of challenges and opportunities \cite{gunduz2022beyond}.

\subsubsection{\textbf{Model selection and adaptation}}
Following this, the framework proceeds to model selection and adaptation. This phase involves choosing an appropriate model architecture. For example, a practitioner might employ Transformers for their strength in handling sequential data in end to end communication systems \cite{choukroun2022error}, or leverage generative models like diffusion models for their capabilities in transceiver design \cite{10104549}. For multimodal tasks such as ISAC, a MLLM might be the most suitable choice \cite{cheng2025large}. The selected model then undergoes domain specific adaptation. Techniques such as parameter efficient fine tuning, including methods like LoRA, are crucial for tailoring the general knowledge of the model to the unique condition and constraints of wireless systems without necessitating complete retraining \cite{hu2021lora, LLM4WM}.

\subsubsection{\textbf{Data integration and multimodal fusion}}
The third stage is data integration and multimodal fusion. Wireless systems generate a rich tapestry of data, including signals, images, and network logs. An effective framework must incorporate mechanisms to process and align these diverse data streams. This is exemplified in applications like ISAC, which fuses sensor data for enhanced environmental awareness \cite{cheng2025large, li2025large}. It is also critical in network operations, where models analyze voluminous data sources such as base station logs and performance metrics to diagnose faults \cite{10829820}. The ability to synthesize multimodal information enables more holistic and context aware decision making.

\subsubsection{\textbf{Interaction and deployment}}
Finally, the framework culminates in interaction and deployment. This involves establishing how the AI model will interface with the network infrastructure. For many applications, this is realized through an agentic paradigm, where the large AI model acts as an intelligent agent that perceives the network state, reasons over complex data, and executes actions via application programming interfaces \cite{xiao2025agenticainetworking6g, dev2025advancedarchitecturesintegratedagentic}. This approach, central to the concept of Agentic AI RAN, enables a dynamic and autonomous loop of monitoring, analysis, and control, paving the way for truly AI-empowered wireless networks.

\subsection{\textbf{Summary and insights}}
The integration of WLAMs across the communication stack represents a paradigm shift from isolated, modular optimizations to a holistic AI-empowered architecture. Analyzing the relationships across these domains reveals a clear functional hierarchy. Physical layer applications demand stringent real-time processing and mathematical adaptability to combat channel impairments, whereas network layer operations and wireless agents prioritize complex logical reasoning, long-term planning, and intent translation. SemCom serve as a critical bridge, transforming raw physical signals into high-level knowledge representations that network agents can comprehend. The core of this ecosystem lies in its bidirectional interaction: physical perception continuously grounds the reasoning of WLAM agents with real-time environmental context, while the agents autonomously orchestrate and reconfigure physical and network layer parameters to fulfill user intents. This cohesive cross-layer integration is essential for evolving from automated components to fully autonomous 6G networks.

\section{Wireless communications for large AI models}\label{section4} 

\begin{figure}[t]
	\begin{center}
		\centerline{\includegraphics[width=0.8\linewidth]{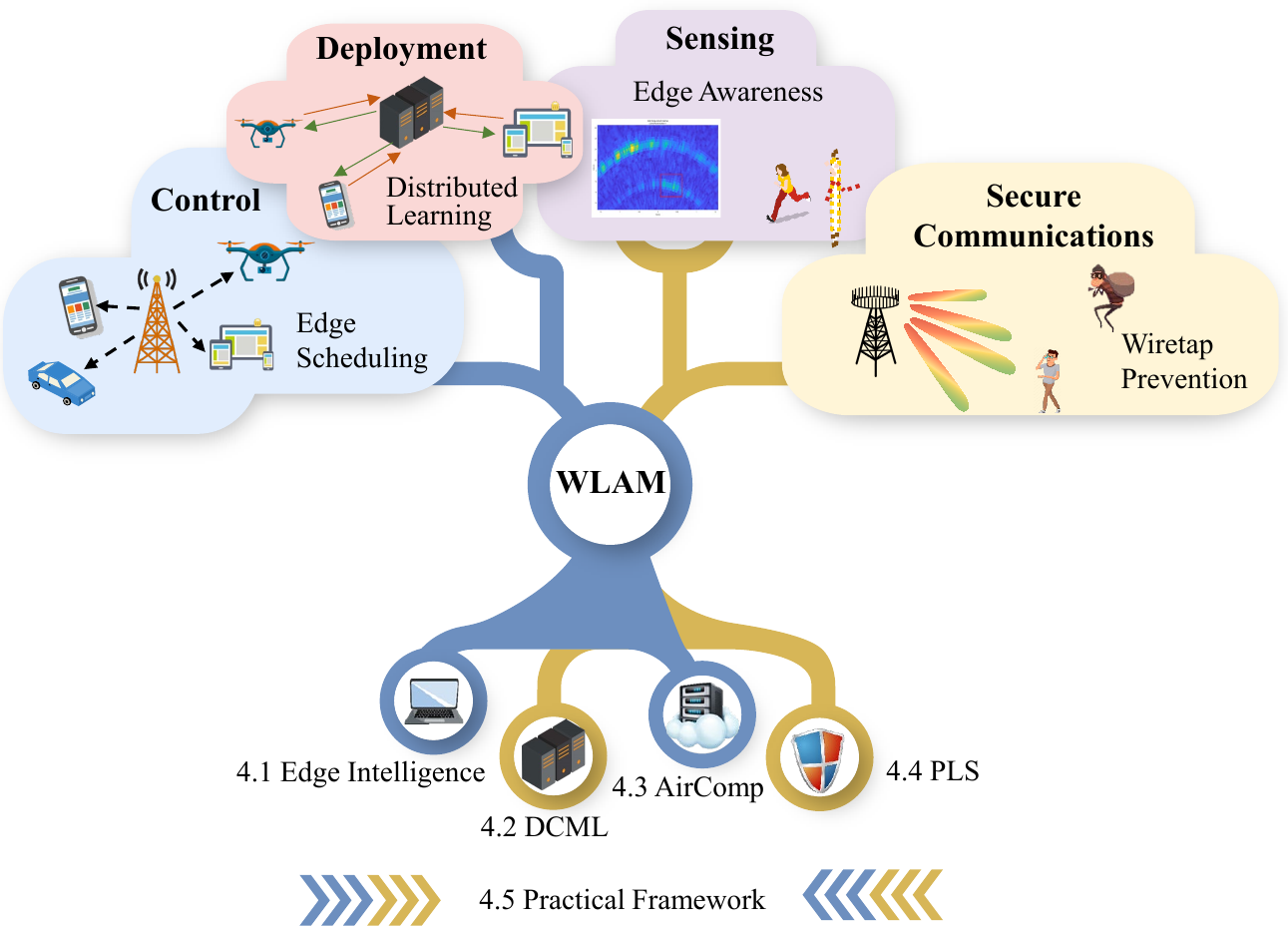}}
		\captionsetup{font=footnotesize, name={Fig.}, labelsep=period} 
		\caption{\, The outline of Section 4.}
		\label{outline:sec4}
		\vspace{-8mm}
	\end{center}
\end{figure}

While Section \ref{section3} has demonstrated the transformative potential of WLAMs in optimizing wireless networks, realizing these applications in practice introduces a new set of challenges. The massive parameter scale that enables the superior performance of WLAMs simultaneously imposes immense pressure on computation, storage, and communication resources. Consequently, the primary research focus shifts from exploring the capacity of AI to optimize networks to addressing the network engineering challenges necessary to efficiently support large-scale AI models. To address these challenges for deployment and training within distributed wireless environments, this section explores the key wireless communication strategies and technologies that enable the efficient operation of large AI models, as illustrated in Fig. \ref{outline:sec4}. To provide a clear and structured overview, our discussion is organized around four foundational pillars, each directly mapping enabling technologies to specific 6G performance indicators. Specifically, we examine how edge intelligence resolves computational and storage bottlenecks for distributed inference on constrained devices; how distributed collaborative learning (e.g., FL, SL) balances massive connectivity with stringent data privacy and training efficiency; how AirComp overcomes communication bottlenecks to minimize data aggregation latency and enhance spectral efficiency; and how physical layer security provides a lightweight, resilient defense against eavesdropping and jamming attacks. These pillars are then synthesized into a practical framework designed to guide the development of robust and efficient systems for supporting WLAM. The related works are summarized in Table \ref{tab:Wireless_for_AI}.

\renewcommand{\arraystretch}{1.2}
\begin{table*}[!t]
\scriptsize\centering
\caption{Summary of related works on wireless communications for large AI models.}
\label{tab:Wireless_for_AI}
\begin{tabular}{|m{0.11\textwidth}<{\centering}|m{0.05\textwidth}<{\centering}|m{0.36\textwidth}<{\raggedright}|m{0.35\textwidth}<{\raggedright}|}
\hline
\textbf{Techniques} &
\textbf{Ref.} &
\textbf{Scenarios} &
\textbf{Contributions} \\ \hline

\multirow{3}{*}[-2em]{\centering\parbox{0.09\textwidth}{\centering Edge Intelligence}} 
  & \cite{lin2023pushing} 
  & Deploying LLMs at the 6G mobile edge for key applications like healthcare and robotics control in cloud-based deployment.
  & Identify key challenges and discuss enabling techniques for efficient LLMs edge training and inference. 
  \\ \cline{2-4} 

  & \cite{zhang2025communication}
  & Distributed on-device LLM inference using tensor parallelism.
  & Propose a communication-efficient framework to accelerate the all-reduce operations in tensor parallelism.
  \\ \cline{2-4} 
  
  & \cite{wdmoe}
  & Splitting MoE LLMs with the gating network on an edge server and expert networks distributed on mobile devices.
  & Propose the WDMoE architecture for collaborative inference, expert selection and bandwidth allocation to minimize latency.
  \\ \hline

\multirow{3}{*}[-2em]{\centering\parbox{0.095\textwidth}{\centering DCML}}
  & \cite{OpenFedLLM}
  & Collaborative and privacy-preserving training of LLMs on decentralized private data using FL.      
  & Propose OpenFedLLM, a concise, integrated, and research-friendly framework/codebase for FL on LLMs.
  \\ \cline{2-4} 

  & \cite{shfMIOT}
  & Federated fine-tuning and collaborative reasoning for large AI models at the network edge.
  & Propose advanced FL paradigms like clustered, hierarchical, and asynchronous FL to address practical wireless challenges.
  \\ \cline{2-4} 

  & \cite{zhao2024fedsllm}
  & Efficiently fine-tuning LLMs over wireless networks by combining data and model parallelism.
  & Propose the FedsLLM framework, which integrates LoRA with FSL to balance the workload between clients and server.

  \\ \hline

\multirow{3}{*}[-2em]{\centering\parbox{0.08\textwidth}{\centering AirComp}}
  & \cite{liu2023over}           
  & Enabling distributed inference of large AI models that require non-linear aggregation functions like Max/Average pooling.
  & Proposing a generalized AirComp framework to support non-linear functions in distributed large AI models.
  \\ \cline{2-4}
  
  & \cite{liu2024over}
  & Efficiently fusing distributed sensor data for real-time inference in large multi-modal AI models.
  & Propose an AirComp framework that exploits feature sparsity to enable scalable data aggregation for large perception models.
  \\ \cline{2-4}
  
  & \cite{shfTWC}
  & Parameter-efficient federated fine-tuning of LLMs over wireless networks.
  & Propose the AirFL-LORA framework using AirComp to achieve communication-efficient aggregation of low-rank gradient updates.
  
  \\ \hline

\multirow{3}{*}[-1em]{\centering\parbox{0.06\textwidth}{\centering PLS}} 
  & \cite{mitev2023physical}              
  & Providing a foundational security layer for distributed large AI model operations in 6G.
  & Provide a comprehensive review of PLS for 6G, paving the way for intelligent networks with large AI models.
  \\ \cline{2-4}
  
  & \cite{infocomfederated}       
  & Defining the fundamental security limits for wireless federated learning, a key challenge for large AI models.   
  & Establish the theoretical limits for securing the high-dimensional parameter updates of large AI models.
  \\ \cline{2-4}
  
  & \cite{federatedPHY}            
  & Ensuring resilient federated training of LLM over wireless channels in the presence of physical layer attacks like jamming.   
  & Establishes a direct mathematical link between the training loss of a LLM and the communication MSE of the physical layer.
   \\ \hline
   
\end{tabular}
\end{table*}

\subsection{Edge intelligence}
6G networks are expected to support in-network, distributed AI capabilities at the edge, facilitating collaborative machine learning across devices with varying computational resources \cite{lin2023pushing, 9797722, 10091939}. However, one of the main challenges is that many edge devices have limited computing power and storage capacity. To address this, wireless technologies are evolving to support the distributed large AI models \cite{zeng2024implementation}.

\subsubsection{Distributed collaborative inference}
While collaborative training is a key application, a critical challenge for deploying pre-trained LLMs is enabling collaborative inference across multiple resource-constrained devices. This avoids the need to store a massive model on a single node and leverages parallel processing to reduce latency. Recent research has focused on two primary paradigms for splitting a single large model for collaborative inference.
\par
One prominent approach is based on architectural model splitting, particularly for MoE models. The authors in \cite{wdmoe} propose the WDMoE framework, where an MoE-based LLM is split by placing the main network body and the gating network on an edge server, while distributing the individual expert networks across multiple mobile devices. This scheme effectively utilizes the parallel processing capabilities of edge devices for the expert layers, which constitute a significant portion of the model's computation.
\par
Another powerful paradigm is tensor parallelism, which partitions the internal weight matrices of an LLM across different devices. While this enables fine-grained parallel computation within each layer, it introduces a communication bottleneck due to the frequent `all-reduce' operations required to aggregate intermediate results. To overcome this, the authors in \cite{zhang2025communication} propose a communication-efficient framework that uses AirComp to accelerate this aggregation step, significantly improving the speed and scalability of on-device collaborative inference.

\subsubsection{Resource allocation in heterogeneous systems}
Effective resource allocation in wireless networks is essential for supporting large AI model training and deployment across edge devices \cite{10091939}. These collaborative inference paradigms transform the traditional resource allocation problem into a more complex, cross-layer optimization challenge. The key task evolves from merely allocating communication resources to the intelligent assignment of model components such as expert networks or tensor shards based on the heterogeneous capabilities of edge devices. This creates a new set of optimization problems, for instance, the joint expert selection and bandwidth allocation required in MoE-based systems to minimize latency \cite{wdmoe}, and the co-optimization of tensor shard placement and communication resources needed to prevent bottlenecks in tensor parallelism frameworks \cite{zhang2025communication}. Successfully solving these emergent problems is central to realizing the full potential of collaborative edge intelligence for large AI models.

\subsection{Distributed collaborative machine learning}
Distributed collaborative machine learning (DCML) gains popularity due to its data privacy advantages \cite{kairouz2021advances}. Unlike traditional methods, data in DCML is accessed collectively without transferring from administrators to any untrusted parties. DCML enables distributed model training on decentralized data. Currently popular DCML methods include FL, split learning (SL) and federated split learning (FSL), and the important properties are summarized in Table \ref{tab:dcml_comparison_final}.

\begin{table*}[!t]
\renewcommand{\arraystretch}{1.2}
\scriptsize\centering
\captionsetup{font=footnotesize, name={Table}, labelsep=period}
\caption{Comparisons of distributed learning frameworks.}
\label{tab:dcml_comparison_final}
\begin{tabular}{|m{0.15\textwidth}<{\centering}|m{0.25\textwidth}<{\raggedright}|m{0.25\textwidth}<{\raggedright}|m{0.25\textwidth}<{\raggedright}|}
\hline
\textbf{Feature} & \textbf{Federated Learning} & \textbf{Split Learning} & \textbf{Federated Split Learning} \\
\hline
\textbf{Core Idea} &
Parallel local training of the full model, followed by central aggregation of the updates. &
Sequential training of a vertically split model, where clients and a server collaboratively process the model layer by layer. &
A hybrid approach combining the parallelism of FL with the low client-side workload of SL. \\
\hline
\textbf{Convergence Analysis} &
The convergence bound reveals a dependency on data heterogeneity $\Gamma$ and the number of local epochs $E$. &
The convergence guarantee is dependent on the number of clients $N$ and local steps $K$ in deep learning scenarios. &
The convergence bound depends on local iterations $\tau$ and data heterogeneity factors such as $\sigma_n^2$ and $\epsilon^2$. \\
\hline
\textbf{System Delay Model} &
Defined by the slowest client's processing time: \newline $\max(T_{\text{comp}} + T_{\text{comm}})$. This is known as the straggler effect. &
Defined by the cumulative time of all clients processing in sequence: \newline $\sum(T_{\text{comp}} + T_{\text{comm}})$. This creates a sequential bottleneck. &
A hybrid model defined by parallel client tasks and sequential server tasks: \newline $\max(T_{\text{client}}) + \sum(T_{\text{server}})$. \\
\hline
\textbf{Client Computation Load} &
The clients must process the entire model architecture, which demands significant local memory and computational resources. &
The computational burden on clients is minimal, as they only process the initial and computationally lighter layers of the model. &
Client computation is significantly reduced by processing only the initial model layers, making the architecture suitable for edge devices. \\
\hline
\textbf{Communication Overhead} &
Content: Full model parameters $S_{\text{model}}$. \newline Frequency: Once per round, at the aggregation stage. &
Content: Smashed data and activations $S_{\text{act}}$. \newline Frequency: Twice per local step (forward and backward). &
Content: Smashed data and activations $S_{\text{act}}$. \newline Frequency: Twice per local step. \\
\hline
\end{tabular}
\end{table*}

\subsubsection{\textbf{Federated learning}}
FL is a DCML paradigm designed to train models across multiple decentralized devices or servers holding local data samples, without exchanging the raw data itself, thereby inherently preserving privacy \cite{10436365}. The typical architecture involves distributed nodes (clients), often mobile or IoT devices, performing training on their local datasets, and a central server (coordinator) responsible for orchestrating the process. The core FL process is iterative: the server distributes a global model, clients train this model locally for a set number of iterations using their data, clients send their updated model parameters or gradients (not their data) back to the server, and the server aggregates these updates to improve the global model. This cycle repeats until the model converges.
\par
FL encompasses several approaches tailored to different data distribution scenarios, particularly relevant in wireless communications. Horizontal FL (sample-based FL) applies when clients share the same feature space but have different user samples, enabling collaboration between entities like telecom companies without sharing user data \cite{yang2020horizontal}. Vertical FL suits scenarios where clients hold different feature sets for the same users, requiring advanced network optimization to combine features effectively \cite{liu2024vertical}. Federated transfer learning addresses situations where both samples and features differ across clients, leveraging shared representations to enhance model training, especially useful for collaboration between diverse wireless service providers or IoT networks with minimal data overlap \cite{liu2020secure}. 

\begin{figure}[t]
	\begin{center}
		\centerline{\includegraphics[width=0.6\linewidth]{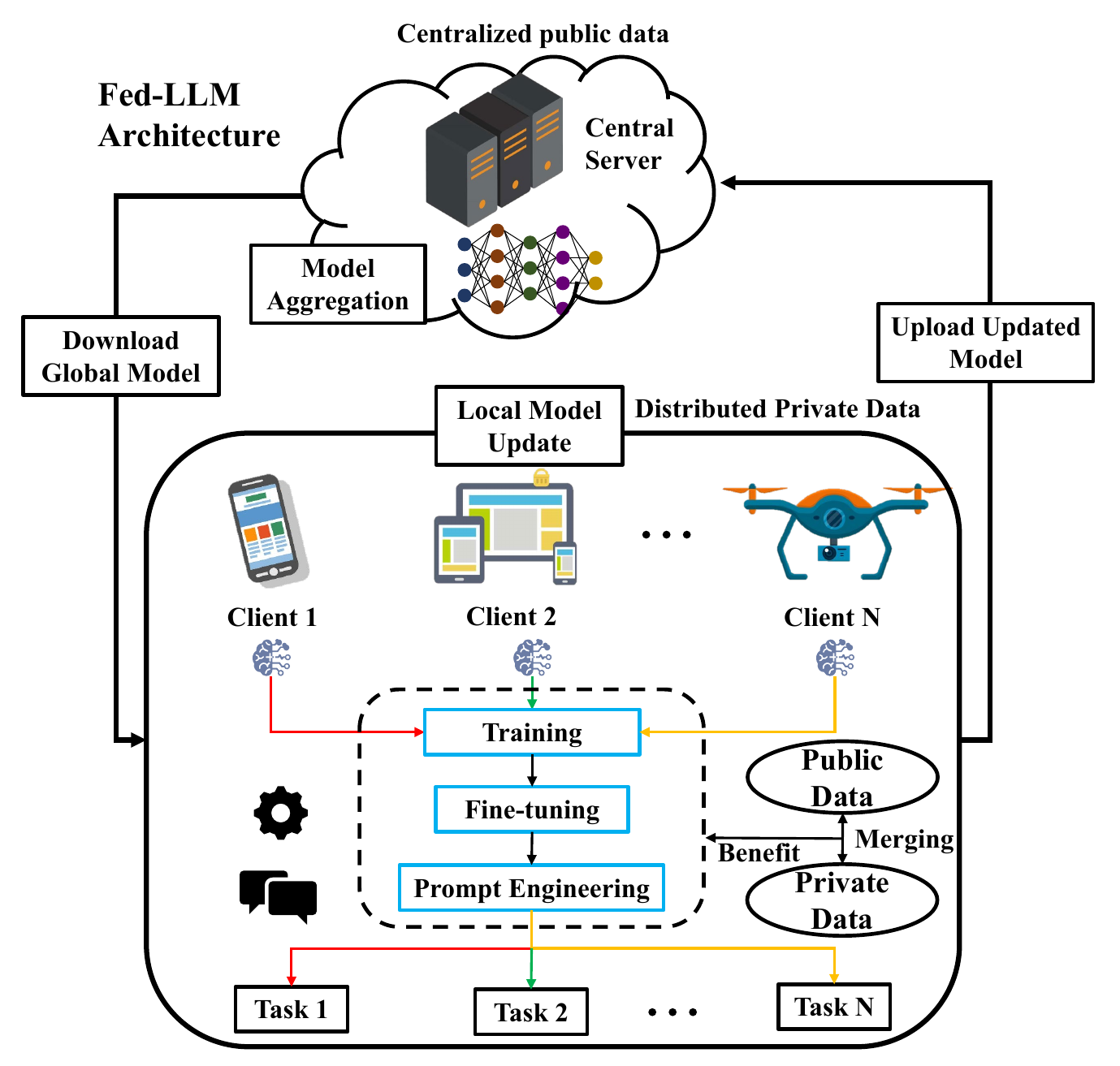}} 
		\vspace{-0mm}
		\captionsetup{font=footnotesize, name={Fig.}, labelsep=period} 
		\caption{\, Workflow of Fed-LLM. By distributing training, fine-tuning, and prompt engineering across edge clients, this architecture leverages local private data for customized model updates. Centralized aggregation then collaboratively refines the global model while strictly preserving data privacy.}
		\label{federated}
		\vspace{-8mm}
	\end{center}
\end{figure}

A crucial step in the FL process is model aggregation, where the server combines the updates received from clients. Various techniques enhance this step. For instance, robust federated aggregation employs the geometric median to resist potential malicious interference \cite{pillutla2022robust}. Selective model aggregation methods evaluate local data quality or computational capacity to choose which client updates to include, optimizing resource use in environments like vehicular edge computing \cite{ye2020federated}. Secure aggregation protocols use secure multi-party computation to protect the privacy of individual client gradients during aggregation \cite{bonawitz2016practical,fereidooni2021safelearn}.
\par
To better adapt federated learning for practical wireless challenges such as device heterogeneity and dynamic connectivity, recent research has introduced more specialized architectures \cite{shfMIOT}, contributing to the broader vision of 'Federated Intelligence'. One such approach is clustered federated fine-tuning, which groups devices with similar characteristics to handle data heterogeneity and train more customized models. Another paradigm, hierarchical federated fine-tuning, employs a multi-level aggregation structure to effectively balance local model personalization with global generalization. Finally, asynchronous federated fine-tuning enhances network efficiency by allowing devices to submit updates independently, thereby accommodating participants with varying processing speeds and intermittent connectivity.
\par
FL also offers methods to address challenges like insufficient labeled data on client devices. Federated few-shot learning (FedFSL) enables models to classify new classes using very few labeled samples \cite{fan2021federated}. Private semi-supervised federated learning leverages abundant unlabeled data alongside scarce labeled data in a privacy-preserving manner \cite{fan2022private}. Personalized FedFSL goes further by identifying optimal collaborating clients for specific tasks, learning personalized feature spaces without data disclosure \cite{zhao2022personalized}. Techniques like FedAffect update feature extractors using disjoint unlabeled private data to learn diverse representations \cite{shome2021fedaffect}. For LLMs, methods like AUG-FedPrompt use minimal labeled data initially and augment it by annotating unlabeled data, though potentially incurring high communication costs due to full parameter tuning \cite{cai2022aug}.
\par
FL deployment enables significant lifecycle enhancements for LLMs, especially within wireless communication applications, as depicted in Fig. \ref{federated}. These enhanced models are termed federated LLMs (Fed-LLMs), and the specific details are outlined as follows.

\paragraph{Pre-training stage}
Fed-LLM can be deployed during the pre-training stage to customize specific needs of end users and improve performance for specialized tasks. The authors in \cite{OpenFedLLM} proposed a framework which integrates centralized public data with distributed private data sources for pre-training. This design tailors the LLM architecture based on pre-training parameter selection and task requirements. Multiple clients initiate pre-training with local data and subsequently perform model sharing. The utilization of diverse computing resources not only enhances model generalization but also maintains data privacy. In the context of wireless communication, this approach allows for robust, efficient, and privacy-preserving model training across various network nodes, improving overall system performance and adaptability.

\paragraph{Fine-tuning stage}
Fed-LLM can also benefit the fine-tuning process of the large AI models in practical wireless environments. In the target deployment scenarios, multiple clients can use local data for federated collaborative parameter-tuning. The fine-tuned model parameters are then uploaded to a server for aggregation and distribution until the model reaches convergence \cite{hilmkil2021scaling}. To reduce the computation and communication costs of full-model fine-tuning, parameter-efficient fine-tuning methods \cite{ding2023parameter} can be integrated into the Fed-LLM framework to minimize parameter gradient calculations and reduce the number of aggregated parameters. This can achieve a balance between maintaining performance and mitigating the computation and communication overhead, which is essential for efficient and scalable model updates. 

\paragraph{Prompt engineering}
Additionally, prompt engineering can take advantage of Fed-LLM. Prompt engineering is the process of guiding a large AI model to produce a desired output. Although large AI models attempt to mimic the behaviors of humans, detailed instructions are still required to create high-quality and relevant output \cite{liu2023pre}. Standard prompt design often relies on public data, limiting applicability for specialized or personalized tasks. The reuse of general prompts from public datasets may also reduce responsiveness of the models. By leveraging FL in wireless communication networks, prompts can be generated using private data while ensuring privacy protection \cite{fedprompt}. This approach improves the generalization ability of the models, enabling the model to handle tasks in specialized fields more proficiently. Moreover, personalized prompts can be created to meet the specific needs of clients, making the AI systems more responsive and contextually aware \cite{cai2022aug}.

\par
To analyze FL performance, we model its distinct communication round. A typical round begins with the central server transmitting the current global model to the $N$ clients of the network. Each client then performs multiple local training iterations on its private dataset before computing a model update. Subsequently, all $N$ clients transmit their updates back to the server. The round concludes when the server aggregates these $N$ updates to produce a refined global model for the next round.
\par
This iterative process directly impacts learning convergence, particularly in the challenging scenario of non-independent and identically distributed (non-IID) client data. For a smooth and strongly convex objective, the convergence of the FedAvg algorithm \cite{Li2020On} is bounded by :
\begin{equation} \label{FedAvgE}
\mathbb{E}[F(\mathbf{w}_T)] - F^* \le \frac{2\kappa}{\gamma+T} \left( \frac{B}{\mu} + 2L\|\mathbf{w}_0 - \mathbf{w}^*\|^2 \right),
\end{equation}
where $B$ is defined as:
\begin{equation}
B = \sum_{k=1}^{N} p_k^2 \sigma_k^2 + 6L\Gamma + 8(E-1)^2 G^2.
\end{equation}
In (\ref{FedAvgE}), $\mathbb{E}[F(\mathbf{w}_T)]$ is the expected loss of the global model weights $\mathbf{w}_T$ after $T$ rounds, and $F^*$ is the optimal loss value. The initial model weights are $\mathbf{w}_0$ and the optimal model weights are $\mathbf{w}^*$. The constants $\mu$ and $L$ represent the strong convexity and smoothness of the loss function, with their ratio defined as $\kappa=L/\mu$. The term $\gamma$ is a constant defined as $\gamma=\max\{8\kappa, E\}$. For the component $B$, $N$ is the total number of clients, $p_k$ is the data proportion for client $k$, and $\sigma_k^2$ is the variance of its stochastic gradients. The term $\Gamma$ quantifies the degree of data heterogeneity across clients. Finally, $E$ is the number of local training epochs each client performs, and $G^2$ is an upper bound on the squared norm of gradients.
\par
The latency of this communication round is also critical. In a synchronous system, the total delay is constrained by the slowest participant \cite{yang2020delay}. This `straggler' effect is a major challenge when many devices are involved. A representative delay model is:
\begin{equation}
T_{FL} = \max_{k \in \{1,...,N\}} \left( E \cdot \frac{D_k C_k}{f_k} \right) + \frac{S_{model}}{B_{dl}} + \frac{S_{model}}{B_{ul}}.
\end{equation}
Here, $T_{FL}$ represents the total time for one round. The first term models the computation delay, where the $\max$ operator is taken over all $N$ clients. For each client $k$, $E$ is the number of local epochs, $D_k$ is its local dataset size, $C_k$ is the computational cycles required per data sample, and $f_k$ is the CPU frequency of the client. The second and third terms represent communication delay, where $S_{model}$ is the model's parameter size, and $B_{dl}$ and $B_{ul}$ are respectively the available downlink and uplink bandwidth.

\subsubsection{\textbf{Split learning}}
SL is a privacy-enhancing machine learning method that enables multiple nodes to collaboratively train a global model without revealing the original data \cite{10843784}. SL typically divides a machine learning model into two parts \cite{gupta2018distributed}: a client-side sub-model and a server-side sub-model, and deploys them on different nodes. Usually, one node acts as the server, hosting the server-side sub-model, while other nodes act as clients, hosting the client-side sub-models. During model training, the client-side and server-side sub-models jointly complete forward propagation and backpropagation in each iteration. Unlike FL, which only employs data parallelism, SL combines data parallelism and model parallelism. This approach is suitable for scenarios with limited client computing and communication capabilities, such as edge computing and the IoT \cite{guo2026accelerating}. Additionally, splitting machine learning models across different nodes in SL helps protect privacy.
\par
Currently, many studies on SL have been conducted. In the field of edge computing, researchers evaluate the learning performance and implementation overhead of SL in real-world IoT scenarios \cite{yansong2020end}. Besides, SL can also be used as a form of distributed learning to achieve URLLC \cite{park2020extreme}. Furthermore, In institutional collaboration, SL can be used to collaboratively train health models across multiple medical institutions \cite{abuadbba2020can}. In terms of privacy protection, some studies have indicated that the activation layer outputs in SL training may lead to data privacy leaks. To address this, various methods and techniques, such as differential privacy, have been proposed to enhance privacy protection \cite{vepakomma2020nopeek}. 
\par
The communication scenario in SL is fundamentally sequential. In a given training round, the server interacts with $N$ clients of the network one by one in a defined sequence. The process begins with the first client, which performs its local training by exchanging activation tensors and gradients with the server. After completing its local steps, this client passes its updated model to the second client in the sequence. This relay-based process continues until the final client, client $N$, has finished its training, marking the end of the round.
\par
This sequential interaction profoundly affects the framework's convergence. For a non-convex objective on non-IID data, the convergence of SL is bounded. As shown in \cite{li2023convergence}:
\begin{equation}
\mathbb{E}[||\nabla f(\overline{x}^{R})||^{2}]\le\frac{4[f(x^{0})-f(x^{*})]}{NK\eta_{g}\eta_{l}R}+12N^{2}K^{2}\eta_{l}^{2}L^{2}G^{2}+6N^{2}K\eta_{l}^{2}L^{2}\sigma^{2}+\frac{4N\eta_{g}\eta_{l}L\sigma^{2}}{K}.
\end{equation}
In this expression, the left side is the convergence metric, where $\overline{x}^{R}$ is the averaged global model over $R$ total rounds. The first term on the right side represents the initialization error, where $f(x^0)$ is the initial loss and $f(x^*)$ is the optimal loss. $N$ is the number of clients, and $K$ is the number of local update steps per client. The terms $\eta_g$ and $\eta_l$ are the global and local learning rates, respectively. The second and third terms collectively represent the client drift error, which depends on the learning rates, the number of clients and local steps, the Lipschitz constant $L$, the bounded gradient norm $G$, and the gradient variance $\sigma^2$. The final term represents the global variance.
\par
The system's latency is a direct reflection of this sequential communication model. The total time for one complete round is the accumulation of the processing times for every single client. This relationship is captured by the delay model:
\begin{equation}
T_{SL} = \sum_{i=1}^{N} \left( K \cdot (T_{fp,i}^c + T_{comm,i}^{act} + T_{server} + T_{comm,i}^{grad} + T_{bp,i}^c) \right).
\end{equation}
In this equation, $T_{SL}$ is the total delay for one round. For each client $i$, the inner parenthesis represents the time for one local step, consisting of client-side forward $T_{fp,i}^c$ and backward $T_{bp,i}^c$ propagation, server processing $T_{server}$, and communication of activations $T_{comm,i}^{act}$ and gradients $T_{comm,i}^{grad}$. This single-step time is multiplied by $K$, the number of local steps. The summation operator $\sum$ over all $N$ clients is the key feature, showing that total latency scales linearly with the number of participants.

\begin{figure}[t]
	\begin{center}
		\centerline{\includegraphics[width=0.6\linewidth]{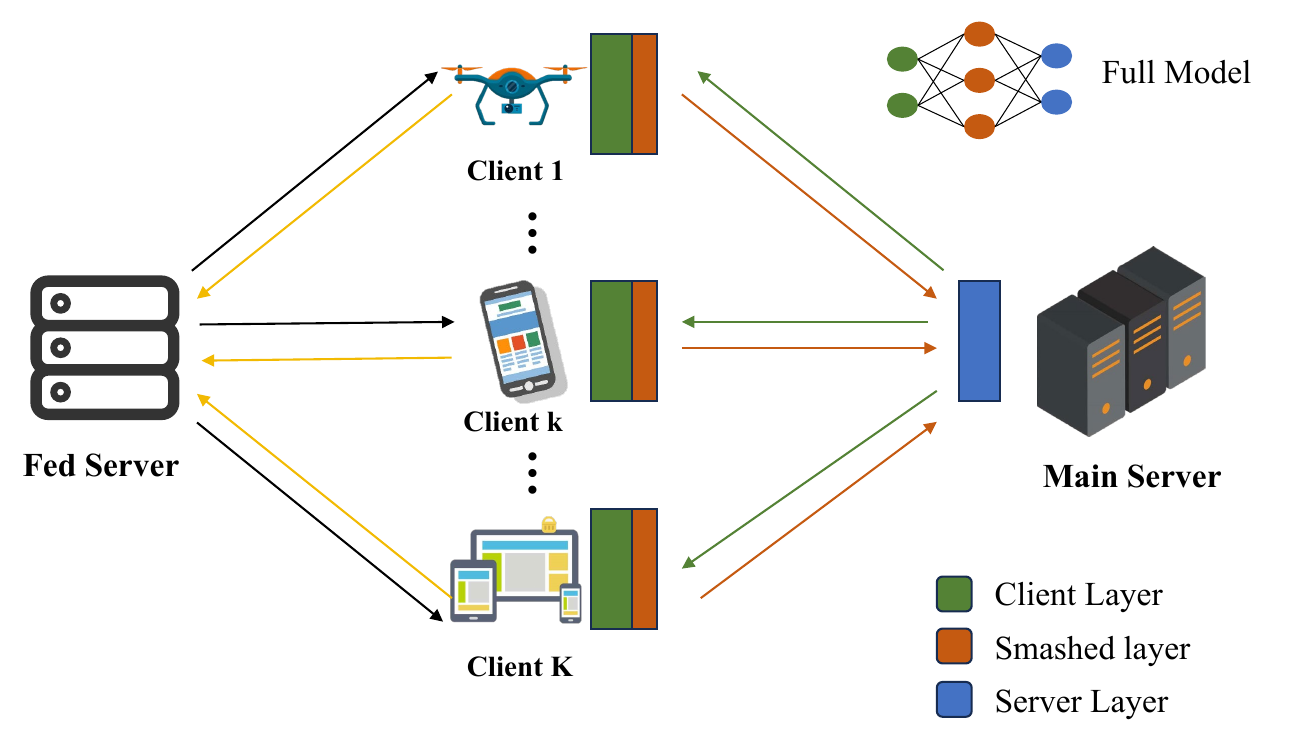}} 
		\vspace{-0mm}
		\captionsetup{font=footnotesize, name={Fig.}, labelsep=period} 
		\caption{\, FSL architecture. Integrating federated and split learning, FSL significantly alleviates client computational burdens. Edge devices process only initial model layers and transmit smashed data to a main server, while a federated server aggregates client-side updates to ensure privacy-preserving collaborative training.}
		\label{fig5}
		\vspace{-8mm}
	\end{center}
\end{figure}

\subsubsection{\textbf{Federated split learning}}
Integrating the benefits of both SL and FL \cite{thapa2022splitfed, zhao2024fedsllm}, FSL has garnered significant research interest, with numerous studies currently exploring its potential. The foundational FSL architecture is depicted in Fig. \ref{fig5}. Research efforts have focused on optimizing FSL, particularly through multi-head SL, which operates without client model synchronization. Experiments confirm this approach is viable and achieves performance comparable to traditional FSL \cite{joshi2021splitfed}. From a security perspective, FSL offers enhanced robustness against model poisoning attacks, as clients handle only partial models and lower-dimensional data \cite{khan2022security}. Studies examining data poisoning attacks further reveal that non-targeted attacks by malicious participants impact global model accuracy more significantly than targeted attacks \cite{gajbhiye2022data}. Additionally, adaptations for specific domains, such as a mobile FSL method for vehicular networks, have demonstrated improved training speeds, substantially reducing training time compared to conventional FSL without sacrificing model accuracy \cite{moon2023split}.
\par
FSL redesigns the communication scenario to introduce parallelism. In an FSL round, a Fed Server coordinates with the network's $N$ clients, while a Main Server handles the server-side model computation. The round begins with the Fed Server distributing the client-side model $x_c$ to all $N$ clients. Each client then performs $\tau$ local iterations. In each iteration, all clients in parallel perform a forward pass and transmit their activation tensors to the Main Server. The Main Server sequentially processes these activations, updates its server-side model $x_s$, and returns gradients to each client. After clients complete their backward passes, they update their local client-side models. At the round's end, all $N$ clients send their updated models to the Fed Server for aggregation.
\par
This architecture's performance has been rigorously analyzed. For a non-convex objective, the convergence of SFL is bounded. As derived in \cite{han2024convergence}:
\begin{equation}
\frac{1}{T}\sum_{t=0}^{T-1}\eta^{t}\mathbb{E}[||\nabla_{x}f(x^{t})||^{2}]\le\frac{4}{T\tau}(f(x^{0})-f^{*})+\frac{8NS\tau}{T}\sum_{n=1}^{N}(a_{n}^{2}+1)(\sigma_{n}^{2}+\epsilon^{2})\sum_{t=0}^{T-1}(\eta^{t})^{2}.
\end{equation}
Here, the left side is the convergence metric averaged over $T$ total rounds. On the right side, the first term represents the initialization error, where $f(x^0)-f^*$ is the initial loss gap and $\tau$ is the number of local iterations. The second term represents the error accumulation due to data heterogeneity. $N$ is the total number of clients, $S$ is the smoothness constant, and $\eta^t$ is the learning rate at round $t$. The term $a_n$ is the weight of client $n$, $\sigma_n^2$ is its local gradient variance, and $\epsilon^2$ bounds the divergence between local and global gradients.
\par
The latency of FSL reflects its hybrid parallel-sequential nature. The per-round delay is determined by the slowest client, combined with the sequential processing at the Main Server. A representative delay model is:
\begin{equation}
T_{FSL} = \max_{k \in \{1,...,N\}} \left( \tau \cdot (T_{fp,k}^c + T_{comm,k}^{act}) \right) + \sum_{k \in \{1,...,N\}} T_{server\_proc,k} + \max_{k \in \{1,...,N\}} \left( \tau \cdot (T_{comm,k}^{grad} + T_{bp,k}^c) \right).
\end{equation}
Here, $T_{FSL}$ is the total round time. The first and third terms represent the parallel computation and communication at the clients, limited by the slowest client among all $N$ participants (the $\max$ operator). The middle term, $\sum_{k \in \{1,...,N\}} T_{server\_proc,k}$, reflects that the Main Server still processes the activations from all $N$ clients sequentially, which can remain a bottleneck.

\subsection{Over-the-air computation}
AirComp represents a paradigm shift from conventional communication protocols by integrating computation directly into the communication process \cite{wang2024over}. Instead of treating concurrent transmissions as interference, AirComp harnesses the inherent waveform superposition property of multiple-access channels to compute desired aggregation functions, such as weighted summations, directly over the air \cite{zhu2021over, wang2024over}. This `compute-when-communicate' approach offers significant advantages in spectral efficiency and latency compared to traditional `compute-after-communicate' strategies.
\par
These benefits are particularly crucial for large AI models, whose distributed operations generate massive data aggregation requirements that can overwhelm conventional networks. By designing transceivers to perform in-network computation, AirComp directly addresses the bottleneck of exchanging high-dimensional gradients in federated fine-tuning, especially when combined with PEFT methods \cite{shfTWC}. Furthermore, it is being adapted to support a wide range of large AI models-centric applications, from enabling scalable federated edge learning (FEEL) \cite{cao2024overview} to executing generalized functions for diverse model architectures \cite{liu2023over} and even efficiently fusing sensor data for large multi-modal perception models \cite{liu2024over}.

\begin{figure}[t]
	\begin{center}
		\centerline{\includegraphics[width=0.5\linewidth]{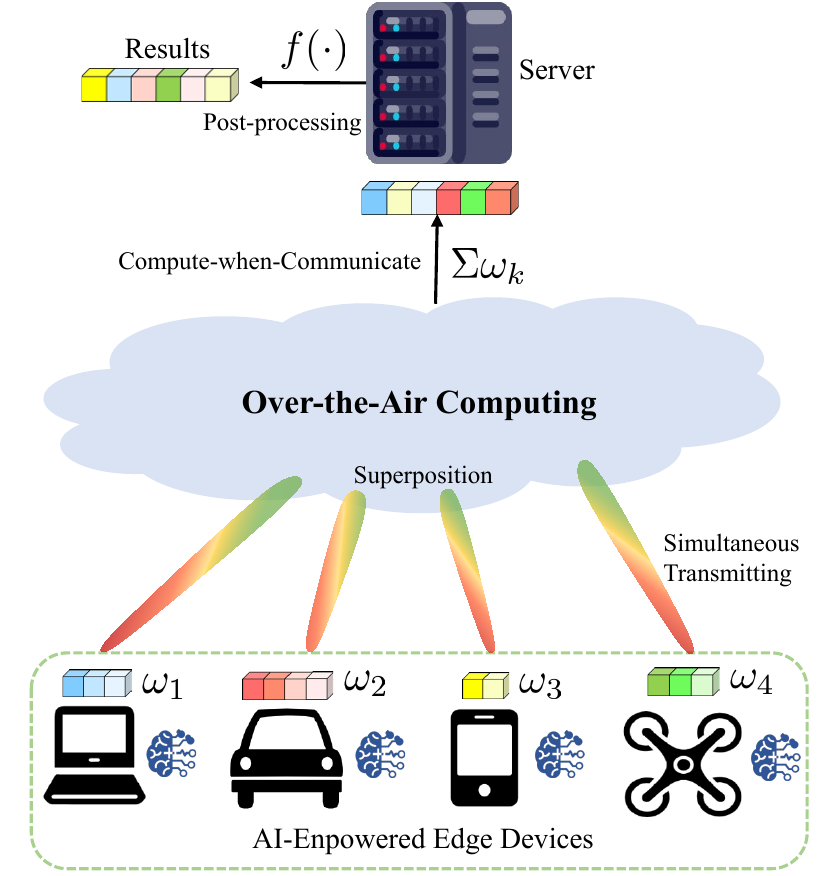}}
		\captionsetup{font=footnotesize, name={Fig.}, labelsep=period} 
		\caption{\, AirComp architecture. By exploiting the inherent waveform superposition of wireless channels, AirComp allows simultaneous data transmission from diverse edge devices. This "compute-when-communicate" paradigm directly aggregates model updates over the air, drastically minimizing latency and bandwidth bottlenecks.}
		\label{sec4_aircomp}
		\vspace{-8mm}
	\end{center}
\end{figure}

\subsubsection{\textbf{AirComp for federated edge learning}}
One of the most significant applications of AirComp is in the domain of FEEL, leading to the concept of scalable FEEL (Air-FEEL) \cite{cao2024overview}. FEEL is a key technique for training large AI models collaboratively across distributed edge devices holding local data, without centralizing sensitive user data. However, in conventional FEEL, the communication overhead associated with frequently exchanging high dimensional model parameters or gradients, characteristic of large AI models between numerous edge devices and a central server constitutes a major bottleneck \cite{XIAO2024}. Air-FEEL addresses this critical challenge by employing AirComp for one shot aggregation of these local updates over the wireless channel \cite{wang2024over}. This approach significantly enhances communication efficiency and reduces training latency, making the distributed training of large models more feasible, compared to methods requiring orthogonal resource allocation or sequential decoding \cite{cao2024overview, wang2024over}. Two primary implementations exist: Air-FedSGD, which aggregates gradients after each local computation, and Air-FedAvg, which aggregates model parameters after multiple local iterations, both facilitating the efficient update of a global large AI model \cite{cao2024overview}. Despite the introduction of aggregation errors, sophisticated techniques analyze and optimize Air-FEEL performance, linking aggregation errors (bias and mean squared error) to the learning optimality gap of the large model \cite{cao2021optimized, cao2022transmission}. Optimized power control strategies further aim to minimize this gap, managing the tradeoff between convergence speed and aggregation errors, thus directly impacting the training efficiency of the large AI model \cite{cao2021optimized, cao2022transmission}. Furthermore, Air-FEEL inherently enhances privacy, a critical concern when dealing with data used to train large models, by masking individual updates within the aggregated signal \cite{cao2024overview}.
\par
Building on this principle, recent works have tailored Air-FEEL specifically for the challenges of fine-tuning large AI models. A key strategy is to combine AirComp with PEFT methods like LoRA. The authors in \cite{shfTWC} propose the AirFL-LORA framework, where instead of transmitting the full, high-dimensional gradients, devices only need to transmit the much smaller low-rank update matrices. This framework designs a multi-input multi-output (MIMO)-OFDM-based AirComp scheme to efficiently aggregate these low-rank adapters over the air. By jointly optimizing the rank of the adapters, resource allocation, and beamforming, the framework minimizes the learning optimality gap, achieving a balance between communication efficiency and model performance that is critical for fine-tuning LLMs on resource-constrained edge devices.

\subsubsection{\textbf{AirComp for large multi-modal models in ISAC}}
The rise of large multi-modal AI models for applications like autonomous driving and collaborative robotics has created new challenges for ISAC. These large AI models often rely on fusing high-dimensional spatial data (e.g., voxel features from LiDAR or camera sensors) from multiple distributed agents to build a comprehensive environmental understanding \cite{liu2024over}. Transmitting these voluminous features to a central fusion node via conventional digital methods presents a significant communication bottleneck, hindering the real-time performance critical for such safety-sensitive systems. Spatial AirFusion is a specialized AirComp framework designed to overcome this challenge. It exploits the unique characteristics of spatial sensing data, such as feature sparsity often found in the intermediate representations of large perception models. By intelligently mapping features to subcarriers and optimizing power, it enables reliable over-the-air aggregation, directly benefiting the accuracy and real-time viability of distributed large multi-modal models in ISAC contexts.

\subsubsection{\textbf{AirPooling for generalized function computation in large AI models}}
Standard AirComp is inherently suited for computing nomographic functions, typically involving summations \cite{wang2024over, zhu2021over}. However, many large AI models, particularly deep neural networks, employ a wider variety of aggregation operations within their architectures. Operations like Max-Pooling, common in CNNs for feature downsampling, or other non-linear aggregation functions used in multi-view sensing or attention mechanisms, are not directly computable via basic AirComp.
\par
To address this limitation, AirPooling extends AirComp capabilities to efficiently realize such crucial non-nomographic functions over the air, directly supporting the distributed implementation or inference of large AI models containing these operations. It specifically addresses Average-Pooling and Max-Pooling by utilizing the mathematical properties of the generalized p-norm function, which smoothly interpolates between summation and the maximum function. AirPooling implements this by decomposing the target computation into appropriate pre-processing at the edge devices and post-processing at the server, leveraging the underlying summation provided by the AirComp mechanism \cite{liu2023over}.
\par
The design optimizes a configuration parameter to carefully balance the accuracy of approximating the desired AI-specific function against the noise amplification inherent in AirComp. Task-oriented optimization, connecting the AirPooling error to final inference accuracy (e.g., using classification margin theory), ensures that the computation directly benefits the performance of the large AI model. AirPooling, therefore, significantly broadens the applicability of AirComp, enabling the efficient wireless execution of essential computational building blocks found within diverse large AI architectures in distributed settings.

\subsection{Physical layer security for large AI models}\label{PLS_for_WLAM}

The deep integration of large AI models into 6G communications fundamentally elevates the importance of security \cite{yang2025privacy}. It creates new attack surfaces and high-value targets. Traditional cryptographic methods, while essential, may face challenges in meeting the stringent latency and scalability demands of distributed AI systems. PLS emerges as a crucial complementary paradigm. It leverages the inherent properties of the wireless channel to provide a lightweight and robust layer of defense for the operations of large AI models \cite{mitev2023physical}.

\subsubsection{\textbf{Fundamental limits of secure federated learning}}
Securing the federated learning process for large AI models over wireless channels introduces a complex interplay of requirements. The system must satisfy the privacy definitions of federated learning. It must also meet the strict confidentiality demands of physical layer security, where information leakage to an external eavesdropper should be negligible. This creates a fundamental trade-off. The work in \cite{infocomfederated} addresses this challenge by proposing a novel architecture for wireless federated learning that incorporates both a privacy mechanism and secure channel coding. The authors formally characterize the theoretical performance boundaries by deriving the capacity-equivocation region. This analysis establishes the fundamental limits on securing the high-dimensional parameter updates that are characteristic of large AI models.

\subsubsection{\textbf{Resilience against physical layer attacks}}
Beyond theoretical limits, the practical performance of a large AI model is directly impacted by the quality of the physical layer. Adversarial attacks like jamming can corrupt the transmission of critical information. This includes the embeddings or gradients exchanged during split federated learning. The work in \cite{federatedPHY} provides a rigorous analysis of this problem. It establishes a direct mathematical connection between the training loss error of a large language model and the communication mean squared error (MSE) over the wireless link. This insight formally demonstrates why the physical layer matters for the convergence of distributed learning. Building on this analysis, the authors propose a resilience-by-design framework. This framework includes methods for beamforming and resource allocation that ensure high task performance and model convergence even under worst-case jamming conditions.

\subsubsection{\textbf{Lightweight security for scalable architectures}}
The distributed nature of many large AI model applications involves a massive number of participating devices. Many of these devices may be resource-constrained. Deploying and managing complex cryptographic protocols across thousands of heterogeneous nodes presents a significant overhead. Physical layer security offers a lightweight solution well-suited for such scalable systems \cite{mitev2023physical, 10850506}. It exploits existing physical channel properties to provide security without requiring extensive computational resources. This low-latency and low-overhead nature makes PLS an ideal component of a defense-in-depth strategy. It can work in concert with traditional cryptography to build a holistic and efficient security architecture for trusted and autonomous networks driven by large AI models in 6G and beyond.

\subsubsection{\textbf{Illustrative privacy and security threats and PLS contributions}}
To illustrate the threats that necessitate the PLS solutions described above, we analyze specific scenarios for both privacy and security, detailing the direct contributions of PLS in mitigating them.
\par
\paragraph{Eavesdropping attack} A primary threat to privacy in distributed learning is posed by inference attacks through passive eavesdropping. Consider a wireless hierarchical federated learning system where an edge server transmits aggregated gradient updates to a cloud server \cite{Zhang2023Finite}. The attack process involves an eavesdropper capturing the signal $Z_{i}(t)$ at each time instant $i$ of communication round $t$. The received signal at the eavesdropper can be modeled as:
\begin{equation}
Z_{i}(t)=gX_{i}(t)+\tilde{g}\tilde{X}_{i}(t)+\eta_{e,i}(t),
\end{equation}
where $X_{i}(t)$ and $\tilde{X}_{i}(t)$ are the legitimate uplink and downlink signals, $g$ and $\tilde{g}$ are the complex channel gains to the eavesdropper, and $\eta_{e,i}(t)$ is additive noise. The consequence is a direct breach of privacy, as the eavesdropper can analyze the captured signals to infer properties of the training data. To counter this, PLS contributes by designing secure channel coding schemes. The achievable secure rate $R_t$ for a given communication round $t$ can be quantified, for example, by the bound:
\begin{equation}
R_{t}=\frac{1}{N_{t}}\log\left(\frac{3\text{SNR}|h|^{2}}{[Q^{-1}(\frac{\tau}{8})]^{2}}\left(1+\frac{\text{SNR}|h|^{2}}{\Psi_{1}\Psi_{2}}\right)^{N_{t}-1}\right).
\end{equation}
Here, $R_t$ is the rate in bits per channel use, $N_t$ is the blocklength, SNR is the signal-to-noise ratio of the legitimate link with channel gain $h$, $Q^{-1}(\cdot)$ is the inverse Gaussian Q-function related to the decoding error probability $\tau$, and $\Psi_1, \Psi_2$ are terms related to the feedback channel quality. This formula demonstrates that PLS can provide a quantifiable positive secure rate, ensuring confidentiality against the eavesdropper.

\par
\paragraph{Jamming attack} On the security front, a direct threat to system integrity is jamming for model performance degradation. In an SFL system fine-tuning an LLM, an adversary can execute a worst-case jamming attack during the uplink \cite{federatedPHY}. The received signal at the base station is modeled as a superposition of the legitimate signals $x_{qnk}$ and a jamming signal $z_{nk}$:
\begin{equation}
y_{nk}=\sum_{q\in\mathcal{Q}}H_{qnk}x_{qnk}+z_{nk},
\end{equation}
where $z_{nk}=G_{nk}u_{nk}+\eta_{nk}$ consists of the jamming signal $u_{nk}$ through the jammer channel $G_{nk}$ and receiver noise $\eta_{nk}$. The consequence is catastrophic, as the jamming directly increases the mean squared error of the communication, causing the server to train the model on corrupted embeddings and preventing convergence. PLS directly addresses this by enabling resilient-by-design communication. The framework in \cite{federatedPHY} proposes designing receive filters based on a surrogate interference covariance matrix that robustly accounts for the jammer:
\begin{equation}
\tilde{C}_{z_{nk}}=\sigma^{2}I_{N_{R}}+\eta A_{B}(\theta_{G})A_{B}^{H}(\theta_{G}),
\end{equation}
where $\tilde{C}_{z_{nk}}$ is the constructed surrogate matrix, $\sigma^2$ is the noise power, $I_{N_R}$ is the identity matrix for $N_R$ receive antennas, $\eta$ is a resilience hyperparameter, and $A_{B}(\theta_{G})$ is the array manifold based on the known directions of arrival $\theta_G$ of the jammer. This allows the system to safeguard the integrity of the training process.

\par
\paragraph{Man-in-the-middle attack} Another critical security threat targets authentication and confidentiality via man-in-the-middle attacks, particularly in protocols that generate secret keys from channel randomness \cite{mitev2023physical}. The attack mechanism involves an adversary transmitting a pre-coded signal $W$ to two legitimate devices, Alice and Bob, during their channel estimation phase. The resulting observations are modeled as:
\begin{equation}
Z_{A}=XH+W+N_{A},
\end{equation}
\begin{equation}
Z_{B}=XH+W+N_{B},
\end{equation}
where $Z_A$ and $Z_B$ are the signals received by Alice and Bob, $X$ is the legitimate probe signal over the reciprocal channel $H$, $N_A$ and $N_B$ are receiver noise, and $W$ is the malicious signal. The consequence is a complete breakdown of confidentiality, as the shared randomness is contaminated by the attacker's known signal $W$. The contribution of PLS lies in redesigning the physical layer protocol. A countermeasure described in \cite{mitev2023physical} involves both users transmitting independent, randomized probe signals, $X$ from Alice and $Y$ from Bob. After post-processing their received signals, their new observations become:
\begin{equation}
\tilde{Z}_{A}=XYH+XW+XN_{A},
\end{equation}
\begin{equation}
\tilde{Z}_{B}=XYH+YW+YN_{B},
\end{equation}
where $\tilde{Z}_{A}$ and $\tilde{Z}_{B}$ are the processed observations. In this new formulation, the shared randomness term is $XYH$, while the attacker contributions $XW$ and $YW$ are no longer identical. This physical layer procedure makes the signal of attacker uncorrelated with the shared secret component, preserving the integrity of the key generation process.

\subsection{Practical framework for enabling large AI models}
The successful deployment and operation of large AI models within distributed wireless environments require a robust framework that systematically addresses computation, communication, and security. This framework provides a guideline for practitioners to build efficient and reliable systems.

\subsubsection{\textbf{Deployment strategy selection}}
The first step is deployment strategy selection, which involves choosing the appropriate distributed collaborative machine learning paradigm \cite{kairouz2021advances}. Depending on data distribution, privacy requirements, and device capabilities, a practitioner might select FL for parallel training on decentralized data \cite{10436365}, SL for scenarios with resource constrained clients \cite{10843784}, or FSL as a hybrid approach that combines the benefits of both \cite{thapa2022splitfed}. This foundational choice determines the overall system architecture and the nature of interactions between nodes.

\subsubsection{\textbf{Resource and model partitioning}}
The second phase is resource and model partitioning. Once a distributed strategy is chosen, the next critical task is to allocate resources and partition the model across the network, which is especially important for edge devices with limited computing power \cite{lin2023pushing}. For heterogeneous systems, this involves the intelligent assignment of model components. For example, in a MoE model, the expert networks can be distributed across multiple mobile devices while a central server manages the gating network \cite{wdmoe}. Alternatively, for other architectures, tensor parallelism can be used to partition the internal weight matrices of a large model across different devices to enable collaborative inference \cite{zhang2025communication}. Successfully solving these joint allocation and partitioning problems is central to realizing the full potential of collaborative edge intelligence.

\subsubsection{\textbf{Communication efficiency optimization}}
In the third phase, the framework must address communication efficiency optimization. The exchange of high dimensional gradients or activations is a major bottleneck in distributed AI. This framework should incorporate advanced wireless techniques to mitigate this challenge. A key technology is AirComp, which leverages the superposition property of wireless channels to perform fast data aggregation \cite{wang2024over, zhu2021over}. In the context of federated learning, this approach, known as Air FEEL, enables one shot aggregation of model updates, significantly enhancing communication efficiency and making the distributed training of large models more feasible \cite{cao2024overview}.

\subsubsection{\textbf{Security integration}}
The final and essential component of the framework is security integration. The distributed nature of these systems introduces new vulnerabilities that must be addressed from the design stage. PLS emerges as a crucial complementary paradigm, offering a lightweight yet robust layer of defense for the operations of large AI models \cite{mitev2023physical}. It can provide confidentiality for model updates against eavesdroppers \cite{infocomfederated} and enhance the resilience of the training process against physical layer attacks like jamming \cite{federatedPHY}. Integrating these security measures is paramount to ensuring the integrity and privacy of distributed AI systems, fostering trust in the autonomous networks of 6G and beyond.

\subsection{\textbf{Summary and insights}}
To support the massive computational and data demands of WLAMs, emerging wireless technologies must operate in tight coordination rather than isolation. Edge intelligence and distributed collaborative learning paradigms, such as FL and SL, primarily tackle the computation and privacy bottlenecks by dispersing heavy training and inference workloads across edge devices. However, this decentralization directly brings a communication bottleneck, which necessitates the integration of AirComp to ensure ultra-fast, concurrent data aggregation. Furthermore, the spread of model parameters across the network expands the attack surface, requiring PLS to provide lightweight, channel-intrinsic defense mechanisms without the severe latency overhead. Ultimately, the successful deployment of WLAMs relies on the intricate balance and synergy of these complementary technologies, where efficient local computation, rapid wireless aggregation, and robust PLS mutually reinforce one another to enable scalable edge intelligence.

\section{Emerging technologies for WLAM}\label{section5}

\begin{table*}[!t]
\footnotesize \centering
\caption{Summary of related works on emerging technologies of WLAM.}
\label{tab:emerging}
\begin{tabular}{|m{0.13\textwidth}<{\centering}|m{0.04\textwidth}<{\centering}|m{0.30\textwidth}<{\raggedright}|m{0.35\textwidth}<{\raggedright}|m{0.09\textwidth}<{\raggedright}|}
\hline
\textbf{Technologies} &
\textbf{Ref.} &
\textbf{Scenarios} &
\textbf{Contributions} &
\textbf{Benefits}
\\ \hline
\multirow{3}{*}[-2em]{\centering\parbox{0.09\textwidth}{\centering Hyper-dimensional Computing}} 
  & \cite{rahimi2016robust}  
  & Execute language classification tasks efficiently and robustly.              
  & Introduce a hardware architecture for a classifier based on hypervector.
  & \multirow{3}{*}[-2em]{\parbox{0.09\textwidth}{Enhanced expressive power}}
  \\ \cline{2-4} 

  & \cite{najafabadi2016hyperdimensional}  
  & Categorizing news articles based on a continuous stream of input letters for text classification.            
  & Demonstrate a software classification framework employing hyperdimensional computing. 
  &
  \\ \cline{2-4}
  
  & \cite{hassan2021hyper} 
  & Execute cognitive tasks efficiently with constrained energy budget and computing resources.
  & Present a overview on the paradigm, algorithms, and applications of hyperdimensional computing.
  &
  \\ \hline

\multirow{6}{*}[-4em]{\centering\parbox{0.09\textwidth}{\centering Quantum Computing}}
  & \cite{botsinis2018quantum}  
  & Develop wireless communication systems that are fast, reliable, secure and energy-efficient.                                  
  & Discuss the quantum algorithms to improve the physical and network layers of wireless communications. 
  & \multirow{6}{*}[-4em]{\parbox{0.09\textwidth}{Powerful computing capability \\and \\improved security}}
  \\ \cline{2-4} 
  
  & \cite{buhrman2003distributed} 
  & Latency-sensitive and computing-intensive tasks with many quantum computers.
  & Introduce the concept of distributed quantum computing and its applications.
  &
  \\ \cline{2-4}
  
  & \cite{narottama2023quantum}            
  & Integrate quantum computing with machine learning for wireless communications.    
  & Discuss the state-of-the-art quantum machine learning algorithms and potential applications in wireless communications.
  &
  \\ \cline{2-4} 
  
  & \cite{narottama2021quantum}            
  & Resource allocation in wireless communication environments.    
  & Present a novel quantum neural network and a reinforcement-enhanced version.
  &
  \\ \cline{2-4}
  
  & \cite{narottama2023federated}              
  & Distributed resource optimization in wireless communication systems. 
  & Present a federated quantum neural network framework utilizing quantum teleportation. 
  &
  \\ \cline{2-4} 
  
  & \cite{chen2021quantum}            
  & Natural language processing with quantum neural networks.    
  & Propose a novel deep neural network model with entanglement embedding module.
  &
  \\ \hline

\multirow{5}{*}[-5em]{\centering\parbox{0.09\textwidth}{\centering Physical Inspired Neural Networks}} 
  & \cite{li2022intelligent}              
  & Fast mapping between codes and radiation patterns for electrically large meta-surfaces in beamforming.           
  & Propose PINN for code-to-pattern mapping and PINN-guided DNN for pattern-to-code mapping, combined for intelligent beamforming.
  & \multirow{5}{*}[-5em]{\parbox{0.09\textwidth}{More \\ responsive, adaptable,
and \\ explainable system}}
  \\ \cline{2-4}
  
  & \cite{10351043}             
  &  Accurate channel prediction in dynamic mobile wireless environments with limited data.          
  & Model channel prediction as ODE problem and design a physics-inspired network with recurrent positioning and Doppler compensation.
  &
  \\ \cline{2-4}

  & \cite{10634040}             
  & Efficient reconstruction of high-resolution radiomaps from sparse samples for wireless network deployment.               
  & Introduce three physics-inspired machine learning methods integrating data-driven AI and model-based radio propagation.
  &
  \\ \cline{2-4} 
 
  & \cite{ncps_beamforming} 
  & MU-MIMO beamforming scenarios in dynamic wireless environments with noise interference. 
  & Introduce a gradient-based liquid neural network framework to effectively perform beamforming. 
  &
  \\ \cline{2-4} 
  
  & \cite{zhu2024robustcontinuoustimebeamtracking}
  & Beam tracking leveraging mmWave to predict best beam index for moving users. 
  & Introduce a robust beam tracking framework employing multi layers of liquid neurons.  
  &
  \\ \hline

\multirow{2}{*}[-1em]{\centering\parbox{0.09\textwidth}{\centering Hyper-Networks}} 
  & \cite{abouamer2023flexible}  
  & Allocate resource in RIS-assisted communication systems with deep neural networks.             
  & Propose a hypernetwork-based approach to generate the beamforming vectors conditioned conditioned on the input user weights. 
  & \multirow{2}{*}[-1em]{\parbox{0.09\textwidth}{Quick adaptation and \\tailoring}}
  \\ \cline{2-4} 
  
  & \cite{tang2024modelgpt} 
  & Develop AI models specifically customized to match the user's provided data or task descriptions.
  & Propose a hypernetwork-based framework to rapidly generate customized AI models.
  &
  \\ \hline
  
\multirow{3}{*}[-2em]{\centering\parbox{0.09\textwidth}{\centering Next-Gen Sequence Modeling Networks}} 
  & \cite{gu2024mamba}  
  &  Long sequence modeling across various modalities (text, audio, picture) where Transformers are inefficient. 
  & Propose Mamba architecture based on selective SSMs with hardware-aware parallel algorithm and simplified design.
  & \multirow{3}{*}[-2em]{\parbox{0.09\textwidth}{Improved inference speed and \\ longer \\ context}}
  \\ \cline{2-4} 

  & \cite{peng-etal-2023-rwkv}
  & General sequence processing tasks aiming for efficiency and performance with linear scaling.
  & Propose RWKV combining Transformer-style parallel training and RNN-style efficient inference using linear attention.
  &
  \\ \cline{2-4}

  & \cite{sun2024learninglearntesttime}
  & Long context sequence modeling needing linear complexity and expressive hidden states.
  & Propose custom layers with self-supervised learning updated hidden states.
  &
  \\ \hline
\end{tabular}
\end{table*}

The communication frameworks discussed in Section \ref{section4} establish the necessary foundation for deploying large AI models. However, realizing the stringent performance and efficiency demands of the 6G era requires moving beyond the optimization of existing architectures. This necessitates a paradigm shift, calling for the integration of novel and deeply integrated methodologies to unlock the full potential of WLAMs. Consequently, emerging technologies become pivotal for the evolution of WLAM, enabling transformative capabilities in wireless communications. This section delves into these technologies, organizing them into two distinct categories to provide a clear overview. First, we explore emerging computing paradigms, and then we investigate emerging neural network architectures. Table \ref{tab:emerging} summarizes related research. Through this exploration, we highlight the profound impact of these technologies in shaping the future of WLAMs.

\subsection{Emerging computing paradigm}\label{Emerging Computing Paradigm}
We first introduce emerging computing paradigms for WLAM, exploring hyperdimensional computing (HDC) for efficient data processing and the revolutionary potential of quantum computing.

\begin{figure}[t]
	\begin{center}
		\centerline{\includegraphics[width=0.5\linewidth]{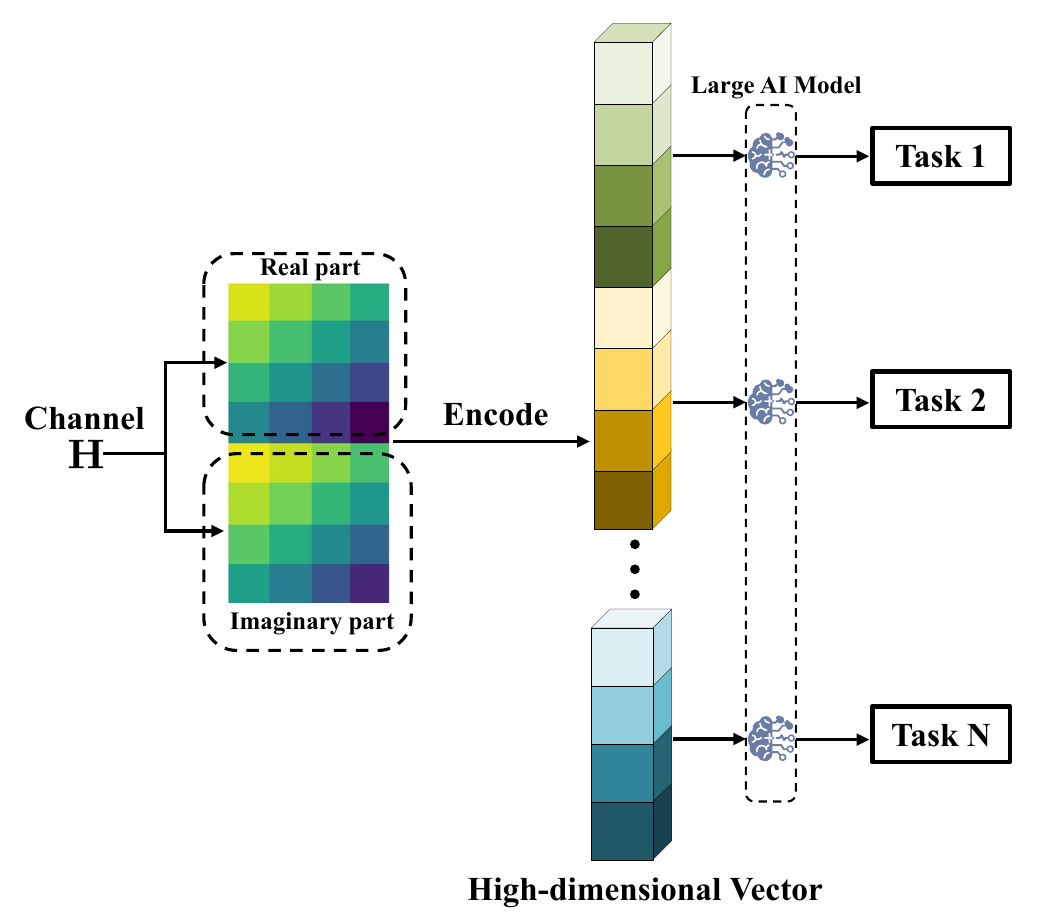}} 
		\vspace{-0mm}
		\captionsetup{font=footnotesize, name={Fig.}, labelsep=period} 
		\caption{\, HDC with large AI models in wireless communications. The pseudo-color plot represents the real and imaginary components of the wireless channel, which are encoded into a high-dimensional vector and then processed by large AI models for different tasks.}
		\label{HDC}
		\vspace{-8mm}
	\end{center}
\end{figure}

\subsubsection{\textbf{Hyperdimensional computing}}
HDC is an emerging technology inspired by the remarkable representation abilities of high-dimensional vector spaces. Unlike traditional computing methods that use low-dimensional data representations, HDC leverages high-dimensional vectors, often comprising thousands of dimensions, to encode information efficiently \cite{kleyko2022survey}. The basic HDC pipeline includes three stages: encoding, training, and comparison \cite{thomas2021theoretical}. At the encoding stage, inputs are mapped into high-dimensional vectors, which are robust to noise and capable of representing complex data structures. During training, these vectors are stored in the associative memory, grouping similar inputs into classes. In the comparison stage, query vectors are matched against stored classes by measuring similarity, typically using the Hamming distance. In the following part we delve into the advantages, various applications, and future prospects of HDC.

\paragraph{Features and advantages of HDC}
HDC has been successfully applied in areas like language recognition \cite{rahimi2016robust}, text categorization \cite{najafabadi2016hyperdimensional}, speech recognition \cite{imani2017voicehd}, etc. It offers several advantages with respect to the implementation in AI \cite{hassan2021hyper}, including significantly lower power consumption and latency compared to traditional deep neural networks, making it suitable for edge computing devices with limited resources. Besides, the high-dimensional nature of HDC provides robustness against noise and uncertainty that is beneficial for real-world applications where data may be incomplete or noisy. Furthermore, the ability of HDC to handle a vast number of unique vectors enhances its scalability across various applications.

\paragraph{HDC in wireless communication}
Fig. \ref{HDC} depicts the application of HDC with large AI models in wireless communication, where the wireless channel is encoded into a high-dimensional vector and then processed by large AI models for different tasks. This richer representation can lead to improved accuracy and efficiency in tasks such as signal processing, security algorithms, and adaptive communication protocols. By leveraging the unique properties of high-dimensional vector spaces, HDC provides a robust, efficient, and scalable solution that meets the growing computational demands of large AI models in wireless communication systems, paving the way for more advanced and capable AI-driven networks.

\paragraph{Future prospects}
Although promising for the benefits mentioned above, HDC is still in its early stages when it comes to its application in large AI models \cite{ge2020classification}. The well-known transformer architecture, which is central to many popular AI models, could significantly benefit from the integration of HDC by utilizing much larger embedding dimensions. High-dimensional vectors in HDC offer richer and more expressive representations, which can enhance the capabilities and performance of AI models, potentially benefiting applications in RL environments.

\subsubsection{\textbf{Quantum computing}}
While moving toward the development of 6G and the era of ubiquitous connectivity, the demand for computational resources is increasing exponentially. To counter this surge of demand, quantum computing has been proposed as a promising solution, offering unprecedented processing capabilities through the principles of quantum mechanics. Quantum computing fundamentally changes how we handle complex computations by utilizing qubits, which can exist in multiple states simultaneously. This unique characteristic enables quantum computers to perform parallel processing on a scale unattainable by classical computers. In the context of WLAM, quantum computing holds the promise of significantly enhancing computational efficiency, optimizing resource allocation, and enabling real-time data processing \cite{wang2022quantum, botsinis2018quantum}. In the following we explore various types, applications, and future prospects of quantum computing in wireless communications.

\paragraph{Types and applications of quantum computing}
Quantum computing can be categorized into several types, each with unique applications. Blind quantum computing, also known as secure quantum computing with privacy preservation, allows a client to delegate computation tasks to remote quantum computers while keeping the source data confidential \cite{barz2012demonstration}. This is achieved by sending transformed qubits to the server, which performs computations and returns temporary results that the client can then convert into the final results, ensuring data privacy throughout the process. Distributed quantum computing involves distributing computational tasks across multiple quantum computers, enhancing data processing speed and decreasing latency. This method is significantly useful for connecting noisy intermediate-scale quantum computers to collaboratively execute complex tasks \cite{buhrman2003distributed}. Additionally, quantum machine learning (QML) leverages quantum computing for machine learning tasks, enabling novel computing services and applications in the context of 6G networks \cite{narottama2023quantum}. Quantum computing offers innovative solutions for meeting the huge computational demands of future wireless communications.

\paragraph{Quantum machine learning}
QML, particularly quantum neural networks (QNNs), is emerging as a powerful force for WLAM in 6G. Addressing the escalating computational demands of WLAM, QML offers pathways to enhance efficiency in critical tasks. For instance, QNN-based frameworks, as explored in \cite{narottama2021quantum}, demonstrate the capability to achieve comparable performance in wireless resource allocation tasks, such as user grouping in non-orthogonal multiple access, but with significantly reduced computational complexity compared to classical neural networks. This efficiency gain is crucial for the real-time operation of WLAM in dynamic 6G environments. Furthermore, the integration of QNNs with FL, as studied in \cite{narottama2023federated}, leverages quantum teleportation to streamline model aggregation in distributed WLAM deployments, potentially accelerating training and enhancing the scalability of intelligent 6G networks.

\paragraph{Future prospects}
The integration of quantum computing with WLAM promises significant advancements in 6G and beyond \cite{chen2021quantum}. As quantum computing progresses, its combination with AI models can vastly improve efficiency and capability in managing the immense data generated by communication networks. Hybrid quantum-classical algorithms could solve complex optimization problems, such as dynamic spectrum management and real-time interference mitigation. Additionally, QNNs combined with advanced AI techniques could enable adaptive, self-optimizing networks that learn and evolve in real time. Quantum-secure communication protocols will enhance data integrity and privacy, meeting the growing computational and connectivity demands of the future. This fusion is poised to revolutionize wireless communication systems, promoting an era of intelligent, efficient, and secure networks.

\subsection{Emerging neural network architecture}
Next, we analyze emerging neural network architectures for WLAM, including physics-informed neural networks (PINNs) for their adaptive learning capabilities, Hyper-networks for dynamic model control and next-generation sequence modeling networks for improved inference speed and efficiency. 
\subsubsection{\textbf{Physics-informed neural networks}}
A promising new research avenue in wireless communication technology is the development of PINNs. This innovative approach effectively bridges the gap between fundamental physical laws and advanced AI techniques. This integration is crucial for tackling the inherent complexities of modern communication systems. Specifically, this research paradigm leverages established principles from wave theory, quantum mechanics, thermodynamics, and related disciplines.

\begin{figure}[t]
	\begin{center}
		\centerline{\includegraphics[width=0.6\linewidth]{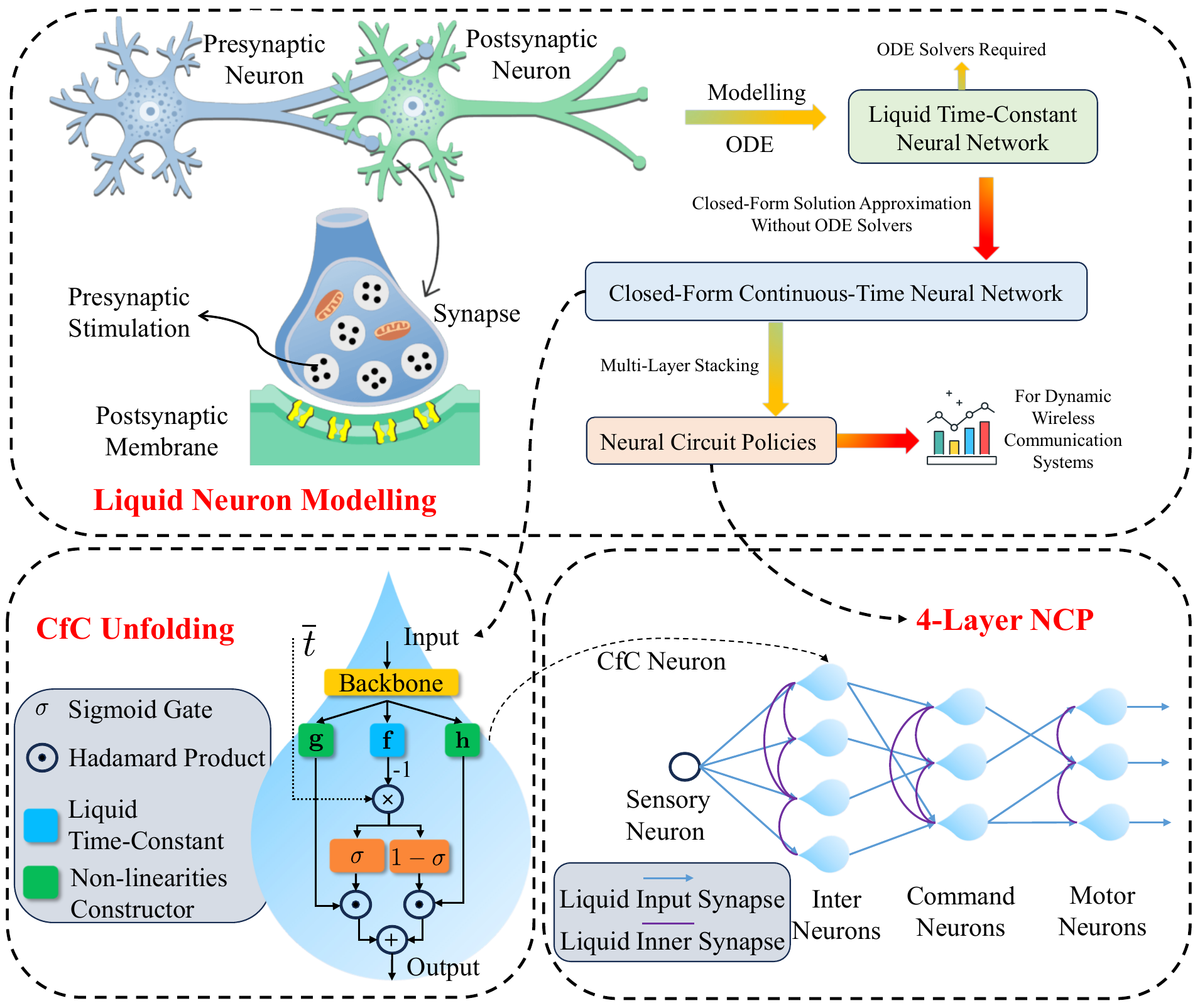}} 
		\vspace{-0mm}
		\captionsetup{font=footnotesize, name={Fig.}, labelsep=period} 
		\caption{\, Liquid neuron and the ODE modeling. This figure illustrates the evolution of liquid neural networks from bio-inspired synaptic interactions to complex control policies. It transitions from Ordinary Differential Equation based Liquid Time-Constant neurons to efficient Closed-form Continuous-time models. These units are stacked into Neural Circuit Policies to enable resilient, adaptive control in dynamic wireless systems.}
		\label{lnn_neuron}
		\vspace{-8mm}
	\end{center}
\end{figure}

\paragraph{ODE based neural networks}
While PINNs are often motivated by scientific research challenges, their fundamental strength lies in solving differential equations. Therefore, ordinary differential equation (ODE) a well-established and popular application area for PINNs. ODE describes the fundamental changes in systems over time, linking physical processes with temporal evolution. By modeling how variables change continuously with respect to time, ODE provides a powerful framework for understanding dynamic systems \cite{ODE}. In the context of neural networks, ODE-based methods leverage this continuous modeling approach to enhance the robustness and efficiency of deep learning models. These networks incorporate ODE to represent the continuous transformation of data, offering a more nuanced and flexible alternative to traditional discrete architectures. The key innovation in ODE-based neural networks is the use of ODE solvers to manage the evolution of hidden states throughout the training and inference processes. This approach enables the network to learn complex, continuous dynamics, which is particularly beneficial for tasks involving temporal or sequential data \cite{lstm-ode}.
\par
Classical examples of ODE-based neural networks include continuous-time (CT) models like CT-RNNs and ODE-long short-term memory (LSTM) networks. CT-RNNs use ODE to represent and capture sequences in continuous time. This makes them particularly adept at handling irregular time intervals and varying sampling rates, which are common in real-world applications. On the other hand, ODE-LSTM networks integrate continuous-time modeling directly into the LSTM framework. This approach enhances the ability of LSTMs to manage continuous dependencies and dynamics, bridging the gap between discrete time steps and continuous-time processes. ODE-LSTMs offer improved flexibility and adaptability in modeling complex temporal sequences compared to traditional LSTMs. Despite these advantages, both ODE-LSTMs and CT-RNNs face notable challenges. The computational complexity associated with solving ODEs and the continuous-time integration can lead to increased training time and resource consumption. Furthermore, since these networks rely on traditional neurons in the bottom layer, they are vulnerable to issues of interpretability and training instability.
\par
Recently, a novel type of ODE-based neural network, known as liquid neural networks (LNNs), has been developed from first principles to address the above shortcomings \cite{ltc}. Unlike traditional AI models, LNNs are grounded in first principles, which involve deriving properties and behaviors from fundamental natural laws to ensure that the design is both innovative and foundational. Inspired by the adaptive and dynamic nature of biological neural systems, LNNs emulate information transmission mechanisms observed in the nematode Caenorhabditis elegans. As illustrated in Fig. \ref{lnn_neuron}, LNNs utilize a nonlinear conductance-based synapse model where stimulation flows from a presynaptic neuron to a postsynaptic neuron. This interaction is described by a first-order ODE, forming the basis of the liquid time-constant neural network (LTC) \cite{ltc}. By approximating the LTC solution with a closed-form expression that uses a few parameters, the closed-form continuous-time neural network (CfC) \cite{cfc} is derived, which circumvents the high overhead of traditional ODE solvers. Multiple LTC or CfC can be stacked to form neural circuit policy (NCP), which has stronger expressive power \cite{ncps}. This approach enables LNNs to emulate the flexibility and resilience of natural neural networks. Unlike static architectures, LNNs can continuously adapt and reorganize in response to new inputs, maintaining high performance and robustness in dynamic and unpredictable environments. This adaptability makes LNNs particularly well-suited for real-world applications where conditions constantly change, such as beam management in dynamic and noisy wireless environments\cite{ncps_beamforming, zhu2024robustcontinuoustimebeamtracking, zhu2025liquidneuralnetworksnextgeneration}. Furthermore, LNNs are able to decompose complex neural dynamics into interpretable and manageable behavioral patterns. By utilizing techniques such as decision trees to analyze neural strategies, LNNs not only provide clear and logical explanations for decision-making processes but also enhance resilience of the system to disturbances, thereby further improving overall robustness \cite{Interpretability}.

\paragraph{Applications of PINNs in wireless}
PINNs are increasingly adopted in intelligent wireless communications to enhance system performance. For instance, PINNs have been instrumental in developing intelligent beamforming schemes. In \cite{li2022intelligent}, researchers introduced a PINN based on the discrete dipole approximation method, for code-to-pattern mapping. Complementarily, a deep neural network trained under PINN guidance is used for pattern-to-code mapping. This integration results in a joint scheduling framework that effectively combines horizontal task distribution and vertical computational enhancement to improve overall system performance. Beyond beamforming, PINNs have also demonstrated successful application in channel prediction. The SCGnet scheme, presented in \cite{10351043}, exemplifies this by innovatively modeling channel prediction as an ODE problem grounded in electromagnetic wave propagation physics and inspired by Neural ODEs. This data-efficient approach requires only historical data for training and minimal fresh measurements for prediction, showcasing superior performance in mobile channel representation, learning, and prediction. Furthermore, radiomap estimation benefits from PINN methodologies, as demonstrated by the physics-informed machine learning methods proposed in \cite{10634040} for high-resolution radiomap reconstruction from sparse samples. Their findings underscore the promising synergy between data-driven AI and model-based radio propagation understanding for this task.

\paragraph{Benefits of PINNs}
PINNs offer a compelling set of advantages that are particularly well-suited to address the complexities and dynamic nature of modern wireless communication systems. These benefits can be broadly categorized into areas that directly enhance system responsiveness, adaptability, and understanding. 
\par
Firstly, PINNs can enhance real-time dynamic learning capability of AI models. Wireless environments are inherently dynamic, characterized by time-varying channels due to user mobility, interference fluctuations, and environmental changes. Traditional machine learning models often require extensive offline training and struggle to adapt quickly to these real-time variations. In contrast, PINNs, by embedding fundamental physical principles, possess a unique capability for real-time dynamic learning. They can leverage incoming data to continuously refine their understanding of the underlying physics governing the wireless channel or system behavior. 
\par
Secondly, PINNs can bring enhanced explainability. Traditional dense neural networks, while powerful, are often criticized for their "black-box" nature, making it difficult to understand their decision-making processes. This lack of explainability can be a barrier to trust and deployment, especially in critical communication infrastructure. PINNs, by incorporating known physical laws as constraints or guiding principles within their architecture, inherently offer enhanced explainability. The learned network is not solely driven by data patterns but also by the imposed physical relationships. This allows researchers and engineers to gain insights into why a PINN is making specific predictions or decisions. By examining the learned parameters and how they interact with the embedded physical models, we can better understand the underlying wireless phenomena being captured by the network. This improved interpretability facilitates debugging, validation, and ultimately, greater confidence in the reliability and robustness of PINN-driven wireless communication systems. 
\par
Lastly, PINNs provide stronger flexible adaptability and generalization. Wireless communication systems are deployed in diverse environments, ranging from dense urban settings to rural areas, and operate across various frequency bands and under differing regulatory constraints. Traditional data-driven models may struggle to generalize effectively across such diverse scenarios, often requiring extensive data collection and retraining for each new deployment environment. PINNs, grounded in fundamental physics, exhibit improved flexible adaptability and generalization capabilities. The embedded physical principles provide a robust inductive bias, allowing PINNs to learn more efficiently from limited data and extrapolate more reliably to unseen conditions or environments.

\paragraph{Integration mechanism with WLAMs}
The technical integration of PINN principles into a WLAM architecture is primarily achieved through the modification of the training objective via a hybrid loss function. A conventional WLAM, such as a large Transformer network for channel prediction, is trained to minimize a data-driven loss, typically the MSE between its predictions and the ground-truth data $\mathcal L_{data}$. A PINN-enhanced WLAM introduces a second component to this objective: a physics-residual loss $\mathcal L_{phys}$.
\par
This $\mathcal L_{phys}$ term is derived from the governing physical laws of the wireless environment, often expressed as partial differential equations (PDEs) like Maxwell's equations. The output of WLAM (e.g., the predicted CSI) is treated as a function and substituted into the PDE. The residual of this equation forms the physics-based loss, and the total loss function for training the WLAM becomes a weighted sum: $\mathcal L_{total} = \lambda \mathcal L_{data} + (1 - \lambda) \mathcal L_{phys}$. By minimizing this combined loss, the WLAM is forced to find a representation that is not only accurate with respect to the training samples but is also consistent with the fundamental principles of wave propagation. This mechanism is the root of its enhanced data efficiency and generalization capabilities, as the physical loss acts as a powerful regularizer, guiding the model towards a valid solution even with sparse data.

\paragraph{Future prospects}
Looking ahead, the integration of PINNs with the burgeoning field of large AI models holds immense promise for revolutionizing wireless communication. As wireless systems become increasingly complex, demanding ultra-high performance and adaptability in dynamic environments, the limitations of purely data-driven large models in terms of generalization, explainability, and sample efficiency become more pronounced. PINNs offer a compelling solution by injecting fundamental physical principles into these large models, potentially leading to a new generation of wireless AI that is not only powerful but also interpretable and robust. It is anticipated that future 6G networks leveraging PINN-enhanced large models for tasks like massive MIMO beamforming, intelligent spectrum management, and network optimization, would achieve unprecedented levels of efficiency and reliability. By grounding large models in the known physics of wireless propagation and system behavior, we can construct more trustworthy, resource-efficient, and ultimately, more capable wireless networks that can seamlessly navigate the complexities of future communication landscapes.

\begin{figure}[t]
	\begin{center}
		\centerline{\includegraphics[width=0.5\linewidth]{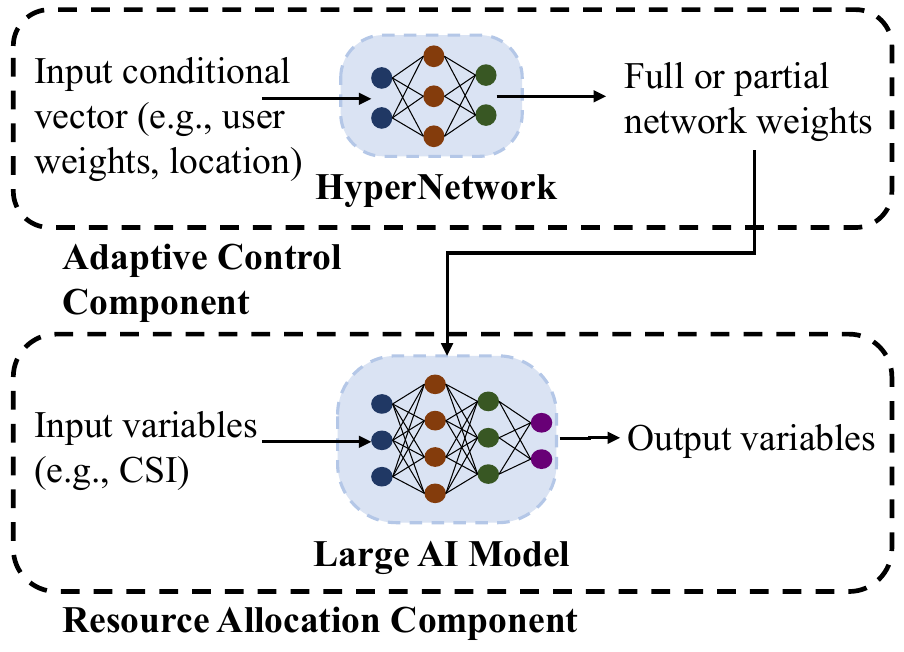}} 
		\vspace{-0mm}
		\captionsetup{font=footnotesize, name={Fig.}, labelsep=period} 
		\caption{\, A resource allocation framework with Hyper-network. A lightweight Hyper-network dynamically generates task-specific weights for the main Large AI Model based on contextual inputs. This decoupled architecture enables rapid, parameter-efficient adaptation to dynamic wireless environments without expensive model retraining.}
		\label{hypernetwork_beamforming}
		\vspace{-8mm}
	\end{center}
\end{figure}

\subsubsection{\textbf{Hyper-networks}}
To address the diverse needs of various applications and users, Hyper-networks (HNs) represent a novel neural network architecture designed to dynamically generate the weights of another network \cite{ha2017hypernetworks}. Traditional neural networks require re-training or fine-tuning to accommodate specific user needs, which is both expensive and time-consuming. HNs overcome this limitation by producing whole or partial weights based on input conditions, thus enabling adaptive and flexible model behavior without the need for re-training or fine-tuning. This dynamic generation process allows for rapid adjustments, making HNs highly suitable for complex and versatile environments. In the following we delve into the applications, benefits, and recent developments of HNs in large AI models for wireless communication.

\paragraph{Applications and benefits of Hyper-networks}
HNs provide substantial benefits for large AI models in wireless communication through dynamic resource allocation. For instance, they can generate adaptive beamforming vectors and configurations based on user weights in reconfigurable intelligent surfaces-assisted systems \cite{10238370}, thus enhancing efficiency without retraining \cite{abouamer2023flexible}. Additionally, they improve non-stationary channel prediction by continuously updating neural network parameters to adapt to changing conditions, which increases CSI prediction accuracy and spectral efficiency \cite{liu2024hypernetwork}. Combined with RNNs, HNs leverage uplink-downlink reciprocity to enhance channel estimation and beamforming performance in frequency division duplex massive MIMO systems \cite{liu2022learning}. This synergy facilitates high-performance processing in environments with limited hardware capabilities, optimizing resource allocation and operational efficiency in wireless applications. HNs reduce the need for extensive retraining by dynamically generating network weights, making them particularly well-suited for rapidly changing environments and diverse user requirements. Their integration with large AI models supports scalable and efficient solutions for modern wireless networks.

\paragraph{Integration mechanism with WLAMs}
The integration of HNs with WLAMs is achieved through a dynamic, two-level architectural design. In this framework, the massive WLAM serves as the `main network' for executing the primary task (e.g., resource allocation), while a significantly smaller HN acts as a `controller network', as shown in Fig. \ref{hypernetwork_beamforming}. The key mechanism is that the HN does not process the primary task data (e.g., CSI); instead, it takes contextual metadata (such as user QoS profiles, device identities, or specific task instructions) as input.
\par
The output of the HN is not a prediction, but rather the weights for specific, adaptable parts of the main WLAM. Most effectively, an HN can be trained to generate the low-rank matrices for PEFT modules, such as LoRA adapters within the WLAM. This creates a highly efficient workflow: when the context changes (e.g., a new user with different requirements connects), only the small HN needs to perform a forward pass to generate a new, customized set of adapter weights. These weights are then injected into the frozen WLAM, instantly tailoring its behavior without the need for any backpropagation through the multi-billion parameter model. This mechanism provides a concrete pathway for achieving rapid and low-cost personalization and task adaptation for WLAMs.

\paragraph{Recent developments in Hyper-networks}
Recent developments in the application of HNs in large AI models showcase innovative approaches to enhancing model efficiency and performance. \cite{ivison2022hint} introduces a novel approach that uses HNs to transform task instructions into parameter-efficient modules, significantly reducing computational costs while improving performance by up to 25\% compared to state-of-the-art methods. Concurrently, \cite{tang2024modelgpt} offers a new framework for generating large AI models tailored to specific tasks or data descriptions, achieving up to 270x speed improvements over traditional fine-tuning approaches. These developments not only highlight the potential for more efficient and effective model inference but also pave the way for future innovations in wireless communication. The integration of HNs and user-customized models, as exemplified above, promises to enhance the adaptability, scalability, and performance of large AI models. By leveraging these advancements, future wireless AI systems can achieve more tailored solutions for diverse communication needs, ultimately advancing the field towards more intelligent and resource-efficient technologies.

\subsubsection{\textbf{Next-generation sequence modeling networks}}

The emergence of 6G communication systems demands advanced AI models capable of processing large-scale sequential data with high efficiency. While Transformer-based architectures have demonstrated high performance, their quadratic complexity with respect to sequence length poses scalability challenges. To address this, next-generation sequence modeling networks, including Mamba \cite{gu2024mamba}, receptance weighted key value (RWKV) \cite{peng-etal-2023-rwkv}, and test-time training (TTT) \cite{sun2024learninglearntesttime}, have been developed, offering improved computational efficiency and adaptability. This subsection examines these architectures, detailing their mechanisms, advantages, and potential applications in large AI model empowered 6G networks.

\begin{figure}[t]
	\begin{center}
		\centerline{\includegraphics[width=0.5\linewidth]{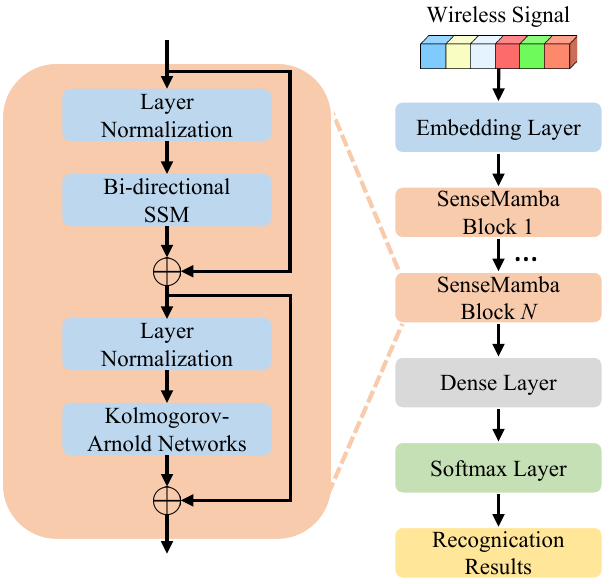}} 
		\vspace{-0mm}
		\captionsetup{font=footnotesize, name={Fig.}, labelsep=period} 
		\caption{\, SenseMamba architecture for wireless human sensing. Integrating Bi-directional State Space Models and Kolmogorov-Arnold Networks, this lightweight architecture efficiently processes long-sequence wireless signals. It achieves linear complexity, enabling real-time, low-overhead human recognition while overcoming the quadratic bottlenecks of traditional Transformers.}
		\label{sensemamba}
		\vspace{-8mm}
	\end{center}
\end{figure}

\paragraph{Mamba}
Mamba is a state-of-the-art sequence modeling architecture rooted in structured state space sequence models. Unlike Transformers, Mamba achieves linear scaling with sequence length through its selective state space model (SSM) mechanism, which dynamically adjusts parameters based on input data. This adaptability enables efficient processing of long and irregularly sampled sequences, overcoming the memory and computational limitations of traditional models.
\par
In 6G systems, Mamba stands out as a powerful solution for reducing the computational overhead of large AI models in communication systems. Its linear complexity allows it to efficiently process vast sequential data, a critical advantage for large models handling extensive datasets without incurring excessive resource costs. This efficiency is particularly valuable for tasks like dynamic channel estimation and adaptive beamforming. Its real-time processing capabilities enable rapid adaptation to changing conditions, which is essential for high-performance, low-latency communication. Besides, Mamba has demonstrated effectiveness in related areas including wireless sensing. A concrete example is the SenseMamba architecture \cite{sensemamba}, illustrated in Fig. \ref{sensemamba}, which applies selective SSM principles within specifically designed lightweight blocks to achieve efficient human sensing using wireless signals. This demonstrates its potential not only for core communication tasks like channel estimation but also for enabling sophisticated, low-overhead sensing applications crucial for future intelligent environments. Its real-time processing capabilities and ability to manage complex temporal dependencies make it particularly suitable for these dynamic scenarios. Furthermore, Mamba is especially beneficial for large AI models for its ability to manage infinite contexts without overwhelming resource demands. Large models often require deep, context-rich understanding to perform effectively, and Mamba delivers this capability seamlessly, decreasing the overhead that typically burdens such systems. 

\paragraph{RWKV}
RWKV stands out as a transformative architecture for large AI models by seamlessly blending the strengths of RNNs and Transformers, particularly in the context of 6G systems. Unlike traditional architectures, RWKV delivers Transformer-level performance with RNN-like efficiency, thanks to its linear time complexity and constant memory usage. This design drastically reduces the computational burden that large models often face, making it ideal for processing extensive sequential data. For 6G applications such as real-time network optimization and time-series prediction, its ability to perform rapid inference with low overhead is a game-changer, enabling scalable and efficient solutions for intelligent wireless systems.
\par
RWKV differs from Mamba with its unique approach. While Mamba relies on SSMs to achieve linear scaling, RWKV leverages a hybrid RNN-Transformer framework with its specialized RWKV layer. This layer allows for parallel training while maintaining sequential processing, offering a distinct edge over both traditional RNNs and Mamba. As a result, RWKV excels at handling vast contextual data, positioning it as a powerful tool for large-scale AI deployments in 6G networks, where agility and precision are paramount.

\paragraph{TTT}
TTT introduces an innovative approach to sequence modeling by encoding data directly into model weights through a process called test-time training. Unlike Transformers, which depend on hidden states to manage sequential data, TTT trains on the input data during inference. This mechanism allows the model to adapt its weights dynamically to new, unseen data while keeping computational demands constant, regardless of the volume of data processed. As a result, TTT avoids the need for larger model sizes or additional computational resources as datasets expand, offering a significant efficiency advantage.
\par
In 6G wireless systems, this efficiency proves particularly valuable for large AI models tasked with handling extensive, multi-modal datasets such as sensor inputs and user activity logs. The ability of TTT to integrate and process data in real time without escalating costs supports adaptive learning applications like anomaly detection, predictive maintenance, and personalized user experience optimization. By maintaining low computational overhead while managing massive datasets, TTT aligns with the resource-constrained requirements of future wireless networks. This makes it an ideal solution for scalable, intelligent communication systems relying on large AI models in 6G.

\paragraph{Integration mechanism with WLAMs}
For next-gen sequence models, the integration with WLAMs is a direct architectural substitution at the core building-block level. Many proposed WLAMs for sequential wireless tasks (e.g., time-series channel prediction, traffic forecasting) are based on the powerful but computationally expensive Transformer architecture. The integration mechanism involves replacing the self-attention blocks of Transformer, which are the source of its $O(n^2)$ complexity, with the more efficient linear-time blocks from these next-gen models.
\par
For example, in a Mamba-based WLAM, the self-attention layer is swapped with a SSM block. This fundamentally changes how the model processes sequential information. Instead of comparing every element of the sequence with every other element, the SSM maintains a compact hidden state that is updated recurrently, allowing for $O(n)$ complexity. This architectural change enables a WLAM to process significantly longer sequences of wireless data (e.g., capturing channel coherence over extended periods) in real-time. This is not merely an efficiency gain; it allows the WLAM to learn long-range temporal dependencies in the wireless channel that are computationally intractable for a standard Transformer, leading to potentially more accurate and robust predictions.

\paragraph{Future prospects}
The advancement of Mamba, RWKV, and TTT marks a paradigm shift towards efficient and scalable sequence modeling, addressing the unique challenges of 6G communication systems. These architectures hold promise for critical tasks such as ultra-reliable URLLC and mMTC. Future research may focus on optimizing their mechanisms for specific wireless applications, such as leveraging the selective processing Mamba for resource allocation or the hybrid design RWKV for signal processing.

\subsection{Tensions and synergies}
While the aforementioned technologies offer individual advantages, their practical integration into 6G systems involves complex inter-dependencies. The following analysis reveals both inherent tensions and potential synergies that define the future architectural roadmap.

\subsubsection{\textbf{Tensions: hardware maturity and algorithmic efficiency}}
A fundamental tension exists between emerging computing paradigms and neural architectures regarding deployment readiness. While computing paradigms like quantum computing and HDC offer theoretical energy efficiency superior to classical AI, they suffer from low hardware maturity and a lack of standardized toolchains. In contrast, emerging neural architectures such as Mamba and TTT are designed to run efficiently on existing GPUs but hit the ``memory wall" on resource-limited edge devices. Consequently, a critical trade-off arises between immediate compatibility and hardware maturity. System designers must balance the deployment of algorithmic innovations that run on existing processors against the pursuit of disruptive computing paradigms, which offer theoretical superiority but currently lack the necessary specialized hardware infrastructure.

\subsubsection{\textbf{Synergies: physics-informed efficiency and dynamic adaptation}}
Conversely, significant synergies emerge when combining these technologies. First, the integration of PINNs with next-generation sequence models creates a powerful duality. While Mamba and RWKV excel at processing long-sequence data with linear complexity, they often lack physical interpretability. By embedding PINN-derived physical constraints into the state-space transitions of Mamba, one can construct ``Physics-Informed State Space Models." These hybrid architectures may achieve high-precision channel prediction with significantly fewer training samples by leveraging physical laws to guide the long-context extrapolation. Second, Hyper-networks offer a synergistic solution to large foundation models. Instead of retraining a massive WLAM for every new user scenario, a lightweight Hyper-network can dynamically generate task-specific adapters for the frozen foundation model. This combination is promising to resolve the conflict between the need for generalist knowledge and the requirement for rapid, user-specific personalization in dynamic wireless environments.

\subsection{Summary and insights}
Emerging technologies are revolutionizing large AI models in wireless communications. These technologies enhance model privacy, efficiency, adaptability, and scalability, addressing critical challenges in dynamic and complex wireless environments. By integrating these advanced methods, future wireless AI systems will achieve higher performance, better resource utilization, and improved user-specific customization, paving the way for more intelligent, efficient, and secure wireless networks.

\section{High-level challenges and future directions}

\begin{table*}[t]
\scriptsize
\centering
\caption{A strategic roadmap: fundamental WLAM bottlenecks to AI-empowered 6G directions.}
\label{tab:vision_roadmap}
\begin{tabular}{|m{0.15\textwidth}<{\raggedright}|m{0.4\textwidth}<{\raggedright}|m{0.38\textwidth}<{\raggedright}|}
\hline

\textbf{Domain} & 
\textbf{Fundamental Bottlenecks} & 
\textbf{The AI-empowered 6G Roadmap} \\ \hline

\textbf{Datasets \& Modality} \newline \textit{(Sec. \ref{sec:highlever_dataset})} & 
\textbf{Heterogeneity \& Quality:} Static benchmarks fail to capture dynamic network realities; severe challenges in accurately labeling adequate noisy real-world data; temporal asynchrony and integration issues in massive multi-modal streams. & 
Synthetic data generation via digital twins, cross-modal soft-alignment and pre-training, active sample selection, and continual learning paradigms to avoid catastrophic forgetting. \\ \hline

\textbf{Architectures \& Protocols} \newline \textit{(Sec. \ref{sec:future_protocols})} & 
\textbf{Scalability \& Standardization:} Rigidity in managing massive nodes across dynamic conditions, compounded by heterogeneous networks lacking unified standards to exchange semantic features, model gradients, and agentic intents. & 
Leveraging graph-based frameworks for dynamic resource allocation, and pioneering self-evolving topologies built on standardized AI-empowered interfaces. \\ \hline

\textbf{Computation, Latency \& Energy} \newline \textit{(Sec. \ref{sec:futures_ee})} &
\textbf{The Memory Wall and Latency Overheads:} Energy and delay bottlenecks driven by memory bandwidth saturation during auto-regressive decoding, compounding KV cache expansion, and thermal throttling on constrained edge devices. &
Accelerating inference via speculative decoding and prompt compression, alongside a paradigm shift to deploy WLAMs as network-wide orchestrators that trade computational loads for massive RF energy savings. \\ \hline

\textbf{Security \& Privacy} \newline \textit{(Sec. \ref{sec:futures_sec})} & 
\textbf{Vulnerabilities and Leakage:} Severe susceptibility to AI-centric attacks and inference leakage, coupled with data sovereignty challenges in cloud deployments and the fundamentally opaque nature of massive models. & 
Establishing comprehensive security guardrails via formal verification for critical infrastructure, and advancing privacy-preserving inference leveraging trusted execution environments and homomorphic encryption. \\ \hline

\textbf{Physical Channel Impact} \newline \textit{(Sec. \ref{sec:futures_channel})} & 
\textbf{Channel Impairments:} The inherent noise, fading, limited bandwidth, and dynamics of wireless mediums fundamentally disrupt WLAMs, causing non-differentiable variables, train-test mismatches, and vulnerabilities to physical errors and jamming. & 
Integrating differentiable surrogate models, physical grounding, and meta-learning, alongside communication-learning protocols that are robust to physical impairments and adversarial jamming. \\ \hline

\textbf{Fundamental Trade-offs} \newline \textit{(Sec. \ref{sec:futures_tradeoff})} & 
\textbf{Systemic Conflicts:} Balancing between performance and efficiency, the conflict between data utility and privacy, and the tension between autonomy and reliability. & 
Integrating hardware software design, deploying model-centric defenses, and ensuring reliability through explainable architectures, formal verification, and human-in-the-loop mechanisms. \\ \hline

\end{tabular}
\color{black}
\end{table*}

The preceding sections have charted the transformative potential of WLAMs across the communication stack, detailed the frameworks for their distributed deployment, and explored the frontiers of emerging technologies. However, translating these theoretical advancements into a robust, AI-empowered 6G reality requires overcoming fundamental systemic barriers that persist despite these innovations. In exploring the high-level challenges and future directions for WLAMs, this section identifies and highlights key areas for advancement, summarized in Table \ref{tab:vision_roadmap}. We begin by examining the foundational challenges related to datasets, telecom network architectures, computational efficiency, and security and privacy. Following this, the analysis delves into the impact of wireless channels on model training. These diverse challenges are then synthesized into a discussion of the fundamental trade-offs that govern the WLAMs design direction. Finally, the section concludes with a summary of key insights.

\begin{table*}[!t]
\scriptsize\centering
\caption{Summary of related works on high-level challenges and future directions.}

\label{tab:challenges}
\begin{tabular}{|m{0.12\textwidth}<{\centering}|m{0.05\textwidth}<{\centering}|m{0.35\textwidth}<{\raggedright}|m{0.35\textwidth}<{\raggedright}|}
\hline
\textbf{Challenges} &
\textbf{Ref.} &
\textbf{Scenarios} &
\textbf{Contributions} \\ \hline

\multirow{4}{*}[-3em]{\centering\parbox{0.09\textwidth}{\centering Dataset}} 
  & \cite{maatouk2023teleqnabenchmarkdatasetassess} 
  & Benchmark the telecommunications knowledge of large language models. 
  & Present the first benchmark dataset designed to evaluate the knowledge of large language models in telecommunications. 
  \\ \cline{2-4} 

  & \cite{deepmimo} 
  & Machine learning for wireless communications with mmWave/massive MIMO channels.
  & Present a ray-tracing wireless channel dataset for machine learning, which is based on virtual maps.
  \\ \cline{2-4} 
  
  & \cite{waird} 
  & Sparse and dense wireless communication scenarios in real-world environments.
  & Present a ray-tracing wireless channel dataset for machine learning, which is based on real maps. 
  \\ \cline{2-4} 
  
  & \cite{Raymobtime} 
  & 3D wireless communication scenarios with mobility and time evolution.
  & Present a multi-modal ray-tracing wireless channel dataset for machine learning for mobility simulation.
  \\ \hline

\multirow{3}{*}[-1em]{\centering\parbox{0.095\textwidth}{\centering Telecom Network Architectures and Protocols}} 
  & \cite{he2021overview}  
  &  Wireless networks with graph structure data.          
  & Introduce the applications of graph neural networks in wireless networks.
  \\ \cline{2-4} 

  & \cite{wang2022learning}  
  & Resource allocation in decentralized wireless networks.               
  & Present a decentralized scheme to allocate resource with graph neural networks.    
  \\ \cline{2-4}
  
  & \cite{kharche2022interoperability} 
  &  Data exchange among the heterogeneous networks in the incoming 6G networks.  
  &  Propose a taxonomy to analyze and solve interoperability challenges in 6G networks for seamless global connectivity. 
  \\ \hline

\multirow{3}{*}[-2em]{\centering\parbox{0.08\textwidth}{\centering Computa-tional Capability and Energy Efficiency}} 
  & \cite{wan2024efficient}  
  & Improve resource management in natural language processing tasks, such as understanding and generation.           
  & Discuss efficient large language models research and offer organized repository to navigate and contribute to this evolving field.
  \\ \cline{2-4} 
  
  & \cite{mu2024learning}            
  & Improve efficiency in language models by compressing prompts to save input space and computation.    
  & Propose a method to train language models to compress prompts into gist tokens, with lower FLOPs and minimal output quality 
    loss.
  \\ \cline{2-4}

  & \cite{razavi2024talescalesreconcilinghorizontal} 
  & Managing resources for model inference serving under variable workloads to maintain performance and efficiency. 
  & Propose a system that reconciles horizontal and vertical scaling by using a two-stage auto-scaling strategy. 
  \\ \hline

\multirow{3}{*}[-2em]{\centering\parbox{0.08\textwidth}{\centering Security and Privacy Issues}} 
  & \cite{yao2024survey}              
  & The deployment of large language models in security and privacy sensitive scenarios.             
  & Discuss the benefits, vulnerabilities and defenses of large language models.
  \\ \cline{2-4}
  
  & \cite{wan2023poisoning}              
  & Data poisoning attack which involves injecting malicious data into the training dataset.              
  & Demonstrate how the strengths of large language models can be exploited as vulnerabilities through poisoning attacks.
  \\ \cline{2-4}

  & \cite{brown2022does} 
  & The training and fine-tuning processes where sensitive data may be leaked.  
  & Discuss the vulnerabilities of the current data privacy protection methods. 
  \\ \hline

  \multirow{3}{*}[-2em]{\centering\parbox{0.1\textwidth}{\centering Channel Impact on Model Training}}
  & \cite{10351043}
  & Accurate channel prediction in highly dynamic and mobile wireless environments.
  & Propose a physics-inspired scheme based on ODE to model and predict the continuous-time evolution of mobile channels.
  \\ \cline{2-4}
  
  & \cite{zhurobustnetwork}
  & Robust communication in the presence of perturbations like channel estimation errors.
  & Propose a model-driven meta-learning algorithm that optimizes the training process itself to find robust solutions.
  \\ \cline{2-4}
  
  & \cite{federatedPHY}
  & Federated training of a LLM over a wireless channel under an adversarial jamming attack.
  & Establishes a direct mathematical link between the  communication MSE of the physical layer and the training loss.
  \\ \hline
   
\end{tabular}
\end{table*}

\subsection{Dataset}\label{sec:highlever_dataset}
Datasets are critical to the development and success of large AI models, particularly in the field of wireless communication where the channels are highly dynamic \cite{waird, deepmimo}. The challenges of datasets in this domain present several opportunities for further advancement.

\subsubsection{\textbf{Data complexity and diversity}}
Wireless communication data is inherently diverse and complex, characterized by a wide range of communication protocols, frequency bands, device types, and usage scenarios \cite{huang2017wireless}. This inherent diversity necessitates AI models that are not only highly adaptable but also capable of processing heterogeneous data sources effectively. In contrast to traditional large models designed for general-purpose applications, which typically rely on more straightforward and less structured datasets, the development of models for telecommunications demands a substantial amount of domain-specific data with complex structures to achieve meaningful results.
\par
To address this challenge, \cite{maatouk2023teleqnabenchmarkdatasetassess} introduced TeleQnA, the first benchmark dataset specifically crafted to evaluate the knowledge of LLMs in the field of telecommunications. TeleQnA consists of 10,000 question-and-answer pairs derived from standards and research articles, developed through an automated question generation framework complemented by human input for quality assurance. Evaluations of GPT-3.5 and GPT-4 using TeleQnA reveal that while these models perform well on general Telecom queries, they encounter difficulties with more complex, standards-related questions. To further assess the capabilities of LLMs in core mathematical modeling and computational tasks specific to wireless communications, \cite{li2025wirelessmathbench} introduced WirelessMathBench. This benchmark addresses the limitations of existing datasets by focusing on mathematical reasoning rather than solely on textual knowledge. However, these static benchmarks fail to capture the dynamic, time-varying nature of real-world wireless networks, where real-time data processing is critical. This underscores the urgent need for constructing comprehensive, multi-modal datasets that go beyond textual knowledge to include raw signals and diverse protocol interactions. Developing such standardized data foundations is a prerequisite for training WLAMs that can generalize across disparate communication protocols (e.g., WiFi, 5G, satellite). Only with such high-quality heterogeneous data can models overcome the current limitations in reasoning and real-time processing.

\subsubsection{\textbf{Volume and scalability}}
The sheer volume of data generated by wireless communication systems presents significant challenges for data storage and processing. This data, including user activity logs, sensor readings, network traffic, and communication logs, is produced continuously and at an increasing rate. Large-scale AI models require vast amounts of training data to achieve high accuracy and robustness, which necessitates effective storage and processing solutions.
\par
To manage this massive influx of data, scalable data management solutions are essential. Distributed storage systems and cloud-based solutions, such as Hadoop distributed file system \cite{shvachko2010hadoop} and cloud storage services, provide the necessary infrastructure for storing and accessing large datasets efficiently. Additionally, advanced data processing frameworks like Apache Hadoop \cite{nandimath2013big} and Apache Spark \cite{meng2016mllib} enable the parallel processing of large-scale data across multiple nodes.
\par
Scalability also involves adaptation to the growing diversity and volume of data generated by a increasing number of connected devices. Technologies such as distributed computing \cite{kshemkalyani2011distributed} and elastic cloud storage solutions are crucial for handling this growth and ensuring that AI models can learn from expanding data sources. Techniques such as data parallelism and mini-batch processing are used to train models efficiently as data volumes increase \cite{bengio2012practical}.
To cope with the exponential growth of data volume, future research should investigate active learning strategies to intelligently select the most informative samples for training, and continual learning paradigms that allow WLAMs to update their knowledge from streaming data without incurring catastrophic forgetting.

\subsubsection{\textbf{Data quality and labeling}}
High-quality data is essential for training effective AI models, especially in the complex field of wireless communication. Datasets such as DeepMIMO \cite{deepmimo} and WAIR-D \cite{waird} have been instrumental in providing large-scale, simulated wireless environments for generating comprehensive training and testing data. By offering rich and realistic simulation environments, these datasets enable the development of various AI applications, including beamforming \cite{zhu2023robust, wgan, wang2023energyefficient, gmml}, localization \cite{wbhVTC}, and other advanced signal processing techniques.
\par
However, despite the availability of these high-quality datasets, practical wireless communication data often presents a range of challenges that can affect model performance. Real-world data is frequently plagued by issues such as noise, missing values, and outliers. These imperfections can significantly degrade the accuracy and reliability of AI models. 
\par
To address these challenges, robust data preprocessing techniques are required. This includes methods for data cleaning, such as filtering out noise, imputing missing values, and detecting and correcting outliers. Techniques like data normalization, standardization, and noise reduction are essential for preparing data that can be effectively used for model training. It is also essential to develop a standardized data format and quality control process for pooling data from different providers into a unified dataset \cite{rappaport2025point,shakya2025standardized}. For instance, methods such as median filtering and statistical outlier detection can help enhance data quality and improve model performance.
\par
Additionally, accurate labeling of data is a crucial but challenging task in wireless communication. The dynamic and often unstructured nature of communication data makes it difficult to consistently label data samples. Automated labeling techniques, which leverage machine learning algorithms, offer a potential solution for this issue. These techniques can include supervised learning methods where labeled data is used to train models that can then predict labels for new data, and semi-supervised or unsupervised methods that help in identifying patterns and classifying data without extensive human intervention \cite{HaoHuangPIER, LuZhuangPIER}. However, to fundamentally resolve the bottleneck of obtaining massive labeled datasets from the physical world, future research should further explore synthetic data generation via digital twins \cite{khan2022digital, bariah2023digital}. By creating high-fidelity virtual replicas of wireless environments, researchers can generate vast amounts of perfectly labeled training data covering rare and extreme scenarios that are difficult to capture in reality.
\par

\subsubsection{\textbf{Multi-modal dataset}}
Multi-modal datasets integrate diverse data types such as CSI, light detection and ranging, camera sensor data, and urban mobility patterns to provide a comprehensive representation of wireless communication environments. These datasets are vital for training large AI models in 6G systems, enabling them to adapt to dynamic, heterogeneous conditions by capturing complex interactions between physical, environmental, and network factors.
\par
A notable example is the Raymobtime project \cite{Raymobtime}, which provides multi-modal datasets tailored for telecommunications. It combines ray-tracing simulations from Wireless Insite, mobility patterns from SUMO, and visual data processed via Blender. Datasets like s007 (Beijing, 2.8/60 GHz) and s008/s009 (Rosslyn, 60 GHz) offer realistic urban scenarios, supporting applications such as beamforming and localization.
\par
However, multi-modal datasets still pose significant challenges. Data integration and synchronization are difficult due to varying formats and sampling rates; for instance, aligning ray-tracing simulations with real-time sensor data requires precise preprocessing \cite{ying2025sitespecificlocationcalibrationvalidation}. The volume and diversity of data also strain storage and processing capabilities, necessitating scalable solutions. Moreover, ensuring data quality and consistent labeling across modalities is complex, as errors in one modality can compromise overall model accuracy.
\par
Future advancements in multi-modal datasets for WLAMs will center on three pivotal enhancements. First, addressing the challenge of temporal asynchrony between RF and visual streams requires researching soft-alignment techniques rather than rigid frame-by-frame synchronization. Second, building robust real-time processing frameworks powered by edge computing will tackle the unprecedented data throughput of 6G networks. Finally, advancing cross-modal learning algorithms is essential; specifically, developing cross-modal pre-training technologies that explicitly encourage the model to learn the semantic correlations between physical signals and environmental context represents a pivotal pathway.

\subsection{Telecom network architectures and protocols} \label{sec:future_protocols}
The integration of large AI models into wireless communication systems necessitates innovative network architectures and protocols to ensure efficient, reliable, scalable, and flexible operation. We present several key challenging areas as follows.

\subsubsection{\textbf{Scalability and flexibility}}
As wireless communication networks become increasingly complex and diverse, constructing scalable and flexible network architectures to efficiently allocate resources becomes inevitable \cite{chi2023mimo, chi2023resource}. Scalability refers to the ability to handle a growing number of nodes, while flexibility denotes the adaptability to varying conditions and requirements. These aspects are crucial for the seamless integration of AI-driven solutions in diverse and dynamic wireless environments. To efficiently enhance the scalability of communication networks with large AI models, embedding nodes in a graph and applying graph-specific techniques, such as graph neural networks \cite{he2021overview}, is a promising future research direction. These techniques can efficiently perform resource allocations while considering the interactions between nodes \cite{wang2022learning}. Additionally, improving flexibility involves designing adaptable large AI models that can dynamically adjust to changing network conditions. Looking forward, a critical research pathway is the evolution towards intent-driven networking architectures. Future work should investigate how WLAMs can serve as the architectural core to autonomously translate high-level service intents into low-level network configurations, enabling a self-evolving topology that scales elastically with user demand.

\subsubsection{\textbf{Interoperability and standardization}}
As wireless heterogeneous networks become more prevalent in wireless communication, ensuring interoperability between different systems and devices is crucial \cite{kharche2022interoperability, yao2024semantic}. These networks often involve multiple wireless communication systems using different access technologies or the same wireless access technology but belonging to different wireless carriers. Standardizing protocols in such heterogeneous networks can facilitate smoother integration and collaboration among various components of the communication infrastructure. Additionally, designing compatible large AI models that can seamlessly integrate with the existing infrastructure is essential for enhancing overall network performance and adaptability. To achieve true seamless integration, future architectural research must focus on defining standardized AI-empowered interfaces. For instance, to facilitate seamless interaction in heterogeneous environments, \cite{li2025llm} proposed the LLM Agent Communication Protocol, emphasizing the urgent need to standardize interaction formats and protocols for diverse LLM agents. Unlike legacy protocols, these new standards must support the efficient and standardized signaling of semantic features, model gradients, and agentic intents across heterogeneous vendor equipment, thereby enabling distributed inference and multi-agent collaboration in a unified 6G ecosystem.

\subsection{Computational capability, energy efficiency, and latency}\label{sec:futures_ee}
The deployment of massive-scale AI models in wireless networks demands unprecedented computational resources, posing severe challenges not only to edge device constraints but also to the overarching sustainability vision of ``Green 6G". To systematically navigate this paradox, this section first provides a granular analysis of the underlying hardware bottlenecks such as the memory wall and thermal limits that impede practical edge deployment. Subsequently, we outline conventional mitigation strategies, ranging from algorithmic model compression to joint horizontal-vertical scheduling. Finally, we elevate the discussion beyond local hardware optimizations to explore a system-level paradigm shift: how AI-empowered 6G architectures can fundamentally resolve this energy dilemma by leveraging WLAMs themselves as network-wide energy orchestrators.

\subsubsection{\textbf{Granular analysis of energy bottlenecks}}
On resource-constrained edge devices, the energy bottleneck exhibits a phase-dependent characteristic. The initial prefill phase is compute-bound, where massive parallel matrix computations for prompt processing result in high instantaneous power consumption dominated by arithmetic logic units. However, as the model enters the auto-regressive decoding phase, the bottleneck shifts to the ``memory wall''. Unlike the compute-bound prefill phase, decoding requires loading the entire model weights from off-chip memory to on-chip processing units for every single token generated. This results in extremely low arithmetic intensity. Since fetching data from off-chip memory consumes orders of magnitude more energy than arithmetic operations, the total energy consumption is dominated by the dynamic power of massive data movement, further exacerbated by the static leakage power from computation units stalling for data. \cite{flashattention, memorywall}. Furthermore, the KV cache, essential for self-attention, grows linearly with sequence length. Although it eliminates redundant computations of historical states, the growing cache must be fetched from memory for every new token generation. In long-context 6G tasks such as continuous channel prediction, this inflating cache saturates memory bandwidth and capacity. Consequently, the energy cost per token escalates dynamically as the inference progresses, rendering standard deployment unsustainable for battery-powered devices \cite{kvcache}. Finally, the high power density of continuous inference often triggers thermal throttling on fanless edge devices, forcing the hardware to operate at suboptimal energy-efficiency points \cite{fromwordstowatts}. To address these specific bottlenecks, diverse strategies ranging from algorithmic compression to system-level scheduling are adopted.

\subsubsection{\textbf{Stringent latency constraints and mitigation}}
Ultra-low latency is widely recognized as a cornerstone of 6G URLLC scenarios. However, integrating WLAMs introduces significant latency overheads compared to traditional lightweight algorithms. To systematically address this issue, the overall latency in networks empowered by WLAMs must be analyzed according to three primary components.
\par
The first component is communication latency. Transmitting high dimensional environmental context to cloud hosted WLAMs and returning decisions incurs severe propagation delays. To mitigate this bottleneck, deploying WLAMs directly at the network edge is beneficial \cite{shen2024large}. Furthermore, leveraging semantic communication can drastically reduce the payload size by transmitting only task relevant features rather than raw data \cite{jiang2024semantic, jiang2024large}, which significantly cuts down the overall transmission time.
\par
The second component involves the prefill latency, commonly known as the time to the first token. During the initial inference phase, the WLAM must process the input prompt, such as historical network logs or specific user intents. This prefill phase is highly computationally intensive. Potential solutions to accelerate this process include prompt compression algorithms \cite{jiang2024longllmlingua}, which distill lengthy network contexts into shorter and densely packed representations. Additionally, prefix caching \cite{pan2025kvflow} allows the model to reuse previously computed hidden states for recurring network optimization tasks, thereby avoiding redundant computations.
\par
The final component is the decoding latency, measured as the time between generated tokens. The autoregressive generation of outputs, including configuration scripts or API calls, is severely constrained by memory bandwidth due to the continuous loading of model weights. Promising solutions to accelerate this phase include speculative decoding \cite{leviathan2023fast}. In this approach, a smaller and faster draft model predicts the subsequent network actions while the large WLAM simultaneously verifies them. Moreover, advanced memory management techniques like PagedAttention \cite{vllm} can be deployed to optimize the KV cache overhead. By systematically mitigating these three latency components, WLAMs can be effectively equipped to meet the strict real time demands of future wireless networks.
\par
Beyond the aforementioned system-level delays, applying WLAMs to URLLC scenarios at the physical layer presents a fundamental feasibility challenge. According to 3GPP specifications, URLLC mandates a stringent user plane latency of 0.5 milliseconds \cite{3gpp.38.913}. This strict budget implies that critical physical layer tasks, such as real time beamforming or channel estimation discussed in Section 3.1, must be executed within a fraction of a millisecond \cite{jiang2025one, jiang2026recursive}. Traditional heuristic algorithms and specialized hardware logic can operate comfortably within these sub-millisecond deadlines. However, they often suffer from limited accuracy and poor generalization in highly dynamic 6G environments. Conversely, deploying standard WLAMs yields exceptional zero-shot accuracy but inherently requires hundreds of milliseconds to seconds for inference due to massive parameter counts and heavy memory bandwidth bottlenecks. Consequently, utilizing WLAMs directly in the physical layer data path for real time URLLC is practically infeasible. This creates a stark latency-accuracy trade-off that must be carefully navigated.
\par
To resolve this extreme trade-off, the application of WLAMs in URLLC must transition from direct execution to an auxiliary and orchestrating role. Specifically, a dual-loop architecture is mandatory for real time physical layer operations. In this paradigm, WLAMs operate entirely in the background edge cloud as the slow loop. They leverage historical data and global context to predict long-term channel statistics, optimize resource allocation policies, or dynamically tune the hyperparameters of traditional mathematical solvers. Meanwhile, the actual physical layer execution is relegated to the fast loop, which consists exclusively of traditional algorithms or ultra-lightweight hardware accelerated logic. By doing so, the fast loop maintains the strict sub-millisecond responsiveness required by URLLC, while the WLAM continuously enhances the baseline accuracy from the background. Alternatively, for scenarios without strict URLLC constraints, such as offline digital twin modeling or delay-tolerant massive connectivity tasks, WLAMs can be employed directly to maximize overall system performance. A detailed comparison of these paradigms regarding their latency and accuracy trade-offs in critical physical layer tasks is summarized in Table \ref{tab:latency_tradeoff}.

\begin{table*}[!t]
\scriptsize\centering
\caption{Latency-accuracy trade-offs and URLLC feasibility for physical layer tasks.}
\label{tab:latency_tradeoff}
\begin{tabular}{|m{0.12\textwidth}<{\centering}|m{0.16\textwidth}<{\centering}|m{0.22\textwidth}<{\raggedright}|m{0.10\textwidth}<{\centering}|m{0.25\textwidth}<{\raggedright}|}
\hline
\textbf{Paradigm} & 
\textbf{Latency Scale} & 
\textbf{Accuracy \& Generalization} & 
\textbf{URLLC Feasibility} & 
\textbf{Role in 6G PHY Layer} \\ \hline

Traditional Algorithms \newline & 
Sub-milliseconds \newline ($10^2~\mu$s scale) & 
Baseline (Poor under non-stationary dynamics) & 
High & 
Direct data path execution for real time processing. \\ \hline

Direct WLAM Inference & 
Milliseconds to Seconds \newline ($10^2$~ms to s scale) & 
Exceptional (Strong zero-shot capabilities) & 
Infeasible & 
Offline analysis and delay-tolerant tasks (e.g., digital twins). \\ \hline

WLAM-Assisted \newline Dual-Loop & 
Sub-milliseconds (Execution) \newline Seconds (Update) & 
High (WLAM-optimized heuristics) & 
High & 
WLAM acts as a background optimizer to tune parameters for traditional solvers. \\ \hline
\end{tabular}
\end{table*}

\subsubsection{\textbf{Algorithmic and system level optimizations}}

To explicitly tackle the ``Memory Wall'' and reduce energy per token, hardware-aware algorithmic optimizations are paramount. Model compression techniques, including quantization, structured pruning, and low-rank approximation, are not merely about reducing parameter counts but are essential for minimizing the data volume transferred between memory and the NPU, thereby drastically lowering memory access energy \cite{rokh2023comprehensive, yeom2021pruning, elsken2019neural}. Furthermore, system-level optimizations, such as operator fusion and graph-level enhancements, reduce the overhead of kernel launching and memory I/O on specific hardware platforms \cite{hu2023convolutional, ren2018sbnet, jia2019taso}, facilitating efficient deployment in resource-constrained wireless environments.

\subsubsection{\textbf{Data centric efficiency strategies}}
Data-centric methods improve efficiency by selecting informative samples, which directly reduces the energy footprint \cite{glass-etal-2020-span, ivison-etal-2023-data}. More importantly, to mitigate the memory pressure from the KV cache during inference, advanced prompt engineering techniques such as prompt compression and gist token learning are employed. By condensing long historical contexts into compact representations, these methods significantly reduce the sequence length, lowering both the static memory consumption of the KV cache and the energy required for attention computation \cite{wei2022emergent, mu2024learning, autoprompt:emnlp20}.

\subsubsection{\textbf{Joint scheduling of horizontal and vertical scalability}}
To address the thermal throttling constraints on single edge devices, the integration of horizontal and vertical scalability offers a promising solution. While vertically boosting the computational power of an edge node (e.g., via dynamic frequency scaling or NPU acceleration) improves single-task acceleration, horizontal expansion offers superior network-wide efficiency. By offloading computation across various devices by using techniques like splitting inference, this method successfully distributes both thermal load and power consumption. A joint scheduling framework coordinates this allocation to achieve balanced workload distribution. This ensures that no single device exceeds its thermal design power limit while maintaining low latency, which is essential for the sustainable deployment of large AI models in dynamic wireless environments \cite{razavi2024talescalesreconcilinghorizontal, li2020elastic}.

\subsubsection{\textbf{Green 6G via WLAM}}
The massive computational energy footprint of WLAMs seemingly contradicts the sustainability imperatives of ``Green 6G" \cite{mao2021ai}. However, achieving true energy efficiency in an AI-empowered 6G era requires moving beyond the narrow optimization of AI models themselves to a holistic, network-wide perspective: leveraging WLAMs to orchestrate and drastically reduce the energy consumption of the entire telecommunications infrastructure. The fundamental philosophy is trading controllable computational energy for vastly more expensive RF radiated energy.
\par
Specifically, WLAMs enable this paradigm shift through two unique capabilities. First, by serving as the engine for semantic and intent-driven communications, WLAMs transition the network from transmitting redundant raw bits to exchanging highly compressed, goal-oriented knowledge \cite{qin2024ai}. This exponentially reduces the required on-air transmission time, allowing power-hungry RF chains and power amplifiers to remain in idle states for significantly longer periods \cite{hu2025energy}. Second, at the infrastructure level, WLAMs operating as autonomous agents can master the intricate spatio-temporal dynamics of network traffic \cite{jiang2026large}. Unlike reactive heuristic algorithms, agentic WLAMs can execute predictive, hyper-granular ``deep sleep" strategies across spatial, temporal, and frequency domains, intelligently powering down specific Massive MIMO antenna elements or entire cell sites milliseconds before traffic drops, without degrading the quality of service. Ultimately, in an AI-empowered 6G ecosystem, the WLAM ceases to be merely a power-consuming computational burden; instead, it transforms into the central energy-saving orchestrator, driving the telecommunications industry towards its carbon-neutral goals.

\subsection{Security and privacy issues}\label{sec:futures_sec}
While Sections \ref{WLAM_for_PLS} and \ref{PLS_for_WLAM} respectively detailed the dual relationship of using AI for PLS and PLS for AI, this section provides a holistic overview of the broader security and privacy challenges confronting the entire WLAM paradigm. In the following parts, we discuss the security and privacy issues, as well as the defenses.

\subsubsection{\textbf{Security issues}} 
Large AI models are vulnerable to various security attacks \cite{yao2024survey}. One well-known attack is the data poisoning attack, which involves injecting malicious data into the training dataset, causing the models to produce unreasonable and harmful results \cite{wan2023poisoning}. Another type of attack is the backdoor attack, which aims to embed a backdoor into the large AI models. When triggered with specific inputs, these backdoors can cause the AI models to perform unethical and illegal actions \cite{li2021hidden}.
Moreover, the explainability of large AI models is crucial for the security of wireless communications. A lack of explainability makes it difficult to understand the underlying operation principles and inference processes, potentially leading to anomalies in communications that are challenging to audit and debug. As wireless communications infrastructure is a foundational building block of society, unexpected and harmful outputs could result in significant and unbearable losses at both social and economic levels.
Therefore, ensuring the security of large AI models is paramount. Beyond patching specific vulnerabilities, future work needs to establish a comprehensive trustworthy AI framework for 6G, integrating formal verification methods to mathematically guarantee the safety and robustness of WLAM decisions in critical infrastructure.

\subsubsection{\textbf{Privacy issues}}
WLAM systems face significant privacy concerns arising from their deployment architecture and interaction modes \cite{brown2022does}. Unlike traditional on-device processing, the massive resource demands of WLAMs often necessitate cloud-based deployment, leading to data sovereignty risks where sensitive user context (e.g., location, channel logs) must be transmitted to third-party model providers. Furthermore, the inference process itself presents unique vulnerabilities; techniques like CoT prompting may inadvertently leak private intermediate reasoning steps or training data snippets in the output \cite{xiang2024badchain}. Additionally, the lack of standardized privacy protocols for agentic AI creates loopholes where private network configurations could be accessed by unauthorized entities during API calls. To address these systemic risks, future research must establish privacy-preserving inference frameworks leveraging trusted execution environments \cite{li2025teeslice} and homomorphic encryption \cite{de2024privacy}, ensuring that high-level intelligence can be delivered without compromising data confidentiality during the service process.

\subsection{Channel impact on model training}\label{sec:futures_channel}
A foundational challenge that permeates the entire WLAM paradigm is the profound and multifaceted impact of the physical wireless channel on AI model training and operation. While many AI algorithms are developed assuming ideal data and reliable connections, the wireless medium is actually characterized by noise, fading, latency, mobility induced dynamics, and limited bandwidth which fundamentally alters their performance. This challenge is not confined to a single domain but is a cross-cutting issue affecting both the applications of large AI models for wireless systems (Section 3) and the wireless frameworks enabling distributed large AI models (Section 4).

\subsubsection{\textbf{Impact on application-specific models}}
The challenges of the wireless channel manifest directly in the training process of large AI models designed for specific communication tasks.
\par
The first challenge is the disruption of gradient-based training by the channel. In tasks like end-to-end transceiver design, the wireless channel acts as a stochastic, non-differentiable layer within the neural network. This physical reality makes standard gradient-based training processes fundamentally nonviable, as it breaks the chain of gradient back-propagation required for optimization \cite{accessdecoding}. Consequently, the entire training paradigm must be re-designed, forcing the adoption of more complex strategies such as learning a differentiable surrogate channel model. These alternative strategies themselves introduce new challenges in terms of convergence stability and computational overhead \cite{10104549}.
\par
Secondly, dynamic environments pose severe challenges to training convergence and generalization. For applications in mobile environments, such as channel prediction and beamforming, the non-stationary nature of the channel poses a severe threat to the training process. A model trained on channel data from one time-instance or location may fail to generalize moments later as the user moves. This dynamic nature means that a static, offline training approach based on simple data fitting is insufficient. It necessitates a fundamental shift in the training methodology itself. Instead of relying solely on massive datasets to implicitly learn the dynamics, more advanced paradigms are required that explicitly model the underlying physical principles of the mobile channel. For example, recent schemes leverage physics-inspired approaches such as ODE to create models whose internal structure is designed to represent and predict the continuous-time characteristics of the channel \cite{10351043}. This transforms a straightforward supervised training task into a much more complex problem of training a physics-grounded dynamical system.
\par
Lastly, the training process is often compromised by the train-test mismatch caused by imperfect CSI. Many AI models for tasks such as resource allocation are trained on datasets generated with the assumption of perfect CSI. However, in practice, CSI is always imperfect due to estimation errors, noise, and feedback delays. A model trained under such idealized assumptions may converge perfectly to a solution that is optimal for the simulation, but it will lead to poor generalization and performance degradation in in reality. Overcoming this issue requires models that are robust to these unseen perturbations. For instance, advanced training strategies like model-driven meta-learning can alleviate this problem by optimizing the search trajectory itself \cite{zhurobustnetwork}, enabling the model to find solutions that are inherently resilient to the impact of channel estimation errors.
\par
To synthesize these solutions, future research must prioritize environment-aware adaptive training frameworks. By integrating differentiable channel modeling with physics-informed constraints directly into the optimization loop, WLAMs can bridge the gap between static model design and dynamic physical realities, ensuring generalization despite the stochastic and non-differentiable nature of the physical layer.

\subsubsection{\textbf{Impact on distributed training frameworks}}
The impact of the channel is equally critical for the performance and stability of the distributed training frameworks.
\par
Firstly, the channel introduces communication bottlenecks that directly impede the training process. Distributed frameworks like FL and SL require frequent exchange of large data payloads such as model parameters, gradients, and activations. The limited bandwidth of wireless channels creates a communication bottleneck, directly increasing the time required for each training round. This is explicitly captured in FL delay models, where transmission time is inversely proportional to bandwidth \cite{yang2020delay}. For SL, the cumulative nature of wireless latency over its sequential communication process creates an even more severe bottleneck for the overall training time.
\par
Secondly, channel impairments introduce errors that corrupt the training process. A clear example is in AirComp, which strategically uses the channel for aggregation. Its accuracy is highly vulnerable to channel impairments, as both channel noise and fading distort the superposition of signals at the receiver, introducing errors into the computed result \cite{cao2021optimized}. This physical layer imperfection directly translates into noise injected into the model training process, which can slow down convergence or increase the optimality gap of the final model.
\par
Lastly, the wireless channel creates vulnerabilities to adversarial attacks that compromise the training process. A malicious actor can exploit the open nature of the wireless medium to launch attacks like jamming. A jamming signal can corrupt the transmission of gradients or model updates, effectively blinding the central server. Research has established a direct mathematical link between the communication MSE over a jammed channel and the training loss of a large language model, formally demonstrating that an attacker can compromise the integrity of the entire training process by manipulating the physical channel \cite{federatedPHY}.
\par
Therefore, the future research pathway lies in establishing robust communication-learning integration. This requires designing aggregation protocols that are not only communication-efficient to bypass bandwidth bottlenecks but also mathematically resilient to physical layer impairments and adversarial jamming, thereby preserving the integrity of the training process against wireless instability.

\subsection{Fundamental trade-offs in WLAM}\label{sec:futures_tradeoff}
The integration of large AI models into wireless communication is not a simple matter of maximizing performance, but rather one of navigating fundamental and often conflicting trade-offs. These inherent tensions define the core challenges and future research landscape for WLAM. The following analysis explores three of the most critical trade-offs: the balance between performance and efficiency, the conflict between data utility and privacy, and the tension between autonomy and reliability.

\subsubsection{\textbf{Performance versus efficiency}}
A primary trade-off exists between the exceptional performance promised by WLAMs and the immense computational and energy demands they impose. While large models can achieve unprecedented accuracy in tasks like channel prediction or semantic understanding, their operational cost challenges deployment on resource-constrained wireless edge devices. Consequently, solutions like model quantization, pruning, or the joint scheduling of horizontal and vertical scaling are proposed. But they represent an inherent compromise, where peak model capability is deliberately traded for the energy efficiency and operational feasibility required in real-world wireless networks. This fundamental conflict indicates that a central future research direction for WLAM should be the pursuit of novel algorithms and hardware-software co-designs that push the Pareto frontier of this trade-off, enabling greater performance with less efficiency cost.

\subsubsection{\textbf{Data utility versus privacy}}
The second fundamental conflict arises between the need for high-fidelity data to ensure model utility and the necessity to protect user privacy. While traditional privacy research like FL focuses on decentralized training without raw data sharing, WLAMs introduce severe intrinsic privacy risks stemming from the models themselves. Specifically, the vast parameter space required for high performance creates a memorization-utility trade-off. The model's capacity to learn complex wireless patterns also enables it to unintentionally memorize sensitive training samples, such as user mobility traces or unique channel fingerprints. This leaves the system vulnerable to model inversion or membership inference attacks, where adversaries can reconstruct private training data solely by querying the deployed model. Furthermore, a new issue emerges during the inference phase of agentic AI. To achieve high utility, agents require rich contextual prompts, which inherently exposes user intent and network state to the model provider. This exposure is compounded by prompt injection vulnerabilities, where malicious actors exploit the model's flexible instruction following ability to bypass safety filters and attain confidential network configurations. Therefore, future research should focus on model-centric defenses such as machine unlearning \cite{li2025machine}, inference-time privacy guardrails \cite{tong2025inferdpt}, and robust alignment \cite{kumar2025robustness} to balance these conflicting demands.

\subsubsection{\textbf{Autonomy versus reliability}}
Finally, a critical trade-off emerges between the drive for network autonomy through agentic AI and the need for absolute reliability in critical infrastructure. The vision of a WLAM acting as an autonomous agent to manage a 6G network must be balanced against the severe risks posed by the flaws of large AI models. The potential for a WLAM to generate hallucinated network configurations, misinterpret high-level intents, or fall prey to backdoor attacks represents a profound threat to the stability and security of communication systems. Ensuring the reliability of these powerful but inherently probabilistic models is therefore a paramount challenge. Future research must prioritize the development of inherently explainable architectures to enable auditing, alongside trustworthy agentic frameworks that integrate formal verification and human-in-the-loop mechanisms to guarantee safe operation in critical infrastructure.

\subsection{Summary and insights}
In summary, this section has systematically analyzed the primary obstacles confronting WLAMs, progressing from specific technical domains to fundamental trade-offs. The challenges in creating datasets, designing network architectures, ensuring efficiency, and safeguarding security are not isolated issues but reflect deeper conflicts between performance and efficiency, data utility and privacy, and autonomy and reliability. Navigating these tensions requires a paradigm shift from isolated technical fixes to holistic, interdisciplinary solutions. Future research must focus on coordinating these conflicting demands by integrating hardware-software co-design, trustworthy agentic frameworks, and intrinsic privacy defenses into a unified architecture. The future of WLAM depends on developing these holistic solutions to bridge the gap between theoretical capabilities and operational realities.

\section{Conclusions}
This survey has provided a comprehensive exploration of the WLAM, charting its fundamentals, applications, and the path toward an AI-empowered 6G. Our key finding is the deeply synergistic and dual relationship where large AI models are revolutionizing wireless systems, and advanced wireless technologies are, in turn, enabling the deployment of large AI models. We have demonstrated that this transformation extends across the entire communication stack. This includes enhancing the physical layer with intelligent transceiver design, optimizing the network layer through automated resource management, and redefining efficiency at the semantic layer, ultimately creating a new agentic layer for autonomous network operations. Furthermore, we identified that future breakthroughs will be driven by the integration of emerging technologies, such as PINNs and next-generation sequence models, which promise to enhance data efficiency and computational performance.
\par
While the potential is immense, significant challenges remain. To guide future work, we highlight several critical research directions. First, there is a need for standardized multi-modal wireless datasets and synthetic data generation methods, which are essential for benchmarking and developing robust models that can fuse diverse data sources like CSI and sensor data. Second, a concerted effort is required to develop edge AI, focusing on novel, highly efficient model architectures and hardware-software co-design to break the ``memory wall'' and overcome the severe resource constraints of wireless devices. Third, as networks become more autonomous, research must prioritize trustworthy autonomy and formal verification, by developing methods to enhance the reliability, explainability, and security of WLAMs against threats like model hallucinations and backdoor attacks. Fourth, addressing intrinsic privacy risks requires exploring model-centric defenses, such as machine unlearning, to prevent the unintended memorization of sensitive user data. Finally, the deeper fusion of physics-informed AI with large AI models presents a fertile ground for creating more generalizable and data-efficient solutions grounded in the fundamental principles of wireless communication. Continued innovation in these areas is essential to harness the full power of WLAMs, shaping the intelligent, adaptive, and efficient future of 6G and beyond.

\Acknowledgements{The work was supported by National Natural Science Foundation of China under Grant 62331023, 62394292,  62425110 and 624B2133, Zhejiang Provincial Natural Science Foundation for Distinguished Young Scholars under Grant LRG26F010001, China National Key R\&D Program under Grant 2025ZD1301900, 2021YFA1000500 and 2023YFB2904804,  Fundamental Research Funds for the Central Universities and Zhejiang University Global Partnership Fund.}

\vspace{12pt}
\end{document}